\documentclass{article}
\usepackage{amsmath,amssymb,amsmath,amsthm,natbib,graphicx,anysize,epstopdf,eurosym,hyperref,lscape,dsfont,pdfsync,comment,color,geometry,subfigure,authblk,makeidx}
\hypersetup{colorlinks = true,
            linkcolor = blue,
            urlcolor  = blue,
            citecolor = blue
}
\newtheorem{theorem}{Theorem}
\newtheorem{condition}{Assumption}
\newtheorem{corollary}{Corollary}

\newtheorem{lemma}{Lemma}
\theoremstyle{definition} % making remark style not italicized text but bold title
\newtheorem{remark}{Remark}

\theoremstyle{plain}
\newtheorem{apptheorem}{Theorem}[section]
\newtheorem{applemma}{Lemma}[section]
 
\usepackage[title]{appendix}

 \usepackage{booktabs} % Table formatting
\usepackage{adjustbox} % Table size.
\usepackage{caption} % Table caption formatting
\usepackage{float} % Table positioning
\usepackage{silence} % Silencing math in section title warnings
\usepackage{pgfplots}
\usepgfplotslibrary{groupplots}
\pgfplotsset{compat=1.18}

\makeatletter
\pgfmathdeclarefunction{normcdf}{1}{%
  \begingroup
    \pgfmathparse{abs(#1)/sqrt(2)}%
    \let\x\pgfmathresult
    \pgfmathparse{1/(1+0.3275911*\x)}%
    \let\t\pgfmathresult
    \pgfmathparse{1-(((((1.061405429*\t-1.453152027)*\t+1.421413741)*\t-0.284496736)*\t+0.254829592)*\t)*exp(-\x*\x)}%
    \let\e\pgfmathresult
    \pgfmathparse{0.5*(1+sign(#1)*\e)}%
    \pgfmath@smuggleone\pgfmathresult
  \endgroup
}
\makeatother

\newcommand{\Phistd}[1]{normcdf(#1)}
\usepackage[symbol]{footmisc}
\renewcommand{\thefootnote}{\fnsymbol{footnote}}
\DeclareMathOperator\supp{supp}

\newcommand{\FYhat}{\widehat{F}_{Y}}
\newcommand{\R}{\mathds{R}}

\begin{document}
\begin{center}
\large \textsc{Estimation of distribution functions, their jumps and interval probabilities under measurement error}\normalsize\\[.2in]

\begin{tabular}{c}
\multicolumn{1}{c}{ \sc Kairat Mynbaev \rm} \\[.1in]
International School of Economics \\
Kazakh-British Technical University \\
Tolebi 59 \\
Almaty 050000, Kazakhstan \\
email: kairat\_mynbayev@yahoo.com \\
\end{tabular}\\[.1in]

\begin{tabular}{c}
\multicolumn{1}{c}{\sc Carlos Martins-Filho \rm} \\[.1in]
Department of Economics \\
University of Colorado  \\
Boulder, CO 80309-0256, USA  \\
email: carlos.martins@colorado.edu  \\
\end{tabular}\\[.1in]

and  \\[.1in]

\begin{tabular}{c}
\multicolumn{1}{c}{\sc Chad Brown \rm} \\[.1in]
Department of Economics \\
University of Manchester  \\
Manchester,  M13 9PL,  UK\\
email: chad.brown@manchester.ac.uk  \\
\end{tabular}\\[.1in]

August 2026\\[.1in]
\end{center}

% ---------------- %
%%% OLD ABSTRACT %%%
% ---------------- %
% \noindent \bf Abstract. \rm In the classical error-in-measurement model, where a random variable $Y$ is observed with error $Z$ that has a known distribution, we propose three new estimators for: a) the distribution function $F_Y$ of $Y$ at its points of continuity; b) interval probabilities $F_Y(y)-F_Y(x)$ for $x<y$; and c) the size of its jumps at points of discontinuity.  
% There exists a small literature that considers estimators for $F_Y$ at continuity points, but no estimator for jump discontinuities has been proposed under the classical error-in-measurement model.  We provide non-asymptotic bounds on the bias and variance of all estimators and establish their asymptotic unbiasedness and consistency.  These results rely on an interesting link between Fourier inversion theorems and the algebraic structure of a class of estimators proposed in \cite{Mynbaev2022}.  A simulation study provides experimental evidence on the finite sample performance of the proposed estimators. \\[.1in]

% ---------------- %
% CHADS ABSTRACT VERSION:
% ---------------- %
\noindent \bf Abstract. \rm
We consider the classical additive measurement-error model $X=Y+Z$,
where the latent random variable $Y$ has unknown distribution $F_Y$ and the 
% independent 
error $Z$ has a known distribution.
We develop direct estimators for three functionals of $F_Y$:
(i) $F_Y(x)$ at continuity points; 
(ii) interval probabilities $F_Y(y)-F_Y(x)$ when $x<y$ are continuity points;
and
(iii) the size of a jump at a prespecified discontinuity.
We derive non-asymptotic bias and variance bounds, and establish asymptotic unbiasedness and consistency.
Unlike previous work, we do not require $F_Y$ to admit a density, have a mixture representation, or satisfy global Sobolev smoothness assumptions.
The framework accommodates arbitrary latent distributions, including those with both discrete and continuous components, and distributions with multiple jumps. 
These results rely on a link between Fourier inversion theorems and the algebraic structure of a class of estimators proposed in \cite{Mynbaev2022}.
A simulation study evaluates feasible tuning procedures and, where available, compares the finite-sample performance of the proposed estimators with existing methods.
\\[.1in]

\noindent \bf Keywords and phrases. \rm Error-in-measurement model; nonparametric deconvolution estimation; distribution function jump estimation; Fourier inversion theorems.\\

\noindent \textbf{MSC 2020}: Primary 60E10,  62G05; secondary 62G20.\\[.25in]

\clearpage
\setlength{\baselineskip}{24pt} 
\pagestyle{plain} 
\setcounter{page}{1}
\renewcommand*{\thefootnote}{\arabic{footnote}}
\setcounter{footnote}{0}

%%%%%%%%%%%%%%%%%%%%%%%%%%%%%%%%%%
\section{Introduction}
%%%%%%%%%%%%%%%%%%%%%%%%%%%%%%%%%%

% ---------------- %
% CHADS INTRO ATTEMPT:
% ---------------- %
% ---------------------------------------------------------%
%%% 1. Model and targets: %%%
% State setup and the three targets.
% ---------------------------------------------------------%
Distribution-function, interval-probability, and point-mass estimation are fundamental statistical problems, but observations are often contaminated by measurement error.
We consider the classical additive measurement-error model $X:=Y+Z$ where $Y$ is a latent random variable with unknown distribution $F_Y$ and the measurement error $Z$ is independent of $Y$ and has a known distribution. 
Based on a random sample from $X$, we develop direct deconvolution estimators for three functionals of $F_Y$:
$F_Y(x)$ when $x$ is a continuity point; 
interval probabilities $F_Y(x,y) := F_Y(y)-F_Y(x)$ when $x<y$ and both are continuity points;
and
the size of a jump $p_x:= F_Y(x)-\underset{\epsilon \downarrow 0}{\lim} \,F_Y(x-\epsilon)$ when $x$ is a known point of discontinuity.

% ---------------------------------------------------------%
%%% 2. Distribution-function literature: %%%
%   Contrast density-based methods with Dattner et al. and 
%   Dattner–Reiser.
% ---------------------------------------------------------%
The deconvolution literature has primarily focused on estimating the \textit{density} of $Y$, when it exists; see \citet{Meister2009} and the references therein. Early estimators of the distribution function $F_Y$ were consequently obtained by integrating suitable density deconvolution estimators, as in \citet{Zhang1990}, \citet{Fan1991}, and \citet{Hall2008}. More recently, \citet{Dattner2011} and \citet{Dattner2013} used Fourier inversion representations of distribution functions (see \citealp{Gil-Pelaez1951}) to construct pointwise estimators of $F_Y$ without first estimating its density. 
These methods are more closely related to ours, but their bias and convergence-rate analyses assume that $F_Y$ possesses a density satisfying global Sobolev smoothness conditions. Thus, even direct distribution-function estimators have generally been studied under global regularity restrictions on the latent distribution.
The literature on interval probabilities remains even more limited; although they can be estimated by differencing existing pointwise estimators, direct estimators and their theoretical properties have received little attention.

% ---------------------------------------------------------%
%%% 3. Jump motivation and literature: %%%
%   Explain applications involving atoms, then discuss 
%   Gugushvili et al.
% ---------------------------------------------------------%
Estimation of jump sizes is important whenever the latent distribution possesses atoms, as occurs naturally for mixed discrete-continuous distributions, censored variables, heaped survey responses, and economic variables with point masses. While jump estimation is rather straightforward in the absence of measurement error, additive noise makes the recovery of jumps a substantially more difficult deconvolution problem.

Estimation of jumps in a latent distribution function under additive measurement error
has previously been considered by \cite{vanEs2008} and \cite{gugushvili2011}, who assume a single atom at the origin together with an absolutely continuous component, and by \cite{Lee2013}, who allow finitely many atoms and continuous components.
These approaches rely on a specified discrete-continuous mixture structure for $F_Y$ and regularity conditions on the continuous component. 
In contrast, we consider an arbitrary jump at any prespecified location of an arbitrary latent distribution. Apart from the target jump, no structural restrictions are placed on  $F_Y$: it may have countably many additional atoms and need not admit a discrete-continuous mixture representation with a globally smooth density.

The three estimators we propose are constructed within a kernel-regularization framework motivated by the generalized Fourier inversion theorem established in \citet{Mynbaev2022}. 
For each estimator, we establish non-asymptotic bias and variance bounds, asymptotic unbiasedness, and consistency.
These results impose no restrictions on $F_Y$ beyond continuity at the evaluation points for $F_Y(x)$ and $F_Y(x,y)$, or a prespecified jump location for $p_x$.
Our non-asymptotic mean squared error (MSE) bounds also yield explicit convergence rates when $F_Y$ satisfies mild local regularity and tail conditions.  These conditions do not require a density or mixture representation and impose no global Sobolev smoothness.
Thus, while much of the previous literature has focused on minimax rates over global smoothness classes, we provide finite-sample and asymptotic guarantees under substantially weaker assumptions on the latent distribution.
% Thus, different from previous work which has pursued minimax optimality over a global smoothness class, we provide finite-sample and asymptotic guarantees under substantially weaker restrictions on the latent distribution.
% Therefore, the principal distinction from previous work is that our results require no density or mixture representation and impose no global Sobolev smoothness conditions. Rather than pursuing minimax optimality over a global smoothness class, we provide finite-sample and asymptotic guarantees under substantially weaker restrictions on the latent distribution.
% Unlike previous work, our general results require no density or mixture representation and impose no global Sobolev smoothness conditions.
% 
% Thus, rather than pursuing minimax optimality over a global smoothness class, we provide finite-sample and asymptotic guarantees under substantially weaker restrictions on the latent distribution.
% This generality comes at the expense of not pursuing
% minimax optimality, but allows the methodology to be applied to a much
% broader class of latent distributions.
% Thus, different from previous work which has pursued minimax optimality over a global smoothness class, we provide finite-sample and asymptotic guarantees under substantially weaker restrictions on the latent distribution

% ---------------------------------------------------------%
%%% 5. Simulation %%%
% ---------------------------------------------------------%
A simulation study evaluates the finite-sample performance of all three estimators using feasible tuning procedures.
For the $F_Y(x)$ and $F_Y(x,y)$ estimators, we select tuning parameters using asymptotic rules and reconvolution cross-validation. For the $p_x$ estimator, we consider an asymptotic rule and a Goldenshluger-Lepski procedure for bandwidth selection. 
We compare our estimator of $F_Y(x)$ with those of \citet{Hall2008} and \citet{Dattner2013}; for interval probabilities $F_Y(x,y)$, we also include the implied estimators obtained by differencing their pointwise estimators.

% We select tuning parameters using asymptotic rules and reconvolution cross-validation for $F_Y(x)$ and $F_Y(x,y)$
% 
% A simulation study evaluates feasible tuning procedures and the finite-sample performance of all three proposed estimators. For the distribution-function and interval-probability estimators, we consider tuning based on asymptotic rules and reconvolution cross-validation; for the jump estimator, we consider an asymptotic rule and Goldenshluger--Lepski bandwidth selection.
% We compare our estimator of $F_Y(x)$ with those of \citet{Hall2008} and \citet{Dattner2013}; for interval probabilities, $F_Y(x,y)$, we also include the implied estimators obtained by differencing their pointwise estimators.

% A simulation study examines feasible implementation procedures and the finite-sample performance of the proposed estimators.
% We compare our estimator for $F_Y(x)$ with those of \citet{Hall2008} and \citet{Dattner2013}; for interval probabilities, $F_Y(x,y)$, we also include the implied estimators obtained by differencing their pointwise estimators.

The rest of the paper is organized as follows. Section~\ref{sec:motiv} provides the theoretical motivation for our estimators. 
Section~\ref{sec:DirDecEst} develops the main results for the pointwise distribution-function, interval-probability, and jump-size estimators.
Section~\ref{sec:sims} considers implementation and presents the simulation study.
Section~\ref{sec:conclusion} concludes and gives directions for future research. 
Appendix~\ref{app:proofs} contains proofs for the theoretical results in the main text,
Appendix~\ref{app:noncompH} provides extensions to non-compact regularization kernels,
and Appendix~\ref{app:sims} provides details on the simulation study and implementation of the proposed estimators.

\noindent \bf Notation. \rm 
We adopt the following notation throughout the paper. 
$F$ denotes a (proper) distribution function, and when necessary we write $F_X$ to denote the distribution function associated with the random variable $X$. 
$\chi_S$ is the indicator function for the set $S$. 
$C(f)$ denotes the points of continuity of the function $f$, and $J(f)$ denotes the points where $f$ has a jump discontinuity.
% 
% For $f:\R\to\R$ and $h>0$ we write $\omega_f(x,h):=\underset{0<u\le h}{\sup} \left| f(x+u)-f(x-u) \right|$.
% 
$C_b$ denotes the space of uniformly bounded continuous functions on $\mathds{R}$ with norm $\|f\|_{C_b}:=\underset{x \in \mathds{R}}{\sup}|f(x)|$; 
$L(\R)$ denotes the space of measurable functions $f:(\R, \mathcal{B}) \to (\R, \mathcal{B})$ with norm $\|f\|_{L}=\int_{\R}|f(t)|dt < \infty$ where $\mathcal{B}$ are the Borel sets of $\R$.
$\mathcal{F}$ and $\mathcal{F}^{-1}$ denote, respectively, the Fourier and inverse Fourier transforms, where
for a function $f \in L(\R)$, $\left(\mathcal{F}f \right)(t):=\int_{\R}e^{ist}f (s)ds$ and $\left(\mathcal{F}^{-1}f \right)(t):=\frac{1}{2\pi }\int_{\R}e^{-ist}f (s)ds$, 
and
for a distribution function $F$, $\left(\mathcal{F}F \right)(t):=\int_{\R}e^{ist}dF(s)$ and $\left(\mathcal{F}^{-1}F \right)(t)=\frac{1}{2\pi }\int_{\R}e^{-ist}dF(s)$.
$f \star g$ denotes the convolution of $f$ and $g$, where $f$ and $g$ may be functions or measures. 
For a function $f:\R \to \R$, $\supp \, f$ denotes its support.
For a set function $G$ defined on intervals $[a,b]\subset \R$, we write
$\|G\|_\infty := \underset{a,b\in\R, a<b}{\sup}|G([a,b])|$,
and
\begin{equation}\label{phiGdef}
    \phi _{G}(N) :=\max \left\{ \underset{b<-N}{\sup}|G([a,b])|,\ \underset{a>N}{\sup}|G([a,b])|,\
\underset{a<-N,\ b>N}{\sup}|G([a,b])-1|\right\}\quad \mbox{for}\;\; N>0.
\end{equation}
For $f:\R\to\R$ and $h>0$ we write $\omega_f(x,h):=\underset{0<u\le h}{\sup} \left| f(x+u)-f(x-u) \right|.$

%%%%%%%%%%%%%%%%%%%%%%%%%%%%%%%%%%
\section{Motivation for the new deconvolution estimators}\label{sec:motiv}
%%%%%%%%%%%%%%%%%%%%%%%%%%%%%%%%%%

The estimators proposed in this paper are motivated by a general class of nonparametric estimators introduced in \cite{Mynbaev2022}, together with a generalized Fourier inversion theorem.  \cite{Mynbaev2022} considered the estimation of $F(x,y):=F(y)-F(x)$ for $x<y$ and $x,y \in C(F)$ based on a random sample $\{X_j\}_{j=1}^n$, $n \in \mathds{N}$, where $F$ is the distribution function associated with the random variable $X$.  In a setting where there is no measurement error they suggested the  nonparametric estimator  
\begin{equation}
\tilde{F}(x,y)=\frac{1}{n}\sum_{i=1}^{n}G\left( \left[ \frac{X_{i}-y}{h},\frac{X_{i}-x}{h}\right] \right),  \label{1}
\end{equation}
where $h>0$ is a bandwidth and $G$ is a function on intervals $[a,b]$, where $a,b \in \R$.  They showed that if $G$ satisfies 
% % \begin{equation*}
% % \lim_{a\rightarrow -\infty ,b\rightarrow \infty
% % }G([a,b])=1,\lim_{a\rightarrow \infty }G([a,b])=0,\lim_{b\rightarrow -\infty
% % }G([a,b])=0 \mbox{ and $G$ is bounded,}
% % \end{equation*}
% \begin{equation*} %\label{Gcond1}
% \lim_{a\rightarrow -\infty ,b\rightarrow \infty}
% G([a,b])=1,
% \quad
% \lim_{a\rightarrow \infty }G([a,b])=0,
% \quad
% \lim_{b\rightarrow -\infty}G([a,b])=0,
% \quad
% \mbox{and}
% \quad
% \left\Vert G\right\Vert _{\infty} < \infty.
% % \mbox{ $G$ is bounded,} 
% \end{equation*}
$\left\Vert G\right\Vert _{\infty} < \infty$
and
$\lim_{N\to\infty}\phi _{G}(N)=0$,
then for any $h>0$ and $\delta \in (0,1)$,  
\begin{equation}
\left\vert E \left(\tilde{F}(x,y)\right)-F(x,y)\right\vert \leq \phi _{G}(h^{\delta
-1})+\left( 1+\left\Vert G\right\Vert _{\infty}\right) \left[ \omega_F (x,h^{\delta
})+\omega_F (y,h^{\delta })\right]\!.  \label{4}
\end{equation}
% where 
% $\phi_G$ is defined as in \eqref{phiGdef}.
% $\phi _{G}(N) =\max \left\{ \underset{b<-N}{\sup}|G([a,b])|,\ \underset{a>N}{\sup}|G([a,b])|,\ \underset{a<-N,\ b>N}{\sup}|G([a,b])-1|\right\}$ for $N>0$.  
It follows immediately from \eqref{4} that, for any distribution function $F$ and any $x,y\in C(F)$ with $x<y$,
\begin{equation}\label{2}
E\left( \tilde{F}(x,y) \right)\rightarrow F(x,y),\quad\text{as }h\rightarrow 0
% \text{ for all }
% F,~\ x,y\in C(F),\ x<y
.  
\end{equation}
Although the estimator \eqref{1} was developed independently of Fourier methods, its expectation admits a representation that coincides with a generalized Fourier inversion formula. This observation provides the key motivation for the deconvolution estimators proposed in this paper.  Specifically, \cite{Mynbaev2022} identified a link between $E(\tilde{F}(x,y))$ and classical Fourier inversion theorems.  Recall that L\'evy's inversion theorem (see \citealp{Lukacs1970}) states that  
\begin{equation}
\frac{1}{2\pi }\int_{-1/h}^{1/h}\frac{e^{-ixt}-e^{-iyt}}{it}(\mathcal{F}F)
(t)dt\rightarrow F(x,y),\mbox{ as } h\rightarrow 0,  \label{7}
\end{equation}
and Borovkov's inversion theorem (see \citealp{Borovkov2013}) gives
\begin{equation}
\frac{1}{2\pi }\int_{\R}\frac{e^{-ixt}-e^{-iyt}}{it}e^{-h^{2}t^{2}}(\mathcal{F}F)
(t)dt\rightarrow F(x,y),\ h\rightarrow 0.  \label{8}
\end{equation}
The left-hand sides of \eqref{7} and \eqref{8} are special cases of
$
\frac{1}{2\pi }\int_{\R}\frac{e^{-ixt}-e^{-iyt}}{it}H(ht)(\mathcal{F}F) (t)dt
$ for $H\in L(\R)$.  The function $H(h \cdot)$ plays the role of a regularization kernel. Different choices of $H$ recover classical Fourier inversion formulas and lead to different deconvolution estimators.  \cite{Mynbaev2022} showed that if $G([a,b])=\frac{1}{2\pi}\int_a^b(\mathcal{F}H)(u)du$, then
\begin{equation}
E\left(\tilde{F}(x,y) \right)=\int_{\R}G\left( \left[ \frac{t-y}{h},\frac{t-x}{h}\right]
\right) dF(t)=\frac{1}{2\pi }\int_{\R}\frac{e^{-ixt}-e^{-iyt}}{it}H(ht)(\mathcal{F}F)
(t)dt.  \label{11}
\end{equation}
The asymptotic unbiasedness of $\tilde{F}(x,y)$ in equation \eqref{2} and the link revealed in equation \eqref{11} provides the motivation for the estimators we propose in subsequent sections.

%%%%%%%%%%%%%%%%%%%%%%%%%%%%%%%%%%
\section{Direct deconvolution estimators}\label{sec:DirDecEst}
%%%%%%%%%%%%%%%%%%%%%%%%%%%%%%%%%%

This section defines and provides the main results for the three direct deconvolution estimators. Throughout,
suppose $X,\,Y,$ and $Z$ are random variables defined on the probability space $(\Omega, \mathcal{F},P)$ with   
\begin{equation}\label{sum}
X=Y+Z
\end{equation}
and let $F_Y$ be the distribution function of $Y$. Estimation is to be conducted based on a random sample $\{X_j\}_{j=1}^n$ of observations on $X$, which has distribution function $F_X$.  $Z$ is an unobserved measurement error, independent of $Y$, with known distribution function denoted by $F_Z$.
% and satisfies $\Phi_{Z}(t) \neq 0$ for all $t \in \R$. 

%%%%%%%%%%%%%%%%%%%%%%%%%%%%%%%%%%
\subsection{Estimation of $F_Y(x)$ for $x \in C(F_Y)$}\label{sec:FYx}
%%%%%%%%%%%%%%%%%%%%%%%%%%%%%%%%%%
 
We consider the estimation of $F_Y(x)$ where $x \in C(F_Y)$.
Independence of $Y$ and $Z$ and equation \eqref{sum} imply that $F_X=F_Y\star F_Z$.  Furthermore, if $\Phi_X:=\mathcal{F}F_X$ denotes the Fourier transform of $F_X$, we have $\Phi _{X}=\Phi _{Y}\Phi_{Z}$.  If $\Phi_{Z}(t) \neq 0$ for all $t \in \R$ we can write $\Phi _{Y}(t)=\Phi_X(t)/\Phi_{Z}(t)$.  If $\Phi_X/\Phi_{Z}$ is integrable we have $F_Y(x) =\left( \mathcal{F}^{-1}\frac{\Phi _{X}}{\Phi _{Z}}\right)(x)$.  In the general case where $\Phi_X/\Phi_{Z}$ is not integrable, we regularize it by multiplying by $H(h\cdot)$ with $h>0$, $H \in L(\R)$ chosen in such a way as to have
\begin{equation}\label{moteq}
F_Y(x)=\underset{h\to 0}{\lim}\,\mathcal{F}^{-1}\left\{\frac{\Phi _{X}}{\Phi _{Z}}H(h\cdot)\right\}(x).
\end{equation}
Equations \eqref{11} and \eqref{moteq} motivate our estimator for $F_Y(x)$.  
% 
% Let $\{X_j\}_{j=1}^n$ be a random sample from $F_X$ and 
For $h,\,\lambda>0$ define 
\begin{equation}\label{decon}
\FYhat(x):=\frac{1}{n}\sum_{j=1}^{n}\! K_{h,\lambda }\!\left(
X_{j}-x\right)\!,
\;\; \mbox{where} \;\;\;
K_{h,\lambda }(t):=\frac{1}{2\pi }\int_{\R}e^{its}\alpha _{h,\lambda}\!\left( s\right) ds
, 
% \;\; \mbox{and} \;\; 
\quad
\alpha _{h,\lambda }(s):=\frac{e^{is/\lambda }-1}{is}\frac{H(hs)}{\Phi _{Z}(s)}. 
\end{equation}
We make the following assumption to ensure that $\FYhat(x)$ is well defined.
%%%%%%%%%%%%%%%%%%%%%%%%%%%%%%%%%
% Assumption A1
%%%%%%%%%%%%%%%%%%%%%%%%%%%%%%%%%
    \begin{condition}\label{A1}
    1. (Identifiability) $\Phi _{Z}(s)\neq 0$ for each $s\in \R$; 2. (Regularity) $\alpha _{h,\lambda } \in L(\R)$ for each $h,\,\lambda >0$; 3. (Symmetry) $H$ is even and real valued.
    \end{condition}
% Given Assumptions \ref{A1}.1 and \ref{A1}.2, for each $h,\lambda >0$ the function $K_{h,\lambda }$ is continuous and bounded. 
% % It follows that $\FYhat(x)$ is continuous and almost surely bounded as a function of $x$, giving $E\left(\FYhat(x)\right)<\infty$.  
% Given Assumption \ref{A1}.3, 
% \begin{equation*}
% \overline{\alpha _{h,\lambda }(s)}=\frac{e^{-is/\lambda }-1}{-is}\frac{H(hs)}{\overline{\Phi _{Z}(s)}}=\frac{e^{-is/\lambda }-1}{-is}\frac{H(-hs)}{\Phi_{Z}(-s)}=\alpha _{h,\lambda }(-s)
% \end{equation*}
% so $K_{h,\lambda }(t)$ is real-valued since
% \begin{equation*}
% \overline{K_{h,\lambda }(t)}=\frac{1}{2\pi }\int_{\R}e^{-its}\overline{\alpha
% _{h,\lambda }\left( s\right) }ds=\frac{1}{2\pi }\int_{\R}e^{-its}\alpha
% _{h,\lambda }\left( -s\right) ds=K_{h,\lambda }(t).
% \end{equation*}
% Hence, under Assumption \ref{A1},  $\FYhat(x)$ is continuous, almost surely bounded,  real-valued, and  $E\left(\FYhat(x)\right)<\infty$.  

Given Assumptions \ref{A1}.1 and \ref{A1}.2, the function $K_{h,\lambda }$ is continuous and bounded for each $h,\lambda >0$. 
% It follows that $\FYhat(x)$ is continuous and almost surely bounded as a function of $x$, giving $E\left(\FYhat(x)\right)<\infty$.  
Given Assumption \ref{A1}.3, 
% $\alpha _{h,\lambda }(t)$ satisfies
\begin{equation*}
\overline{\alpha _{h,\lambda }(s)}
% =\frac{e^{-is/\lambda }-1}{-is}\frac{H(hs)}{\overline{\Phi _{Z}(s)}}
=\frac{e^{-is/\lambda }-1}{-is}\frac{H(-hs)}{\Phi_{Z}(-s)}
=\alpha _{h,\lambda }(-s),
\quad \text{ and } \quad
\overline{K_{h,\lambda }(t)}
=\frac{1}{2\pi }\int_{\R}e^{-its}\alpha_{h,\lambda }\left( -s\right) ds
=K_{h,\lambda }(t),
\end{equation*}
so $K_{h,\lambda }(t)$ is real-valued.
Hence, under Assumption \ref{A1},  $\FYhat(x)$ is continuous, almost surely bounded,  real-valued, and  $E\left(\FYhat(x)\right)<\infty$.

The algebraic structure of $\FYhat(x)$ is explained by the following lemma. 
This 
makes precise the connection between
$\FYhat(x)$ and the generalized Fourier inversion representation
in Section~\ref{sec:motiv}.
For $H\in L(\R)$, define
$$
G_H([a,b])
    :=\frac{1}{2\pi}\int_a^b(\mathcal FH)(v)\,dv,
    \qquad \mbox{for $a,b\in \R$,\, $a<b$.}
$$
% We note that since $H \in L(\R)$, by standard Fourier transform theory, $\mathcal{F}H\in C_b(\R)$ and $G_H([a,b])$ is well defined as a finite Lebesgue and Riemann integral.
We note that $H \in L(\R)$ implies $\mathcal{F}H\in C_b(\R)$ so $G_H([a,b])$ is well defined.
% as a finite Lebesgue and Riemann integral.

%%%%%%%%%%%%%%%%%%%%%%%%%%%%%
%% lemma
%%%%%%%%%%%%%%%%%%%%%%%%%%%%%
\begin{lemma}\label{lem:algebra_FYx}\label{lem1}
Let  $H\in L(\R)$. 
For any distribution function $F$, $h,\lambda>0$, and $x\in\R$,
\[
\frac{1}{2\pi}\int_{\R}
    \frac{e^{-it(x-1/\lambda)}-e^{-itx}}{it}
    H(ht)(\mathcal FF)(t)\,dt
=
\int_{\R}
G_H\!\left(
\left[
\frac{u-x}{h},
\frac{u-(x-1/\lambda)}{h}
\right]\right)dF(u).
\]
Under the measurement-error model and Assumption~\ref{A1}, taking $F=F_Y$
gives
\[
E\left(\FYhat(x)\right)
=
\int_{\R}
G_H\!\left(
\left[
\frac{u-x}{h},
\frac{u-(x-1/\lambda)}{h}
\right]\right)dF_Y(u).
\]
\end{lemma}

Building on the approach of \cite{Mynbaev2022} discussed in Section \ref{sec:motiv}, the next assumption is used to obtain a finite-sample bias bound and asymptotic unbiasedness. 

%%%%%%%%%%%%%%%%%%%%%%%%%%%%%
%% Assumption
%%%%%%%%%%%%%%%%%%%%%%%%%%%%%
\begin{condition}\label{AG}
(Inversion kernel)
$H\in L(\R)$ and the interval function $G_H$ satisfies 
$\left\Vert G_H\right\Vert _{\infty} < \infty$
and
$\lim_{N\to\infty}\phi _{G_H}(N)=0$ where
% , for any interval function $G$ and $N>0$,
% \begin{equation}\label{phiGdef}
%     \phi _{G}(N) :=\max \left\{ \underset{b<-N}{\sup}|G([a,b])|,\; \underset{a>N}{\sup}|G([a,b])|,\;
% \underset{a<-N,\ b>N}{\sup}|G([a,b])-1|\right\}
% % \quad \mbox{for}\;\; N>0
% .
% \end{equation}
$\phi _{G}$ is defined in \eqref{phiGdef}.
% \[
% \|G_H\|_\infty<\infty
% \qquad\text{and}\qquad
% \varphi_{G_H}(N)\longrightarrow0
% \quad\text{as }N\to\infty.
% \]
\end{condition}

%%%%%%%%%%%%%%%%%%%%%%%%%%%%%
%% Remark
%%%%%%%%%%%%%%%%%%%%%%%%%%%%%
\begin{remark}\label{rem:inversion_kernel}
For generality, the conditions in Assumption~\ref{AG} are stated in terms of $G_H$.
% because it is weaker than the usual absolute-integrability conditions
% on $\mathcal FH$. 
A convenient sufficient condition stated directly in terms of $H$ is
\[
    H\in L(\R),\qquad
    \mathcal FH\in L(\R),\qquad
    H(0)=1,
\]
with $H$ continuous at zero. These conditions are not necessary. For
example, 
% the Fourier cutoff 
$H=\chi_{[-1,1]}$ satisfies
Assumption~\ref{AG}, although
$
    (\mathcal FH)(u)
    =
    % \frac{2\sin u}{u}
    {2\sin (u)}/{u}
    \notin 
    L(\R).
$
Examples of kernels that satisfy Assumption \ref{AG} 
include the Gaussian, Laplace, and Bartlett kernels, 
% include: 
% $H(t)=\exp(-\frac{1}{2}t^2)$ (Gaussian), $H(t)=\exp(-|t|)$ (Laplace), $H(t)=(1-|t|)\chi_{[-1,1]}(t)$ (Bartlett),
as well as $H(t)=(1-t^2)^4\chi_{[-1,1]}(t)$
which will be used for the simulation study in Section~\ref{sec:sims}.
The assumption that $G_H$ is a bounded function of intervals $[a,b]$ is added since it does not follow from $\mathcal{F}H \in C_b(\R)$.
If, in addition, $\mathcal{F}H \ge 0$ then $\frac{1}{2\pi}(\mathcal{F}H)$ is a density function associated with the probability measure $G_H$. 

\end{remark}

%%%%%%%%%%%%%%%%%%%%%%%%%%%%%%%%%%%%%%%%
% Theorem 1
%%%%%%%%%%%%%%%%%%%%%%%%%%%%%%%%%%%%%%%%
\begin{theorem}\label{thm7}
Suppose Assumptions \ref{A1} and \ref{AG} hold. 
Then, for any $F_{Y},$ and $x\in C(F_{Y})$
\begin{equation} \label{unqua2}
\left|E\left(\FYhat(x)\right)-F_{Y}(x)\right|\rightarrow 0,
\quad \mbox{ as $h,\,\lambda \rightarrow 0$.  
% for all $F_{Y},\ x\in C(F_{Y})$
} 
\end{equation}
Furthermore, for any $\delta \in (0,1)$, $h \in (0,1]$ and $\lambda >0$ such that $\lambda h^\delta \le 1/2$, we have the following non-asymptotic bound
\begin{equation}\label{gen_bound2}
\left|E\left(\FYhat(x)\right)-F_{Y}(x)\right| \leq \phi _{G_H}(h^{\delta -1})+(1+\left\Vert
G_H\right\Vert _{\infty})\omega_{F_Y} (x,h^{\delta })+(2+\left\Vert G_H\right\Vert _{\infty})F_{Y}\left(x-\frac{1}{\lambda} +1\right).
\end{equation}
% where $\phi _{G_H}(N) =\max \left\{ \underset{b<-N}{\sup}|G_H([a,b])|,\ \underset{a>N}{\sup}|G_H([a,b])|,\ \underset{a<-N,\ b>N}{\sup}|G_H([a,b])-1|\right\} $.
\end{theorem}

We note that the bound on the bias of $\FYhat(x)$ does not rely on any assumption on the distribution $F_Y$. Unlike \cite{Dattner2011} and \cite{Dattner2013}, whose bias analysis requires Sobolev smoothness of $F_Y$, inequality \eqref{gen_bound2} holds without assuming that $F_Y$ possesses a density.

%%%%%%%%%%%%%%%%%%%%%%%%%%%%%
%% Remark
%%%%%%%%%%%%%%%%%%%%%%%%%%%%%
\begin{remark}[Generalized Fourier inversion]\label{r2lem2}%\label{rem:point_inversion}
The proof of Theorem~\ref{thm7}, applied to an arbitrary
distribution function $F$, also gives
\[
F(x)
=
\lim_{h,\lambda\to0}
\frac{1}{2\pi}\int_{\R}
    \frac{e^{-it(x-1/\lambda)}-e^{-itx}}{it}
    H(ht)(\mathcal FF)(t)\,dt,
    \qquad x\in C(F).
\]
and the non-asymptotic bias bound in Theorem~\ref{thm7} provides a rate of convergence.
\cite{Adell2003} also obtained a rate but under smoothness conditions on $F$, viz., second-order moduli of smoothness for continuous functions or functions of bounded variation.  Our bound is expressed in terms of the local oscillation of the distribution function and applies when no density exists.
% non-asymptotic bound on the approximation error in this generalized
% Fourier inversion identity. Unlike bounds based on global smoothness
% conditions, its dependence on $F$ is through its local oscillation at
% $x$ and its lower tail.
\end{remark}

%%%%%%%%%%%%%%%%%%%%%%%%%%%%%
%% Remark
%%%%%%%%%%%%%%%%%%%%%%%%%%%%%
\begin{remark}[Necessity]\label{remk1}%\label{rem:necessity_FYx}
Suppose Assumption \ref{A1} holds, $H\in L(\R)$, and
$\mathcal FH\geq0$. Then the following statements are equivalent:
\begin{enumerate}
    \item Assumption~\ref{AG} holds;
    \item
    $
    |E(\FYhat(x))-F_Y(x)|\rightarrow 0
    $
    as $h,\,\lambda \rightarrow 0$,
    for every distribution function $F_Y$ and every
    $x\in C(F_Y)$;
    \item $H$ is such that
    % \[
    % \mathcal FH\in L(\R),
    % \qquad
    % \frac{1}{2\pi}\int_{\R}(\mathcal FH)(u)\,du=1.
    % \]
    \begin{equation}\label{73.1}
        \mathcal{F}H\in L(\R),
        \quad \mbox{ and }  \quad
        \frac{1}{2\pi }\int_{\R}(\mathcal{F}H)(u)du=1.
    \end{equation}
\end{enumerate}

The implication $1\Rightarrow2$ follows from Theorem~\ref{thm7}, while
$3\Rightarrow1$ follows from the nonnegativity of $\mathcal FH$.
To prove $2\Rightarrow3$, take $F_Y$ to be the distribution function
of a point mass at zero and let $x=1$.
Then, 
\[ 
\int_{\R} G_H\left( \left[ \frac{t-1}{h},\frac{t-1+1/\lambda }{h}\right] \right)
dF_Y(t)=G_H\left( \left[ -\frac{1}{h},\frac{-1+1/\lambda}{h}\right]
\right).
\]
By Lemma~\ref{lem1}, choosing
$h_n=\lambda_n=1/n$ gives
% \[
$
G_H([-n,n^2-n])
=
E[\FYhat(1)]
\rightarrow 1
$
as $n\to \infty$.
% \quad 
% \mbox{as $n\to \infty$}.
% \]
Because 
% $[-n,n^2-n]\uparrow\R$ 
$\lim_{n\to\infty}[-n,n^2-n] = \R$
and $\mathcal FH\geq0$,
the monotone convergence theorem yields
$
\frac{1}{2\pi}\int_{\R}(\mathcal FH)(u)\,du=1,
$
and hence $\mathcal FH\in L(\R)$.
\end{remark}

The next theorem provides a bound for the variance of $\FYhat(x)$ and gives a sufficient condition for the convergence of $\FYhat(x)$ in quadratic mean.  This, in turn, implies consistency of $\FYhat(x)$.
%%%%%%%%%%%%%%%%%%%%%%%%%%%%%%%%%
%% Theorem 8
%%%%%%%%%%%%%%%%%%%%%%%%%%%%%%%%%
\begin{theorem}\label{thm8} 
Suppose Assumption \ref{A1} holds. Then, 
\begin{equation}
V\left(\FYhat(x)\right) \le \frac{1}{n}\frac{v_{h,\lambda}^2}{4 \pi^2}, \quad \mbox{ where } \quad v_{h,\lambda }:=\int_{\R}|\alpha _{h,\lambda }(t)|dt.
\end{equation} 
If, in addition, Assumption \ref{AG} holds, and 
$h:=h_{n} \to 0$, $\lambda :=\lambda _{n} \to 0$ are chosen such that $v_{h,\lambda }/\sqrt{n}\rightarrow 0$ as $n\rightarrow \infty $,
% then for $x\in C(F_{Y})$
then,
for any distribution function $F_Y$ and any $x\in C(F_{Y})$, we have 
$\FYhat(x)\overset{p}{\rightarrow }F_{Y}(x)$.
% Then, if the conditions on Theorem \ref{thm7} hold, $\FYhat(x)\overset{p}{\rightarrow }F_{Y}(x)$.
\end{theorem}
% 
% \begin{theorem}\label{thm8} 
% Under Assumption \ref{A1}, 
% \begin{equation}
% V\left(\FYhat(x)\right) \le \frac{1}{n}\frac{v_{h,\lambda}^2}{4 \pi^2}, \quad \mbox{ where } \quad v_{h,\lambda }:=\int_{\R}|\alpha _{h,\lambda }(t)|dt.
% \end{equation} 
% Suppose $h:=h_{n} \to 0$, $\lambda :=\lambda _{n} \to 0$ are chosen so that $v_{h,\lambda }/\sqrt{n}\rightarrow 0$ as $n\rightarrow \infty $.  Then, if the conditions on Theorem \ref{thm7} hold, $\FYhat(x)\overset{p}{\rightarrow }F_{Y}(x)$.
% \end{theorem}
%%%%%%%%%%%%%%%%%%%%%%%%%%%%%%%%%%%%%%%%%%%
The usefulness of Theorem \ref{thm8} depends on obtaining conditions under which $v_{h,\lambda}/\sqrt{n} \to 0$ as $n \to \infty$.  The following lemmas give conditions to obtain the order of $v_{h,\lambda}$.  
 Lemma \ref{lem6a} provides results when $H$ has compact support, and Lemma \ref{lem6a_NoncompH} in Appendix \ref{app:noncompH} provides an analogous result when $H$ does not have compact support.
Lemmas \ref{lem6a} and \ref{lem6a_NoncompH} both obtain orders that rely solely  on conditions imposed on $H$.

%%%%%%%%%%%%%%%%%%%%%%%%%%%%%%%%%%%%%%%%%%
\begin{lemma}\label{lem6a} 
Suppose Assumption \ref{A1} holds, $\supp H \subset [-\gamma,\gamma]$ for some $\gamma >0$, and 
% $\underset{t \in [-\gamma,\gamma]}{\sup}|H(t)|<\infty$.  
$C_H:=\underset{t \in [-\gamma,\gamma]}{\sup}|H(t)|<\infty$.
Then, there exist constants $C>0$ and $\beta\in(0,\pi/2)$, depending only on $C_H$ and $\Phi_Z$, such that for all $\lambda \in (0,\beta/\pi)$ and $h\in(0,\gamma/\beta)$, 
$$
 v_{h,\lambda} 
 \leq
 C\left[ \log \lambda^{-1}+  \underset{ \beta < t\le\gamma/h}{\int} \frac{1}{t|\Phi_Z(t)|}dt\right].
 $$ 
\end{lemma}
%%%%%%%%%%%%%%%%%%%%%%%%%%%%%%%%%%%%%%%%%%%
The order of $\underset{ \beta < t\le\gamma/h}{\int} \frac{1}{t|\Phi_Z(t)|}dt$
% and $\underset{ \beta < t\le t(h)}{\int} \frac{|H(ht)|}{t|\Phi_Z(t)|}dt$ 
depends on the behavior of $|\Phi_Z(t)|$ as $|t| \to \infty $.  Following \cite{Fan1991}, \cite{Dattner2011} and \cite{Dattner2013} we consider two cases:
    \begin{subequations}\label{eq:error_smoothness}
    \begin{align} 
    \text{(super-smooth errors)} &\qquad  |\Phi_Z(t)| \asymp \exp(-\tau |t|^\rho), \qquad \text{as } |t|\to\infty, \label{eq:error_smoothness_super} \\ 
    \text{(ordinary-smooth errors)} & \qquad |\Phi_Z(t)| \asymp |t|^{-\rho}, \qquad \text{as } |t|\to\infty, \label{eq:error_smoothness_ordinary} 
    \end{align} 
    \end{subequations}
where $\tau>0$ and $\rho>0$ in \eqref{eq:error_smoothness_super}, and $\rho>0$ in \eqref{eq:error_smoothness_ordinary}.
% The next lemma considers the order of $\underset{ \beta < t\le\gamma/h}{\int} \frac{1}{t|\Phi_Z(t)|}dt$ under both smoothness conditions.
% %%%%%%%%%%%%%%%%%%%%%%%%%%%%%
% %% lemma
% %%%%%%%%%%%%%%%%%%%%%%%%%%%%%
% \begin{lemma}\label{lem7a}
% Let $0<\beta<\xi_n$ for $n \in \mathds{N}$ with $\xi_n \to \infty$ as $n \to \infty$ and 
% $I(\xi_n):=\int_{\beta <t \le {\xi_n}}\frac{1}{t|\Phi_Z(t)|}dt$.  
% % $\underset{ \beta <t \le {\xi_n}}{\int}\frac{1}{t|\Phi_Z(t)|}dt$.
% \begin{enumerate}
% \item[(a)]
% % a)  
% If there exist $\tau, \rho>0$ such that 
% \eqref{eq:error_smoothness_super} holds,
% % $|\Phi_Z(t)| \asymp \exp (-\tau |t|^{\rho})$ as $|t| \to \infty$,  
% then $I(\xi_n)=O\left( \frac{ \exp(\tau \xi_n^\rho)}{\tau \xi_n^\rho} \right)$ and if $\xi_n \le \left( \frac{\log n}{2 \tau} \right)^{1/\rho}$, $I(\xi_n)=o(n^{1/2})$. 
% % b)
% \item[(b)]
% If there exists $\rho>0$ such that 
% % $|\Phi_Z(t)| \asymp |t|^{-\rho}$ as $|t| \to \infty$, 
% \eqref{eq:error_smoothness_ordinary} holds,
% then $I(\xi_n) = O\left( \xi_n^\rho  \right)$ and if $\xi_n= o(n^{1/2\rho})$, $I(\xi_n)=o(n^{1/2})$. 
%  \end{enumerate}
% \end{lemma}
Theorem \ref{thm10a} gives conditions on $\lambda$ and $h$ that ensure $\frac{v_{h,\lambda}}{\sqrt{n}}=o(1)$ when $H$ has compact support under ordinary or super-smooth errors. 
% This follows by applying Lemma \ref{lem7a} with $\xi_n=h_n/\lambda_n$.
Theorem \ref{thm11a} in Appendix \ref{app:noncompH} considers the case where $H$ is not compactly supported and errors are super-smooth.
Combined with Theorem \ref{thm8}, these provide sufficient conditions to ensure consistency. 
% for any distribution function $F_Y$ and any $x\in C(F_{Y})$ we have $\FYhat(x)\overset{p}{\rightarrow }F_{Y}(x)$.

% function $F_Y$ and any $x\in C(F_{Y})$
% When Assumptions \ref{A1} and \ref{AG} hold, these will also imply $\FYhat(x)\overset{p}{\rightarrow }F_{Y}(x)$ for any $x\in C(F_Y)$ by Theorem \ref{thm8}.

% \noindent For compactly supported $H$ on $[-\gamma,\gamma]$ and super-smooth
% error $|\Phi_Z(t)|\asymp\exp(-\tau|t|^{\rho})$, we replace the requirement
% $h^{\rho}\log n\to\infty$ by
% \begin{equation}
%  h=h_n\to0
%  \qquad\text{and}\qquad
%  \frac{\gamma}{h_n}\le\Big(\frac{\log n}{2\tau}\Big)^{1/\rho}
%  \ \text{for all large }n,
%  \tag{$\star$}
% \end{equation}
% equivalently $\displaystyle\liminf_{n\to\infty}h_n^{\rho}\log n\ge2\tau\gamma^{\rho}$,
% i.e.\ $h_n\ge\gamma(2\tau)^{1/\rho}(\log n)^{-1/\rho}$ eventually. The previous
% hypothesis $h^{\rho}\log n\to\infty$ is the special case
% $\liminf_n h_n^{\rho}\log n=+\infty$.

%%%%%%%%%%%%%%%%%%%%%%%%%%%%%%
% Theorem thm10a
%%%%%%%%%%%%%%%%%%%%%%%%%%%%%%
\begin{theorem}\label{thm10a}
% Suppose $\supp H \subset [-\gamma,\gamma]$ for some $\gamma >0$ and let $C_H:=\underset{t \in [-\gamma,\gamma]}{\sup}|H(t)|$.  
% Let $h:=h_{n}$ and $\lambda:=\lambda_{n}$.
Suppose Assumption \ref{A1}.1 holds,
$\supp H \subset [-\gamma,\gamma]$ for some $\gamma >0$, and $\underset{t \in [-\gamma,\gamma]}{\sup}|H(t)|<\infty$. 
Let 
% $C_H:=\underset{t \in [-\gamma,\gamma]}{\sup}|H(t)|$, 
% $h:=h_{n}>0$, and $\lambda:=\lambda_{n}>0$,
$h:=h_{n} \to 0$, $\lambda :=\lambda _{n} \to 0$
% be such that 
% $\displaystyle \lim_{n\to\infty} h_{n} = 0$,
% $\displaystyle \lim_{n\to\infty} \lambda_{n} = 0$,
% $h_{n} \to 0$, $\lambda _{n} \to 0$,
and
% $\lambda_n=\exp(-o(\sqrt{n}))$.
 suppose $\log(\lambda_n^{-1})=o(\sqrt{n})$.
\begin{enumerate}
\item[(a)]
% a) 
If 
 there exist $\tau, \rho>0$ such that 
 \eqref{eq:error_smoothness_super} holds,
% % $|\Phi_Z(t)| \asymp \exp (-\tau |t|^{\rho})$ as $|t| \to \infty$ 
% \eqref{eq:error_smoothness_super} holds.
% % Let $h:=h_{n}$ and $\lambda:=\lambda_{n}$.
% If 
% $\lambda_n=\exp(-o(\sqrt{n}))$,
% $\displaystyle \lim_{n\to\infty} h_{n} = 0$,
and
$h_n^{\rho}\log n\ge2\tau\gamma^{\rho} $
for all $n$ sufficiently large,
% $$
% \lambda_n=\exp(-o(\sqrt{n}))
% ,\quad
% \lim_{n\to\infty} h_{n} = 0
% , \quad \mbox{and} \quad
% h_n^{\rho}\log n\ge2\tau\gamma^{\rho} 
% \;\;\; \forall n \, \text{ sufficiently large},
% $$
% and if $h \to 0,\,\displaystyle\liminf_{n\to\infty}h_n^{\rho}\log n\ge2\tau\gamma^{\rho}$ and $\lambda=\exp(-o(\sqrt{n}))$, 
then $\frac{v_{h,\lambda}}{\sqrt{n}}=o(1)$. 
% b) 
\item[(b)] 
If there exists $\rho>0$ such that
\eqref{eq:error_smoothness_ordinary} holds,
and
% $\lambda_n=\exp(-o(\sqrt{n}))$,
% $\displaystyle \lim_{n\to\infty} h_{n} = 0$,
% and
$\displaystyle \lim_{n\to\infty} n h_n^{2 \rho} = \infty$,
% $$
% \lambda_n=\exp(-o(\sqrt{n}))
% ,\quad
% \lim_{n\to\infty} h_{n} = 0
% , \quad \mbox{and} \quad
% \lim_{n\to\infty} n h_n^{2 \rho} = \infty,
% $$
% and if  $n h^{2 \rho}  \to \infty$ and $\lambda=\exp(-o(\sqrt{n}))$, 
then $\frac{v_{h,\lambda}}{\sqrt{n}}=o(1)$. 
\end{enumerate}
\end{theorem}

\begin{remark}The condition
% $\lambda=\exp(-o(\sqrt{n}))$ 
$\log(\lambda_n^{-1})=o(\sqrt{n})$
in Theorems \ref{thm10a} and \ref{thm11a} is mild and met if $\lambda \asymp n^{-a}$ for any $a>0$.  The conditions on $h$ are more demanding.  For super-smooth errors and compactly supported $H$, a slowly vanishing sequence such as $h_n=(\log n)^{-a}$ with $0< a <\rho^{-1}$ suffices.  In the case of ordinary-smooth errors a polynomial rate of decay, such as $h_n=n^{-a}$ for $0< a<\frac{1}{2\rho}$ suffices.  
% In addition, it should be noted that 
The condition on $h$ in Theorem \ref{thm11a} for non-compact $H$ is strictly stronger than that in part (a) of Theorem \ref{thm10a}.
\end{remark}

It follows directly from Theorems \ref{thm7} and \ref{thm8} that the mean squared error (MSE) for $\FYhat(x)$ satisfies 
\begin{align*}
MSE(\FYhat(x)) &\le  \Big[\phi _{G_H}(h^{\delta -1})+(1+\left\Vert G_H\right\Vert _{\infty})\omega_{F_Y} (x,h^{\delta })+(2+\left\Vert G_H\right\Vert _{\infty})F_{Y}(x+1-1/\lambda ) \Big]^2
+
\frac{1}{n}\frac{v_{h,\lambda}^{2}}{4\pi^{2}}.
\end{align*}
The following theorem provides bounds on the 
MSE
of $\FYhat(x)$.  
% It is stated without proof since it is a direct consequence of the bound on $v_{h,\lambda}$ obtained in Lemma \ref{lem6a}.
%%%%%%%%%%%%%%%%%%%%%%%%%
% Theorem 5
%%%%%%%%%%%%%%%%%%%%%
\begin{theorem}\label{thm:FYx_MSEbnd}
% Suppose the conditions stated in Theorem \ref{thm7} hold.
Suppose Assumptions \ref{A1} and \ref{AG} hold,
$\supp H \subset [-\gamma,\gamma]$ for some $\gamma >0$, and  $\underset{t \in [-\gamma,\gamma]}{\sup}|H(t)|<\infty$. 
% Let $C_H:=\underset{t \in [-\gamma,\gamma]}{\sup}|H(t)|$.
% Fix $\delta \in (0,1)$.
Then, there exists $C>0$ and $\beta\in \left(0,\min\{\gamma,\pi/2\}\right)$
such that, for every $\delta \in (0,1)$, $h\in (0,1]$, and $\lambda\in(0,\beta/\pi)$,
the following bound holds for any distribution function $F_Y$ and any $x\in C(F_{Y})$
% then for any $F_Y$ and $x\in C(F_{Y})$
% \begin{align*}
% MSE(\FYhat(x)) &\le \frac{1}{n} \frac{C}{ 4\pi^{2}}  \left( \log \lambda^{-1}+  \underset{ \beta < t\le\gamma/h}{\int} \frac{1}{t|\Phi_Z(t)|}dt\right)^2+ \left[\phi _{G_H}(h^{\delta -1})+(1+\left\Vert
% G_H\right\Vert _{C_b})\omega_{F_Y} (x,h^{\delta }) \right.\\&\left.+(2+\left\Vert G_H\right\Vert _{\infty})F_{Y}(x+1-1/\lambda ) \right]^2
% \end{align*}
\begin{align*}
MSE(\FYhat(x)) 
& \le 
\Big[\phi _{G_H}(h^{\delta -1})+(1+\left\Vert
G_H\right\Vert _{C_b})\omega_{F_Y} (x,h^{\delta }) +(2+\left\Vert G_H\right\Vert _{\infty})F_{Y}(x+1-1/\lambda ) \Big]^2
\\&
\quad + 
\frac{1}{n} \frac{C}{ 4\pi^{2}}  
\left( \log \lambda^{-1}+  
% \underset{ \beta < t\le\gamma/h}{\int} 
\int_\beta^{\gamma/h}\!\frac{1}{t|\Phi_Z(t)|}dt\right)^2
\end{align*}
\end{theorem}

The next result illustrates the rates implied by Theorem \ref{thm:FYx_MSEbnd} under additional
local regularity and tail assumptions on $F_Y$. These assumptions are not
needed for consistency, but they make the dependence of the bound on $n$
explicit.

\begin{corollary}\label{coro1}
Suppose the conditions of Theorem \ref{thm:FYx_MSEbnd} a) hold and that, for some
$k,s,q>0$ and $C<\infty$,
\[
\phi_G(N)\le CN^{-k}, \qquad
\omega_{F_Y}(x,u)\le Cu^s,\qquad
F_Y(x+1-u_1)\le C u_1^{-q},
\]
for all sufficiently large $N,\,u_1>0$ and small $u>0$.
Define $r=\frac{ks}{k+s}$.

\begin{enumerate}
\item[(a)]
If there exist $\tau, \rho>0$ such that 
    \eqref{eq:error_smoothness_super} holds
then, for any constant
$
A>\gamma (2\tau)^{1/\rho},
$
with
$
h_n
=
A(\log n)^{-1/\rho},
\,
\lambda_n=h_n^{\,r/q},
$
we have
\[
\operatorname{MSE}(\widehat F_Y(x))
=
O\!\left(
(\log n)^{-2r/\rho}
\right).
\]

\item[(b)]
If there exists $\rho>0$ such that
    \eqref{eq:error_smoothness_ordinary} holds,
and $
h_n\asymp n^{-1/\{2(\rho+r)\}},
\,\lambda_n=h_n^{\,r/q},
$ then
% If there exist $\rho>0$ such that $|\Phi_Z(t)|\asymp |t|^{-\rho},\, |t|\to\infty$, then, with
% $
% h_n\asymp n^{-1/\{2(\rho+r)\}},
% \,\lambda_n=h_n^{\,r/q},
% $ we have
\[
\operatorname{MSE}(\widehat F_Y(x))
=
O\!\left(
n^{-r/(\rho+r)}
\right).
\]
\end{enumerate}
\end{corollary}

\cite{Dattner2011} and \cite{ Dattner2013} derive minimax optimal convergence rates under global Sobolev smoothness assumptions on $F_Y$. In contrast, Corollary \ref{coro1} establishes explicit rates under substantially weaker assumptions consisting only of a local modulus of continuity for the distribution function together with mild local tail conditions. Although the resulting rates need not match the minimax Sobolev rates when both sets of assumptions hold, they apply to a considerably broader class of distributions, including settings where no global smoothness of the distribution is available.

%%%%%%%%%%%%%%%%%%%%%%%%%%%%%%%%%%%%%
\subsection{Estimation of $F_{Y}(x,y)$ for $x,y \in C(F_Y)$}\label{sec:FYxy}
%%%%%%%%%%%%%%%%%%%%%%%%%%%%%%%%%%%%%
Although the interval probabilities $F_Y(x,y):=F_Y(y)-F_Y(x)$ for $x<y$ and $x,y \in C(F_Y)$ can be estimated by differencing pointwise estimators of $F_Y$, a direct estimator has several advantages. It avoids differencing two noisy estimators, yields sharper non-asymptotic bounds, and provides a natural analogue of the generalized inversion estimator developed in Section \ref{sec:FYx}.

The estimator we propose is obtained by replacing the interval $(x-1/\lambda,x)$ used in Section \ref{sec:FYx} by the fixed interval $(x,y)$.  Let $\{X_j\}_{j=1}^n$ be a random sample from $F_X$. For $h>0$ define 
% $\FYhat(x,y)=\frac{1}{n}\sum_{j=1}^nK_{x,y,h}\left(X_j-x\right)$, with 
% \begin{equation*}
% K_{x,y,h}(t)=\frac{1}{2\pi }\int_{\R}e^{its}\beta _{x,y,h}(s)ds \mbox{ and } \beta _{x,y,h}(s)=\frac{1-e^{-i(y-x)s}}{is}\frac{H(hs)}{\Phi_{Z}(s)}.
% \end{equation*}
\begin{equation*}
\FYhat(x,y):=\frac{1}{n}\sum_{j=1}^{n}\!K_{x,y,h}\!\left(X_j-x\right)\!,
\;\, \mbox{where} \;\,
K_{x,y,h}(t)=\frac{1}{2\pi }\!\int_{\R\!}\!e^{its}\beta _{x,y,h}(s)ds
, 
% \;\; \mbox{and} \;\; 
% \quad
\;\;
\beta _{x,y,h}(s)=\frac{1-e^{-i(y-x)s}}{is}\frac{H(hs)}{\Phi_{Z}(s)}\!. 
\end{equation*}
Assumptions \ref{A1}.1 and \ref{A1}.2 imply $\beta_{x,y,h} \in L(\R)$ for each $h>0$ and $x,y \in \R$,
so $K_{x,y,h}$ is bounded and continuous for each $h>0$.
Furthermore, under Assumption~\ref{A1}.3
\begin{equation*}
\overline{\beta_{x,y,h}(s)}=\frac{1-e^{i(y-x)s}}{-is}\frac{H(-hs)}{\Phi _{Z}(-s)}=\beta _{x,y,h}(-s), \;\;\; \mbox{ and } \;\;\; \overline{K_{x,y,h}(t)}=\frac{1}{2\pi }\int_{\R}e^{-its}\beta_{x,y,h}(-s)ds=K_{x,y,h}(t),
\end{equation*}
so $K_{x,y,h}$ is real-valued. 
Hence, under Assumption \ref{A1},  $\widehat{F}_{Y}(x,y)$ is continuous, almost surely bounded,  real-valued, and  $E\left( \FYhat(x,y) \right)<\infty$.
% is bounded, continuous and real-valued, and $E\left( \FYhat(x,y) \right)<\infty$.

% Assumptions \ref{A1}.1 and \ref{A1}.2 imply $\beta_{x,y,h} \in L(\R)$ for each $h>0$ and $x,y \in \R$.
% Hence, $K_{x,y,h}$ is bounded and continuous for each $h>0$.  It follows that $E\left( \FYhat(x,y) \right)<\infty$.  Furthermore, under Assumption~\ref{A1}.3
% \begin{equation*}
% \overline{\beta_{x,y,h}(s)}=\frac{1-e^{i(y-x)s}}{-is}\frac{H(-hs)}{\Phi _{Z}(-s)}=\beta _{x,y,h}(-s) \mbox{ and } \overline{K_{x,y,h}(t)}=\frac{1}{2\pi }\int_{\R}e^{-its}\beta_{x,y,h}(-s)ds=K_{x,y,h}(t).
% \end{equation*}
% Hence, $K_{x,y,h}$ exists, is bounded, continuous and real-valued and consequently so is $\widehat{F}
% _{Y}(x,y)$. 
The following result gives a finite-sample bias bound and asymptotic unbiasedness under similar conditions as Theorem \ref{thm7}. Notably, this result places no assumptions on the distribution $F_Y$.
%%%%%%%%%%%%%%%%%%%%%%%%%%%%%%%%%%%%%%%%%%%
% Theorem Bias of \hat{F}(x,y)
%%%%%%%%%%%%%%%%%%%%%%%%%%%%%%%%%%%%%%%%%%%
\begin{theorem}\label{thm13n} 
% Suppose Assumption \ref{A1} holds and $H$ and $G$ are defined as in Theorem \ref{thm7}.  
Suppose Assumptions \ref{A1} and \ref{AG} hold. 
Then, for any $F_{Y},$ and $x,y\in C(F_{Y})$
% Then
\begin{equation} \label{unqual}
\left|E\left(\FYhat(x,y)\right)-F_{Y}(x,y)\right|
\rightarrow 0,
\quad \mbox{ as $h,\,\lambda \rightarrow 0$.  }
% \ h \rightarrow 0\text{ for all }F_{Y},\ x,y \in C(F_{Y}). 
\end{equation}
Moreover, for $\delta \in (0,1)$, $h \in (0,1]$ and $h^\delta \le \frac{y-x}{2}$
\begin{eqnarray}
\left|E\left(\FYhat(x,y)\right)-F_{Y}(x,y)\right| &\leq &\phi _{G}(h^{\delta
-1})+(1+\left\Vert G\right\Vert _{\infty})[\omega_{F_Y}(x,h^{\delta })+\omega_{F_Y}
(y,h^{\delta })].  \label{qual1}
\end{eqnarray}
\end{theorem}
%%% %%%%%%%%%%%%%%%%%%%%%%%%%%%
%%REMARK 2
%%%%%%%%%%%%%%%%%%%%%%%%%%%%
\iffalse
\begin{remark}\label{remk2} 
If the convergence in equation \eqref{unqual} holds, then by Lemma \ref{lem1} 
\[
\int_{\R}G\left( \left[ \frac{t-y}{h},\frac{t-x}{h}\right] \right) dF_Y(t)\to F_Y(x,y),\, \mbox{ as } h\to 0 \mbox{ for all } F_Y, \, x,y \in C(F_Y).
\]
Taking $F_Y$ to be the Heaviside function, $y=1$, $x=-1$ and the fact that $\underset{a\rightarrow -\infty ,b\rightarrow \infty}{\lim}G([a,b])=1$ we obtain
\[ 
\int_{\R} G\left( \left[ \frac{t-1}{h},\frac{t+1}{h}\right] \right)
dF_Y(t)=G\left( \left[ -\frac{1}{h},\frac{1}{h}\right]
\right).
\]
For any sequence $h:=h_n \to 0$ as $n \to \infty$ we have $G\left( \left[ -\frac{1}{h_n},\frac{1}{h_n}\right] \right)\to 1$ since $-1/h_n\rightarrow -\infty$ and $1/h_n \to \infty$.  Denoting $\Delta_n =\left[ -1/h_n,1/h_n\right]$, we get  
$$
\frac{1}{2\pi }\int_{\R}(\mathcal{F}H)(u)\chi _{\Delta _{n}}(u)du\rightarrow 1,
$$
from the fact that $G([a,b])=\frac{1}{2 \pi}\int_a^b(\mathcal{F}H)(v)dv$. 
\end{remark}
\fi 
Following the same reasoning as in Remark \ref{remk1}, we have that if $\mathcal{F}H \geq 0$, then  $\FYhat(x,y)$ is asymptotically unbiased if, and only if, Assumption \ref{AG} holds, which is, in turn, equivalent to the conditions in equation \eqref{73.1}.

\iffalse
Let $\varepsilon \geq 1$.  The function $H=H_{\varepsilon ,h}$ can be chosen in such a way that $\mathcal{F}H\geq 0$ and the bound
\begin{equation*}
|E\FYhat(x,y)-F_{Y}(x,y)|\leq \omega (x,h^{\varepsilon })+\omega(y,h^{\varepsilon }),\ h\in (0,1)
\end{equation*}
holds for all $F_{Y},\ x,y\in C(F_{Y})$.
\fi
The next theorem provides a bound for the variance of $\FYhat(x,y)$ and gives a sufficient condition for the convergence of $\FYhat(x,y)$ in quadratic mean.  This, in turn, implies consistency of $\FYhat(x,y)$.
%%%%%%%%%%%%%%%%%%%%%%%%%%%%%%%%%
%% Theorem consistency of \FYhat(x,y)
%%%%%%%%%%%%%%%%%%%%%%%%%%%%%%%%%
\begin{theorem}\label{thm8a} 
Suppose Assumption \ref{A1} holds. Then,  
\begin{equation}
V\left(\FYhat(x,y)\right) \le \frac{1}{n}\frac{v_h(x,y)^2}{4 \pi^2}, 
\quad \mbox{ where } \quad 
v_h(x,y):=\int_{\R}|\beta _{x,y,h}(t)|dt
% \mbox{ where $v_h(x,y):=\int_{\R}|\beta _{x,y,h}(t)|dt$}
.
\end{equation}
If, in addition, Assumption \ref{AG} holds, and 
$h:=h_{n} \to 0$ is chosen so that $v_{h}(x,y)/\sqrt{n}\rightarrow 0$ as $n\rightarrow \infty $,  
% Then, if the conditions on Theorem \ref{thm13n} hold 
then, for any distribution function $F_Y$ and any $x\in C(F_{Y})$, we have
$\FYhat(x,y)\overset{p}{\rightarrow }F_{Y}(x,y)$.
\end{theorem}
%%%%%%%%%%%%%%%%%%%%%%%%%%%%%%%%%%
The following lemma is similar to Lemma \ref{lem6a} and gives the order of $v_h(x,y)$ under conditions on $H$. Lemma~\ref{lem6ab_NoncompH} in Appendix~\ref{app:noncompH} gives a similar result for $H$ with non-compact support. 
%%%%%%%%%%%%%%%%%%%%%%%%%%%%%%%%%%%
% Lemma 6-\beta
%%%%%%%%%%%%%%%%%%%%%%%%%%%%%%%%%%
\begin{lemma}\label{lem6ab} 
Suppose Assumption \ref{A1} holds, $\supp H \subset [-\gamma,\gamma]$ for some $\gamma >0$, and 
$C_H:=\underset{t \in [-\gamma,\gamma]}{\sup}|H(t)|<\infty$.
Then, there exist constants $C,\beta>0$, depending only on $C_H$ and $\Phi_Z$, such that 
for all $h\in(0,\gamma/\beta)$ and $x,y\in\R$ with $x<y$, 
$$
v_{h}(x,y) \leq C\left[ \log(1+y-x) + \underset{ \beta < t\le\gamma/h}{\int} \frac{1}{t|\Phi_Z(t)|}dt\right]\!.
$$ 
% Suppose $\supp H \subset [-\gamma,\gamma]$ for some $\gamma >0$ and $\underset{t \in [-\gamma,\gamma]}{\sup}|H(t)|<\infty$.  Then, there exists $\beta>0$ such that for all $h\in(0,\gamma/\beta)$, 
% $$
%  v_{h}(x,y) =O\left( \underset{ \beta < t\le\gamma/h}{\int} \frac{1}{t|\Phi_Z(t)|}dt\right)
%  $$ 
\end{lemma}
As in the case of $v_{h,\lambda}$, the order of $v_h(x,y)$ 
% depends on the order of 
% $\underset{ \beta < t\le\gamma/h}{\int} \frac{1}{t|\Phi_Z(t)|}dt$ 
% % and $ \underset{ \beta < t\le t(h)}{\int} \frac{|H(ht)|}{t|\Phi_Z(t)|}dt$, 
% which in turn 
depends on the behavior of $|\Phi_Z(t)|$ as $|t| \to \infty $. 
% The next two theorems give conditions on $h$ to assure that $\frac{v_{h}(x,y)}{\sqrt{n}}=o(1)$ when $H$ has compact and non-compact support.
The next theorem gives conditions on $h$ to ensure that $\frac{v_{h}(x,y)}{\sqrt{n}}=o(1)$,
which can be used with Theorem \ref{thm8a} to obtain consistency.
% when $H$ has compact support.
Theorem \ref{thm11ab} gives a similar result when $H$ has non-compact support.

%%%%%%%%%%%%%%%%%%%%%%%%%%%%%%
% Theorem thm10ab
%%%%%%%%%%%%%%%%%%%%%%%%%%%%%%
\begin{theorem}\label{thm10ab}
Suppose $\supp H \subset [-\gamma,\gamma]$ for some $\gamma >0$ and $\underset{t \in [-\gamma,\gamma]}{\sup}|H(t)|<\infty$.
% let $C_H=\underset{t \in [-\gamma,\gamma]}{\sup}|H(t)|$.  
\begin{enumerate}
\item[(a)]
If there exist $\tau, \rho>0$ such that 
% $|\Phi_Z(t)| \asymp \exp (-\tau |t|^{\rho})$ as $|t| \to \infty$ 
\eqref{eq:error_smoothness_super} holds,
and if $h \to 0,\,\displaystyle\liminf_{n\to\infty}h_n^{\rho}\log n>2\tau\gamma^{\rho}$, then $\frac{v_{h}(x,y)}{\sqrt{n}}=o(1)$.
\item[(b)]
If there exists $\rho>0$ such that
% $|\Phi_Z(t)| \asymp |t|^{-\rho}$ as $|t| \to \infty$ 
\eqref{eq:error_smoothness_ordinary} holds,
and if  $n h^{2 \rho}  \to \infty$, then $\frac{v_{h}(x,y)}{\sqrt{n}}=o(1)$. 
\end{enumerate}
\end{theorem}

It follows directly from Theorems \ref{thm13n} and \ref{thm8a} that the MSE for $\FYhat(x,y)$ at $x,y \in C(F_Y)$ satisfies
$$
MSE(\FYhat(x,y)) \le  \Big[ \phi _{G}(h^{\delta
-1})+(1+\left\Vert G\right\Vert _{\infty})(\omega_{F_Y}(x,h^{\delta })+\omega_{F_Y}
(y,h^{\delta })) \Big]^2
+
\frac{1}{n}\frac{v_h(x,y)^2}{4 \pi^2}.
$$
With this, the following theorem follows immediately from the bound on $v_h(x,y)$ obtained in Lemma \ref{lem6ab}.
% Given the bound on $v_h(x,y)$ obtained in Lemma \ref{lem6ab} we have the following theorem, that is stated without proof.
%%%%%%%%%%%%%%%%%%%%%%%%%%%%%%%%%%%%%%%%%%
\begin{theorem}\label{thm:FYxy_MSEbnd}
Suppose Assumptions \ref{A1} and \ref{AG} hold,
$\supp H \subset [-\gamma,\gamma]$ for some $\gamma >0$, and  $\underset{t \in [-\gamma,\gamma]}{\sup}|H(t)|<\infty$. 
% Suppose the conditions stated in Theorem \ref{thm13n} hold.
% If $\supp H \subset [-\gamma,\gamma]$ for some $\gamma >0$ and 
% $\underset{t \in [-\gamma,\gamma]}{\sup}|H(t)|<\infty$, 
% then, for $h \in (0,1]$ there exists $C, \, \beta>0$ such that 
Then, there exist constants $C>0$ and $\beta \in (0,\gamma)$ such that, for any distribution function $F_Y$, and any $\delta \in (0,1)$, any $x,y\in C(F_{Y})$ with $x<y$ and any $h \in (0,1]$ satisfying $h^\delta\leq (y-x)/2$,
the following bound holds 
\begin{align*}
MSE(\FYhat(x,y)) 
&\le 
\Big[ \phi _{G_H}(h^{\delta
-1})+(1+\left\Vert G_H\right\Vert _{\infty})(\omega_{F_Y}(x,h^{\delta })+\omega_{F_Y}
(y,h^{\delta })) \Big]^2
\\&
\quad +
\frac{1}{n} \frac{C}{ 4\pi^{2}} 
\left( \log(1+y-x) + \underset{ \beta < t\le\gamma/h}{\int} \frac{1}{t|\Phi_Z(t)|}dt\right)^2\!
.
\end{align*}
\end{theorem}

%%%%%%%%%%%%%%%%%%%%%%%%%%%%%%%%%%%%%%%%%%%%
\subsection{Estimation of a jump $p_{x}$ of $F_{Y}$}\label{sec:px}
%%%%%%%%%%%%%%%%%%%%%%%%%%%%%%%%%%%%%%%%%%%%
Unlike our estimators for $F_Y(x)$ and $F_Y(x,y)$, estimation of jump sizes does not follow from a straightforward application of the generalized inversion formula.
% 
% Whereas our estimators for $F_Y(x)$ and $F_Y(x,y)$ are constructed in a similar manner, our estimator for $p_x$ requires some additional work.  Estimation of jump sizes does not follow directly by a straightforward application of the generalized inversion formula. 
Instead, it requires a different approximation argument based on localized kernels.
Since \(p_x\) can be written as \(\mathbb P(Y=x)\), the idea is to approximate the singleton indicator \(u\mapsto\chi_{{x}}(u)\) by a localized kernel \(W((u-x)/h)\) for $h>0$.

% Unlike the estimators of \(F_Y(x)\) and \(F_Y(x,y)\), which follow directly from generalized Fourier inversion identities, estimation of the jump size \(p_x\) requires a different construction. The inversion identities used in the preceding sections do not directly isolate the probability mass at a discontinuity. Since \(p_x=\mathbb P(Y=x)\), the basic idea is instead to approximate the singleton indicator \(u\mapsto\mathbf 1_{{x}}(u)\) by a localized kernel \(W((u-x)/h)\). As \(h\downarrow0\), this expression remains equal to one at \(u=x\) and converges to zero for every \(u\neq x\), thereby isolating the mass at \(x\).

% Since \(p_x\) can be written as \(\mathbb P(Y=x)\), the idea is to approximate the singleton indicator \(u\mapsto\chi_{{x}}(u)\) by a localized kernel \(W((u-x)/h)\) for $h>0$.
To motivate our estimator, 
consider a distribution function $F$ with a jump of size $p_x=F(x)-\underset{\varepsilon \downarrow 0}{\lim}\,F(x-\varepsilon)$ at $x$.
% 
% If a random variable $X$ has distribution $F$, then \(p_x=\mathbb P(X=x)\) so the idea is to approximate the singleton indicator \(u\mapsto\chi_{{x}}(u)\) by a localized kernel \(W((u-x)/h)\) for $h>0$.
% Since \(p_x=\mathbb P(Y=x)\), the idea is to approximate the singleton indicator \(u\mapsto\chi_{{x}}(u)\) by a localized kernel \(W((u-x)/h)\) for $h>0$.
Suppose $W: \R \to \R$ such that
\begin{equation}\label{A3}
W\text{ is continuous on }\mathbb R,\qquad
W(0)=1,\qquad
\lim_{|u|\to\infty}W(u)=0.
\end{equation}
Equation \eqref{A3} implies 
$W$ is bounded and
$\underset{h \to 0}{\lim}W\left(\frac{u-x}{h}\right)= \chi_{{x}}(u)$,
% for fixed $u$, \(W\left(\frac{u-x}{h}\right)\to 1_{{x}}(u)\) as \(h\to 0\), 
so bounded convergence gives 
% $\underset{h \to 0}{\lim}\int_{\R}W\left(\frac{u-x}{h}\right)dF(u) = p_x$.
\begin{equation}\label{jumpmot}
    \lim_{h\to 0}
    \int_{\R}W\left(\frac{u-x}{h}\right)dF(u)
    % \to 
    =
    p_x
    % \quad
    % \text{ as }
    % h\downarrow0
    .
\end{equation}
% as \(h\downarrow0\).
To quantify the approximation, define
\begin{equation*}
\omega _{W}(\varepsilon )
:=
\sup_{\left\vert v\right\vert \leq \varepsilon}\left\vert W(v)-1\right\vert 
,\qquad 
\phi _{W}(N)
:=
\sup_{\left\vert v\right\vert\geq N}\left\vert W(v)\right\vert
,\quad\mbox{and}\quad 
\delta _{F}(x,\varepsilon):=\int_{\left\vert t-x\right\vert <\varepsilon }dF(t)-p_{x}\geq 0.
\end{equation*}
When equation \eqref{A3} holds, Lemma 3 in \cite{Mynbaev2022} showed that, for any $h, \,\varepsilon _{1}\in (0,1)$ and $\varepsilon_{2}>\varepsilon_{1}$
\begin{equation}\label{25n}
\left\vert \int_{\R\!}\!W\!\left(\! \frac{u-x}{h}\right)\!dF(u)-p_{x}\right\vert 
\leq 
\omega _{W}(h^{\varepsilon_{2}-\varepsilon _{1}})\!\left[ p_{x}+\delta _{F}(x,h^{1-\varepsilon _{1}})\right]  
+
\left( 1+\left\Vert W\right\Vert _{C_b}\right) \delta _{F}(x,h^{1-\varepsilon_{1}}) 
+
\phi _{W}(h^{-\varepsilon _{1}}),  
\end{equation}
and we also have 
% If equation \eqref{A3} holds, then
$\underset{\varepsilon \rightarrow 0}{\lim}\omega _{W}(\varepsilon )= 0,$
$\underset{N \rightarrow \infty}{\lim}\phi _{W}(N)= 0,$
and
$\underset{\varepsilon \rightarrow 0}{\lim}\delta_{F}(x,\varepsilon )=0$
% for every $x$
by continuity of probability measures.
Thus, unlike the bounds in Sections \ref{sec:FYx} and \ref{sec:FYxy}, the approximation error is governed by the local probability mass around the discontinuity rather than by a modulus of continuity.  

We now connect this to a Fourier representation. 
Let $H \in L(\R)$ be real-valued and even, and set $W:=\mathcal{F}H$.
% Set $W:=\mathcal{F}H$ for $H \in L(\R)$.
If $H$ is a kernel, i.e., $\int_{\R}H(t)dt=1$, then $\mathcal{F}H$ satisfies equation \eqref{A3} since $(\mathcal FH)(0)=1$ and $\mathcal FH$ is continuous and vanishes at infinity by the Riemann-Lebesgue lemma. 
In addition, Fubini's Theorem gives the following relationship
\begin{equation}\label{e77n}
\int_{\R}W\left(\frac{u-x}{h}\right)dF(u)
=
\int_{\R}(\mathcal{F}H)\left(\frac{u-x}{h}\right)dF(u)
=
\int_{\R}e^{-isx}(\mathcal{F}F)(s)hH(hs)ds.
\end{equation}

Equation \eqref{e77n} motivates the following estimator for $p_x:=F_Y(x)-\underset{\varepsilon \downarrow 0}{\lim}\,F_Y(x-\varepsilon)$.
Let $\{X_j\}_{j=1}^n$ be a random sample from $F_X$, and for $h>0$ define
% Let $p_x:=F_Y(x)-\underset{\varepsilon \downarrow 0}{\lim}\,F_Y(x-\varepsilon)$, and $\{X_j\}_{j=1}^n$ be a random sample from $F_X$. We define the following estimator of $p_x$
\begin{equation*}%\label{decon}
\hat{p}_{x}=\frac{1}{n}\sum_{j=1}^nK_{h}\!\left(X_{j}-x\right)\!,
\quad \mbox{where} \quad
K_{h}(t):=\int_{\R}e^{ist}\gamma_{h}(s)ds
,
\quad
\gamma _{h}(s)=\frac{hH(hs)}{\Phi _{Z}(s)}. 
\end{equation*}
% 
% Inspired by equation \eqref{e77n}, if $\{X_j\}_{j=1}^n$ is a random sample from $F_X$,     we define the estimator for $p_x:=F_Y(x)-\underset{\varepsilon \downarrow 0}{\lim}\,F_Y(x-\varepsilon)$ by $\hat{p}_{x}=\frac{1}{n}\sum_{j=1}^nK_{h}(X_{j}-x)$ for $h>0$, with 
% \begin{equation*}
% \gamma _{h}(s)=\frac{hH(hs)}{\Phi _{Z}(s)} \mbox{ and } K_{h}(t)=\int e^{ist}\gamma
% _{h}(s)ds.
% \end{equation*}
% 
To ensure $\hat{p}_{x}$ is well defined, we use Assumptions \ref{A1}.1 and \ref{A1}.3 again, but we replace Assumption \ref{A1}.2 with the following assumption.
% We make the following assumption to assure that $\FYhat(x)$ is well defined.
% As in Assumption \ref{A1}, we assume that $\Phi _{Z}(s)\neq 0$ for each $s\in \R$ and take $H$ to be even and real valued.  In addition, we assume 
\begin{condition}\label{A2}
(Regularity)
$\gamma _{h} \in L(\R)$ for each $h>0$.
\end{condition}
Then, under Assumptions \ref{A1}.1, \ref{A1}.3  and \ref{A2}, $K_{h}$ is bounded, continuous and real valued with
$
\overline{K_{h}(t)}=\int_{\R}e^{-ist}\frac{hH(-hs)}{\Phi _{Z}(-s)}ds=K_{h}(t).
$
Consequently, $\hat{p}_{x}$ is also bounded, continuous and real valued.

The next theorem characterizes the asymptotic unbiasedness of $\hat{p}_x$. 
Remarkably, asymptotic unbiasedness of $\hat{p}_x$ is equivalent to $\int_{\R}H(t)dt=1$.
% Remarkably, for the estimator $\hat{p}_x$, asymptotic unbiasedness is equivalent to $\int_{\R}H(t)dt=1$. 
In addition, we obtain a non-asymptotic bound on the bias, since (see the proof of Theorem \ref{thm11}) 
$$
E(\hat{p}_{x})=\int_{\R}(\mathcal{F}H)\left( \frac{u-x}{h}\right)dF_Y(u).
$$
  
%%%%%%%%%%%%%%%%%%%%%%%%%%%%%%%%%%%%%%%%%%
% \newpage
\begin{theorem}\label{thm11} 
Suppose $H\in L(\R)$, and Assumptions \ref{A1}.1, \ref{A1}.3, and \ref{A2} hold. Then the following statements are equivalent:
\begin{enumerate}
\item $\int_{\R}H(t)dt=1$;

\item $E(\hat{p}_{x}) \rightarrow p_{x}$ as $h \to 0$ for any $F_{Y}$ and $x\in J(F_{Y})$;

\item 
$W=\mathcal{F}H$ satisfies equation \eqref{A3}.
\end{enumerate}
Moreover, if these equivalent conditions hold, then, for every distribution function $F_{Y}$ and $x\in J(F_{Y})$, the following bias bound holds whenever
$h,\,\varepsilon _{1}\in (0,1)$ and $\varepsilon_{2}>\varepsilon _{1}$,
\begin{eqnarray*}
\left\vert E(\hat{p}_{x}) - p_{x} \right\vert
&\leq &
\omega _{\mathcal{F}H}(h^{\varepsilon_{2}-\varepsilon _{1}})\left[ p_{x}+\delta_{F_Y}(x,h^{1-\varepsilon _{1}})\right]  
+
\left( 1+\left\Vert \mathcal{F}H\right\Vert _{C_b}\right) \delta_{F_Y}(x,h^{1-\varepsilon_{1}}) 
+
\phi _{\mathcal{F}H}(h^{-\varepsilon _{1}}). 
\end{eqnarray*}
\end{theorem}
%%%%%%%%%%%%%%%%%%%%%%%%%%%%%%%%%%%%%%%%%%
% \begin{theorem}\label{thm11} 
% Suppose $H\in L(\R)$, Assumptions \ref{A1}.1, \ref{A1}.3, and \ref{A2} hold and $h,\,\varepsilon _{1}\in (0,1)$, $\varepsilon_{2}>\varepsilon _{1}$.  Then following statements are equivalent:
% \begin{enumerate}
% \item $\int_{\R}H(t)dt=1$,

% \item $E(\hat{p}_{x}) \rightarrow p_{x}$ as $h \to 0$ for all $F_{Y}$ and $x\in J(F_{Y})$

% \item 
% $W=\mathcal{F}H$ satisfies equation \eqref{A3} and the inequality
% \begin{eqnarray*}
% \left\vert \int_{\R}(\mathcal{F}H)\left( \frac{u-x}{h}\right)dF_Y(u)-p_{x}\right\vert
% &\leq &
% \omega _{\mathcal{F}H}(h^{\varepsilon_{2}-\varepsilon _{1}})\left[ p_{x}+\delta_{F_Y}(x,h^{1-\varepsilon _{1}})\right]  
% +
% \left( 1+\left\Vert \mathcal{F}H\right\Vert _{C_b}\right) \delta_{F_Y}(x,h^{1-\varepsilon_{1}}) 
% \notag\\
% &+&
% \phi _{\mathcal{F}H}(h^{-\varepsilon _{1}}). 
% \end{eqnarray*}
% holds for any $F_{Y}$,  $x\in J(F_{Y})$.
% \end{enumerate}
% \end{theorem}
%%%%%%%%%%%%%%%%%%%%%%%%%%%%%%%%%%%%%%%%%%%%%
% Theorem~\ref{thm11} shows that asymptotic
% unbiasedness depends only on the normalization of the regularization
% kernel and does not require the existence of a density, a mixture
% representation, or global smoothness assumptions on the latent
% distribution.
% Theorem~\ref{thm11} shows that asymptotic when $\int_{\R}H(t)dt=1$
% % Thus, once the normalization in part 1 is imposed, 
% asymptotic unbiasedness holds
Theorem~\ref{thm11} establishes that $\hat p_x$ is asymptotically unbiased as $h\to 0$ under conditions only on the error distribution and localization kernel $H$. These place no additional restrictions on $F_Y$ and do not require the existence of a density,
a mixture representation, or global smoothness assumptions on the latent distribution.

\begin{remark}\label{rem5}
The normalizations for $H$ differ across the three estimators.
% For $\FYhat(x)$ and $\FYhat(x,y)$ it follows from Theorem \ref{thm7} and Remark \ref{rem:inversion_kernel}
% that sufficient conditions for asymptotic unbiasedness include $H(0)=1$. 
For $\FYhat(x)$ and $\FYhat(x,y)$ the sufficient conditions given in Remark \ref{rem:inversion_kernel}
% that sufficient conditions for asymptotic unbiasedness 
include $H(0)=1$. 
By contrast, Theorem~\ref{thm11} shows $\int_{\R}H(t)dt=1$, i.e. $(\mathcal{F} H)(0)=1$, is necessary and sufficient for asymptotic unbiasedness of $\hat p_x$ for any distribution function $F_Y$ at every $x\in J(F_Y)$.
% By contrast, $\hat p_x$ requires $\int_{\R}H(t)dt=1$, i.e., $(\mathcal{F} H)(0)=1$.
Note, e.g., that the Gaussian kernel
$H(t)=e^{-t^{2}/2}$ satisfies the first normalization but not the second.  Relatedly, $K_{h,\lambda}$ and
$K_{x,y,h}$ carry the factor $1/(2\pi)$ while $K_h$ does not.
\end{remark}

The next theorem provides a bound for the variance $V(\hat{p}_x)$. This, together with Theorem \ref{thm11}, gives sufficient conditions for convergence in quadratic mean of $\hat{p}(x)$, which implies consistency.
\begin{theorem}\label{thm12} 
Given Assumptions \ref{A1}.1, \ref{A1}.3 and \ref{A2}
\begin{equation}
V(\hat{p}_x)\le \frac{1}{n}u_h^2, \mbox{ where $u_{h}=\int_{\R}|\gamma _{h}(t)|dt.$}
\end{equation}  
Suppose $h=h_{n}\to 0$ is chosen so that $u_{h}/\sqrt{n}\rightarrow 0$ as $n\rightarrow \infty$.  Then, under the conditions of Theorem \ref{thm11}, $\hat{p}_{x} \overset{p}{\rightarrow} p_x$.
\end{theorem}
Similar to results in the previous two sections, the next theorem gives conditions on $h$ to ensure that $\frac{u_{h}}{\sqrt{n}}=o(1)$ under ordinary and super smoothness, but in this case, only for compactly supported $H$.

%%%%%%%%%%%%%%%%%%%%%%%%%%%%%%%%%%%%%%%%%%%%
% Theorem thm10j
%%%%%%%%%%%%%%%%%%%%%%%%%%%%%%%%%%%%%%%%%%%%
\begin{theorem}\label{thm10j}
Suppose Assumption \ref{A1}.1 holds, $\supp H \subset [-\gamma,\gamma]$ for some $\gamma >0$, and $\underset{t \in [-\gamma,\gamma]}{\sup}|H(t)|<\infty$. Let $h:=h_{n} \to 0$. 
\begin{enumerate}
    \item[(a)]
    If there exist $\tau, \rho>0$ such that 
    % $|\Phi_Z(t)| \asymp \exp (-\tau |t|^{\rho})$ as $|t| \to \infty$
    \eqref{eq:error_smoothness_super} holds, 
    and $h_n^{\rho}\log n\ge2\tau\gamma^{\rho}$ for all $n$ sufficiently large, 
    % and if $h \to 0,\,\displaystyle\liminf_{n\to\infty}h_n^{\rho}\log n\ge2\tau\gamma^{\rho}$, 
    then $\frac{u_{h}}{\sqrt{n}}=o(1)$.
    \item[(b)]
     % If $|\Phi_Z(t)| \asymp |t|^{-\rho}$ as $|t| \to \infty$ 
     If there exists $\rho>0$ such that
    \eqref{eq:error_smoothness_ordinary} holds,
     and $\displaystyle \lim_{n\to\infty} n h_n^{2 \rho}  = \infty$, then $\frac{u_{h}}{\sqrt{n}}=o(1)$.
\end{enumerate}
% Suppose $\supp H \subset [-\gamma,\gamma]$ for some $\gamma >0$ and let $C_H=\underset{t \in [-\gamma,\gamma]}{\sup}|H(t)|$.  a) If there exist $\tau, \rho>0$ such that $|\Phi_Z(t)| \asymp \exp (-\tau |t|^{\rho})$ as $|t| \to \infty$ and if $h \to 0,\,\displaystyle\liminf_{n\to\infty}h_n^{\rho}\log n\ge2\tau\gamma^{\rho}$, then $\frac{u_{h}}{\sqrt{n}}=o(1)$. b) If $|\Phi_Z(t)| \asymp |t|^{-\rho}$ as $|t| \to \infty$ and if  $n h^{2 \rho}  \to \infty$, then $\frac{u_{h}}{\sqrt{n}}=o(1)$. 
\end{theorem}
%%%%%%%%%%%%%%%%%%%%%%%%%%%%%%%%%%%%%%%%%%%%
It follows directly from Theorems \ref{thm11} and \ref{thm12} that the mean squared error for $\hat{p}_x$ at $x \in J(F_Y)$ satisfies
$$
MSE(\hat{p}_x) 
\le 
\Big[ \omega _{\mathcal{F}H}(h^{\varepsilon_{2}-\varepsilon _{1}})\left( p_{x}+\delta _{F_Y}(x,h^{1-\varepsilon _{1}})\right)  +\left( 1+\left\Vert \mathcal{F}H\right\Vert _{C_b}\right) \delta _{F_Y}(x,h^{1-\varepsilon_{1}}) \notag\\
+\phi _{\mathcal{F}H}(h^{-\varepsilon _{1}}) \Big]^2
+
\frac{1}{n}u_h^2
.
$$
Given the bound on $u_h$ and equation \eqref{eq10j} from the proof of Theorem \ref{thm10j} we have the following theorem, which is stated without proof.
%%%%%%%%%%%%%%%%%%%%%%%%%%%%%%%%%%%%%%%%%%
\begin{theorem}\label{thm14}
Suppose the conditions of Theorem \ref{thm11} hold, $\supp H \subset [-\gamma,\gamma]$ for some $\gamma >0$, and 
% $C_H:=\underset{t \in [-\gamma,\gamma]}{\sup}|H(t)|$, 
$\underset{t \in [-\gamma,\gamma]}{\sup}|H(t)|<\infty$. 
% Suppose $H\in L(\R)$, and Assumptions \ref{A1}.1, \ref{A1}.3, and \ref{A2} hold.
% If the equivalent conditions stated in Theorem \ref{thm11} hold, $\supp H \subset [-\gamma,\gamma]$ for some $\gamma >0$ and 
% $C_H:=\underset{t \in [-\gamma,\gamma]}{\sup}|H(t)|$, 
% $\underset{t \in [-\gamma,\gamma]}{\sup}|H(t)|<\infty$,
Then,
% then 
there exist constants $C, \, \beta>0$ such that, 
for every distribution function $F_{Y}$ and $x\in J(F_{Y})$, the following holds whenever
$h,\,\varepsilon _{1}\in (0,1)$ and $\varepsilon_{2}>\varepsilon _{1}$,
% $h \in (0,1]$ 
\begin{align*}
MSE(\hat{p}_x) 
&\le 
\Big[ 
    \omega _{\mathcal{F}H}(h^{\varepsilon_{2}-\varepsilon _{1}})\left( p_{x}+\delta _{F_Y}(x,h^{1-\varepsilon _{1}})\right)
    +
    \left( 1+\left\Vert \mathcal{F}H\right\Vert _{C_b}\right) \delta _{F_Y}(x,h^{1-\varepsilon_{1}})
    +\phi _{\mathcal{F}H}(h^{-\varepsilon _{1}}) 
\Big]^2
\\&
\quad
+
C \frac{h^2}{n} \left(  \underset{ \beta < t\le\gamma/h}{\int} \frac{1}{|\Phi_Z(t)|}dt \right)^2
.
\end{align*}
% \begin{align*}
% MSE(\hat{p}_x) &\le C \frac{h^2}{n} \left(  \underset{ \beta < t\le\gamma/h}{\int} \frac{1}{|\Phi_Z(t)|}dt \right)^2+ \left[ \omega _{\mathcal{F}H}(h^{\varepsilon_{2}-\varepsilon _{1}})\left( p_{x}+\delta _{F_Y}(x,h^{1-\varepsilon _{1}})\right)\right.\\  
% &+\left.\left( 1+\left\Vert \mathcal{F}H\right\Vert _{C_b}\right) \delta _{F_Y}(x,h^{1-\varepsilon_{1}})
% +\phi _{\mathcal{F}H}(h^{-\varepsilon _{1}}) \right]^2.
% \end{align*}
\end{theorem}
\noindent Under additional assumptions on $\mathcal{F}H$ and $\delta_{F_Y}$, the following corollary obtains explicit rates of decay for $MSE(\hat{p}_x)$.

% ----------------------------------------------------------------------
% Corollary 2
% ----------------------------------------------------------------------
\begin{corollary}\label{coro2}
Suppose the conditions of Theorem \ref{thm11} hold, $\supp H \subset [-\gamma,\gamma]$ for some $\gamma >0$, and 
$\underset{t \in [-\gamma,\gamma]}{\sup}|H(t)|<\infty$.
Fix a distribution function $F_Y$ and $x\in J(F_Y)$. Assume that, for some $k,m>0$ and $C<\infty$
$$
\varphi_{\mathcal FH}(N)\leq CN^{-k},
\qquad
\omega_{\mathcal FH}(\varepsilon)\leq C\varepsilon^m,
\quad \text{and}\quad
\delta_{F_Y}(x,\eta)\leq\psi(\eta),
$$
for all sufficiently large $N$ and all sufficiently small $\varepsilon,\eta>0$, where $\psi$ is nondecreasing with $\psi(0^+)=0$.
Fix $\varepsilon_1\in(0,1)$ and $\varepsilon_2>\varepsilon_1$, and define
$c:=\min\!\left\{m(\varepsilon_2-\varepsilon_1),\,k\varepsilon_1\right\}.$
\begin{enumerate}
    \item[(a)]
    Suppose there exist $\tau, \rho>0$ such that \eqref{eq:error_smoothness_super} holds,
    and take $h_n=A(\log n)^{-1/\rho}$ for some $A^{\rho}>4\tau\gamma^{\rho}$. Then, 
\[
 \mathrm{MSE}(\hat p_x)
 =O\!\Big((\log n)^{-2c/\rho}
       +\psi\big(A^{1-\varepsilon_1}(\log n)^{-(1-\varepsilon_1)/\rho}\big)^{2}\Big).
\]
In particular, if $\psi(\eta)= C\eta^{s}$ for some $s>0$, take $\varepsilon_1=\tfrac{s}{k+s}$,
$\varepsilon_2=1$. If $m\ge s$, then $c=r:=\tfrac{ks}{k+s}$, and
\[
 \mathrm{MSE}(\hat p_x)=O\!\big((\log n)^{-2r/\rho}\big).
\]
\item[(b)]
    % \emph{a) Ordinary-smooth error.} 
    % Suppose $|\Phi_Z(t)|\asymp|t|^{-\rho}$ as $|t|\to\infty$, 
Suppose there exists $\rho>0$ such that \eqref{eq:error_smoothness_ordinary} holds and that $\psi(\eta)= C\eta^{s}$ for some $s>0$.
Take $\varepsilon_1=\tfrac{s}{k+s}$, $\varepsilon_2=1$, and $r:=\tfrac{ks}{k+s}$. 
If $m\geq s$ and $h_n\asymp n^{-1/\{2(r+\rho)\}}$ then 
\[
 \mathrm{MSE}(\hat p_x)=O\!\big(n^{-r/(r+\rho)}\big).
\]
\end{enumerate}
\end{corollary}

The estimator proposed in this section extends the scope
of existing work on the estimation of an arbitrary jump under contaminated sampling.
Whereas prior work (e.g., \citealp{vanEs2008,gugushvili2011,Lee2013})
has largely relied on a specified discrete-continuous mixture structure for $F_Y$ and global smoothness conditions, the present estimator applies to an arbitrary jump of a completely general
distribution function. 
We do not pursue pursue minimax optimality; in exchange, our methodology applies to a much broader class of latent distributions.

% Estimation of jumps in a latent distribution function under additive measurement error
% has previously been considered by \cite{vanEs2008} and \cite{gugushvili2011}, who assume a single atom at the origin together with an absolutely continuous component, and by \cite{Lee2013}, who allow finitely many atoms and continuous components.
% These approaches rely on a specified discrete-continuous mixture structure for $F_Y$ and regularity conditions on the continuous component. 
% In contrast, we consider an arbitrary jump at any prespecified location of an arbitrary latent distribution. Apart from the target jump, no structural restrictions are placed on  $F_Y$: it may contain countably many additional atoms and need not admit a discrete-continuous mixture representation with a globally smooth density.

%%%%%%%%%%%%%%%%%%%%%%%%%%%%%%%%%%%%%
\section{Implementation and simulation study}\label{sec:sims}
%%%%%%%%%%%%%%%%%%%%%%%%%%%%%%%%%%%%%
\newcommand{\horiz}{\noindent\makebox[\linewidth]{\rule{\paperwidth}{0.4pt}}}

\newcommand{\FHaLa}{\widehat{F}_{HL}}%
\newcommand{\FDaRe}{\widehat{F}_{DR}}%
\newcommand{\Chx}{\Phi_{Z}}%
\newcommand{\FYhatAR}{\FYhat^{AR}}%
\newcommand{\FYhatCV}{\FYhat^{CV}}%
\newcommand{\coloneqq}{:=}%
\newcommand{\sampavg}{\frac{1}{n}\sum_{j=1}^n}%

\newcommand{\pxhat}{\hat{p}_{x}}%
\newcommand{\pxhatGL}{\hat{p}_{x}^{GL}}%
\newcommand{\pxhatAR}{\hat{p}_{x}^{AR}}%
\newcommand{\tabnote}[1]{
    \smallskip 
    % \justifying
    % \small
    \footnotesize
    \noindent
    \textit{Note:} #1 
}% % LEAVE EMPTY LINE BETWEEN end{tabular} AND \tabnote{}!

\newcommand{\cvvarpen}{\kappa}

In this section we conduct simulation studies for the estimators proposed in this paper: $\widehat{F}_Y(x)$, $\widehat{F}_Y(x,y)$, and  $\hat{p}_x$.
For each estimator, we consider multiple feasible methods for tuning-parameter selection and compare performance across different sample sizes and distributions for $Y$ and $Z$.
% In each case we suggest several methods for tuning parameter selection and compare finite sample performance across different distributions for $Y$ and $Z$.

%%%%%%%%%%%%%%%%%%%%%%%%%%%%%%%%%%
\subsection{Estimation of ${F}_Y(x)$}\label{sec:FYxSims}
%%%%%%%%%%%%%%%%%%%%%%%%%%%%%%%%%%
This subsection studies the finite-sample performance of the proposed estimator of $F_Y(x)$ for $x \in C(F_Y)$.
% The following setup is used in our simulation study.

We consider five distributions for $Y$: two smooth distributions and three non-smooth distributions with kink points. The smooth distributions are the standard normal distribution $Y\sim N(0,1)$ and the Gamma distribution $Y\sim\Gamma(3, 1/\sqrt{3})$, as considered by  \citet{Dattner2013} and \citet{Hall2008}.
The non-smooth distributions are the one-kink design $Y\sim K_1$, the gap-kink design $Y\sim K_2$, and the asymmetric piecewise-uniform design $Y\sim K_3$, shown in Figure \ref{fig:kink-dgps}.
The precise definitions of these distributions are given in Appendix~\ref{app:FYkink_def}.

\begin{figure}[h]
\centering
\makebox[\textwidth][c]{%
\begin{tikzpicture}

% The three CDFs below match the kink designs used in the simulations.
%
% K_1: a standard-normal density multiplied by 3 to the left of zero
%      and by 1 to the right. The normalizing constant is 2.
\def\Cone{2}

% K_gap: standard-normal tails with no density on [-1,1].
\pgfmathsetmacro{\PhiMinusOne}{\Phistd{-1}}
\pgfmathsetmacro{\Cgap}{2*\PhiMinusOne}

\begin{groupplot}[
    group style={group size=3 by 1, horizontal sep=1.35cm},
    width=0.35\textwidth,
    % height=0.27\textwidth,
    ymin=0, ymax=1,
    samples=200,
    axis lines=box,
    grid=major,
    grid style      =   {gray!30, line width=0.15pt},
    tick style      =   {gray!30, line width=0.20pt},
    axis line style =   {gray!75, line width=0.20pt},
    every axis plot/.append style={
        black,
        line width=0.55pt,
        mark=none,
        smooth
    },
    xlabel={$x$},
    ylabel={$F_Y(x)$},
    title style={font=\small},
    label style={font=\tiny},
    ticklabel style={font=\tiny},
    ytick={0,0.2,0.4,0.6,0.8,1},
    scaled ticks=false,
    enlargelimits=false,
    clip=true
]

% ------------------------------------------------------------------------
% One-kink design: edges = [-inf,0,inf], c = [3,1].
% ------------------------------------------------------------------------
\nextgroupplot[
    title={$Y\sim K_1$},
    xmin=-3, xmax=3,
    xtick={-3,-1.5,0,1.5,3}
]

% x <= 0: 3 Phi(x)/2.
\addplot[domain=-3:0] {3*\Phistd{x}/\Cone};

% x >= 0: [3 Phi(0) + Phi(x)-Phi(0)]/2.
\addplot[domain=0:3] {(1+\Phistd{x})/\Cone};

% ------------------------------------------------------------------------
% Gap-kink design: edges = [-inf,-1,1,inf], c = [1,0,1].
% ------------------------------------------------------------------------
\nextgroupplot[
    title={$Y\sim K_{\mathrm{2}}$},
    xmin=-3, xmax=3,
    xtick={-3,-1,0,1,3}
]

% Left normal tail.
\addplot[domain=-3:-1] {\Phistd{x}/\Cgap};

% No probability density on [-1,1].
\addplot[domain=-1:1] {\PhiMinusOne/\Cgap};

% Right normal tail.
\addplot[domain=1:3] {
    (\PhiMinusOne + \Phistd{x} - \Phistd{1})/\Cgap
};

% ------------------------------------------------------------------------
% Asymmetric piecewise-uniform design:
% edges = [-2,-0.5,0.25,2], p = [0.20,0.60,0.20].
% ------------------------------------------------------------------------
\nextgroupplot[
    title={$Y\sim K_{3}$},
    xmin=-2.25, xmax=2.25,
    xtick={-2,-0.5,0.25,2},
    xticklabels={$-2$,$-0.5$,$0.25$,$2$}
]

% Outside the left support endpoint.
\addplot[domain=-2.25:-2] {0};

% Mass 0.20 uniformly distributed over [-2,-0.5].
\addplot[domain=-2:-0.5] {0.20*(x+2)/1.5};

% Mass 0.60 uniformly distributed over [-0.5,0.25].
\addplot[domain=-0.5:0.25] {0.20 + 0.60*(x+0.5)/0.75};

% Mass 0.20 uniformly distributed over [0.25,2].
\addplot[domain=0.25:2] {0.80 + 0.20*(x-0.25)/1.75};

% Outside the right support endpoint.
\addplot[domain=2:2.25] {1};

\end{groupplot}
\end{tikzpicture}%
}
\caption{Kink distribution functions used in the simulation study.}
\label{fig:kink-dgps}
\end{figure}
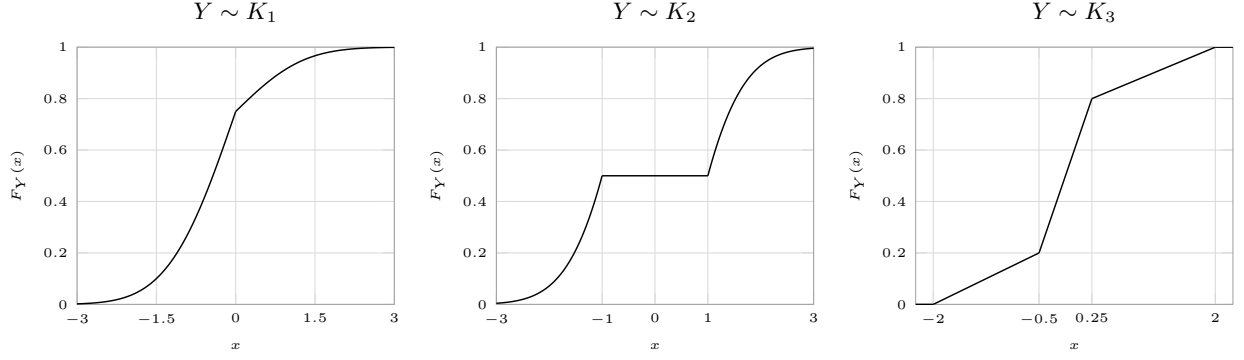

We consider three measurement-error distributions for $Z$, using rescaled
versions of error families considered in the simulation studies of
\citet{Dattner2013} and \citet{Hall2008}. The first is the normal distribution
$Z\sim N(0,1/5)$, where $1/5$ denotes the standard deviation. The second is
the Gamma distribution
$Z\sim\Gamma\left(2,1/(5\sqrt{2})\right)$, with shape $2$ and scale
$1/(5\sqrt{2})$.
The third is the symmetrized Gamma distribution from
\citet{Hall2008}.
We denote this as $Z\sim \text{S}\Gamma\left(2,1/(5\sqrt{2})\right)$ when $Z$ is distributed as $U_1 - U_2$ where $U_1$, $U_2$ are independent 
$\Gamma\left(1,1/(5\sqrt{2})\right)$ random variables.

These distributions allow us to vary error asymmetry and smoothness while
holding the error variance fixed. The normal and symmetrized-Gamma errors are
symmetric, whereas the Gamma errors are asymmetric; the normal errors are
supersmooth, whereas the two Gamma-based errors are ordinary-smooth.
% 
% In particular, if $Z\sim N(0,1/5)$, then
% \eqref{eq:error_smoothness_super} holds with $\rho=2$ and $\tau=1/50$.
% If
% $Z\sim\Gamma\left(2,1/(5\sqrt{2})\right)$ or
% $Z\sim\mathrm{S}\Gamma\left(2,1/(5\sqrt{2})\right)$, then
% \eqref{eq:error_smoothness_ordinary} holds with $\rho=2$.
% 
All three error distributions have standard deviation $\sigma_Z=1/5$.
Consequently, for a fixed $Y$ distribution, the contamination ratio $\sigma_Z/\sigma_Y$ is the same across the three $Z$ distributions.
Across the $Y$ distributions, this ratio is
$0.20$ for $Y\sim N(0,1)$ and
$Y\sim\Gamma(3,1/\sqrt{3})$, $0.22$ for $Y\sim K_1$, $0.13$ for
$Y\sim K_2$, and $0.24$ for $Y\sim K_3$.
% 
% Consequently, for any fixed distribution of $Y$, they generate the same level
% of contamination, measured by $\sigma_Z/\sigma_Y$. 
% The resulting contamination
% ratios are $0.20$ for $Y\sim N(0,1)$ and
% $Y\sim\Gamma(3,1/\sqrt{3})$, $0.22$ for $Y\sim K_1$, $0.13$ for
% $Y\sim K_2$, and $0.24$ for $Y\sim K_3$.

The five distributions of $Y$ and three distributions of $Z$ yield 15
measurement-error designs. We consider each design at sample sizes
$n\in\{200,1000,5000\}$, giving 45 simulation settings, and generate 500
independent samples for each setting. For every sample, we estimate $F_Y(x)$
at a collection of evaluation points. For the smooth distributions of $Y$, we
use the five quantiles satisfying
$
F_Y(x)\in\{0.1,0.3,0.5,0.7,0.9\}.
$
For the non-smooth distributions, we focus on their kink points: $x=0$ for
$Y\sim K_1$; $x\in\{-1,1\}$ for $Y\sim K_2$; and
$x\in\{-1/2,1/4\}$ for $Y\sim K_3$. 
The non-smooth designs therefore assess estimator performance at points where
$F_Y$ is not differentiable, whereas the smooth designs assess performance at
points where $F_Y$ is continuously differentiable.

In all of the above designs we report two feasible implementations of $\FYhat(x)$, which choose $(h,\lambda)$ by different procedures.
Both implementations use the Fourier kernel
    \begin{equation}\label{eq:Hsimsdef}
        H(t) \;=\;
        \begin{cases}
        (1-t^2)^{4}, & |t| \leq 1, \\[1.2em]
        0, & |t| > 1.
        \end{cases}
    \end{equation}

% In all of the above designs we report two feasible implementations of $\FYhat(x)$, both of which use the  Fourier kernel
%     \begin{equation}\label{eq:Hsimsdef}
%         H(t) \;=\;
%         \begin{cases}
%         (1-t^2)^{4}, & |t| \leq 1, \\[1.2em]
%         0, & |t| > 1.
%         \end{cases}
%     \end{equation}
The first implementation, denoted by $\FYhatAR(x)$, chooses $(h,\lambda)$ using a simple asymptotic rate rule motivated by Corollary \ref{coro1}. 
Applying the
corollary with $k=4$ and $s=q=1$ gives
% Using Corollary \ref{coro1} with $s=1$ and $q=1$ therein, $\FYhatAR(x)$ uses
\[
(h_n,\lambda_n)=
\begin{cases}
\left(\dfrac{1}{4}n^{-5/28},\,h_n^{4/5}\right),
    & 
    % \text{under ordinary-smooth errors},
    \text{if } 
    Z\sim\Gamma\left(2,\frac{1}{5\sqrt{2}}\right)  
    \text{ or }
    Z\sim\mathrm{S}\Gamma\left(2,\frac{1}{5\sqrt{2}}\right)\!,
    \\[0.8em]
\left(\dfrac{1.01}{5}(\log n)^{-1/2},\,h_n^{4/5}\right),
    & 
    \text{if } Z\sim N\left(0,\frac{1}{5}\right)\!
    % \text{under super-smooth errors}
    ,
\end{cases}
\]
where the factor $1.01$ ensures that the super-smooth-error rule satisfies 
$
h_n
>\gamma (2\tau/\log(n))^{1/\rho}
.
$

The second implementation, denoted $\FYhatCV(x)$, chooses $(h,\lambda)$ using a 
% variance penalized 
reconvolution cross-validation procedure. Since $Y$ is latent, each candidate $(h,\lambda)$ is evaluated by reconvolving $\FYhat$ with the known distribution of $Z$ and comparing the resulting estimate of $F_X$ with the empirical distribution of the observed $X$-sample. The criterion includes a
variance penalty term based on Theorem \ref{thm8}. 
We fix the variance-penalty coefficient for $\FYhatCV(x)$ at $\cvvarpen=0.0025$ across all DGPs and sample sizes. 
A detailed description of $\FYhatCV(x)$ is given in Appendix~\ref{app:FYx_CV}.
Table~\ref{tab:FYx_CV_sensitivity} in Appendix~\ref{app:Full_sim_results} reports results for $\cvvarpen\in\{0.001,0.0025,0.005\}$, which indicate the overall performance is relatively stable across alternative $\cvvarpen$ choices.

For comparison, we include two deconvolution estimators from the existing
literature. The first, $\FDaRe(x)$, is the adaptive estimator of
\citet[Section~2.2]{Dattner2013}. 
The second, $\FHaLa(x)$, is the estimator of
\citet{Hall2008}, with bandwidth selected using the normal-reference procedure
described in their Section~4.2. 
% The second, $\FHaLa(x)$, is the distribution estimator with normal-reference bandwidth selection 
% using the normal-reference procedure
% described in their Section~4.2.
Implementation details for
% $\FYhatCV(x)$, 
$\FDaRe(x)$ and $\FHaLa(x)$ are provided in
Appendix~\ref{app:SimsEstImplDet}.

%%%%%%%%%%%%%%%%%%%%%%%%%%%%%%%%%%
\subsubsection{Numerical results}
%%%%%%%%%%%%%%%%%%%%%%%%%%%%%%%%%%
{
\setlength{\tabcolsep}{9pt}
\begin{table}[h!t]
\caption{
    Monte Carlo MSE$\times 100$ and variance$\times 100$ (in parentheses) for estimators of $F_Y(x)$. 
    }\label{tab:FYx_main}
\adjustbox{max width=\textwidth, center = \textwidth}{ 
\small
% Requires \usepackage{booktabs}
% Displayed MSE x 100; displayed variance x 100.
% The difference between MSE and variance is squared bias.
% Boldface marks the smallest unrounded MSE in each row.
\begin{tabular}{lcccc}
\toprule
&
$\FYhatCV(x)$ &
$\FYhatAR(x)$ &
$\FDaRe(x)$ &
$\FHaLa(x)$ \\
\midrule
\multicolumn{1}{l}{\rule[-.10ex]{0pt}{3.25ex}\!Smooth $F_Y$, $Z\!\sim\! N\!\left(0,\frac{1}{5}\right)\!$} &  &  &  &  \\
\quad $n=200$ & 0.095 (0.082) & 0.080 (0.075) & \textbf{0.079 (0.074)} & 0.091 (0.090) \\
\quad $n=1000$ & 0.018 (0.017) & 0.017 (0.015) & \textbf{0.015 (0.014)} & 0.018 (0.018) \\
\quad $n=5000$ & 0.005 (0.004) & 0.005 (0.003) & \textbf{0.004 (0.003)} & 0.004 (0.004) \\
\multicolumn{1}{l}{\rule[-.10ex]{0pt}{3.25ex}\!Smooth $F_Y$, $Z\!\sim\! \Gamma\!\left(2,\frac{1}{5\sqrt{2}}\right)\!$} &  &  &  &  \\
\quad $n=200$ & 0.087 (0.074) & 0.071 (0.063) & \textbf{0.069 (0.064)} & 0.079 (0.078) \\
\quad $n=1000$ & 0.018 (0.017) & 0.017 (0.015) & \textbf{0.015 (0.014)} & 0.017 (0.017) \\
\quad $n=5000$ & 0.004 (0.004) & 0.004 (0.003) & \textbf{0.004 (0.003)} & 0.004 (0.004) \\
\multicolumn{1}{l}{\rule[-.10ex]{0pt}{3.25ex}\!Smooth $F_Y$, $Z\!\sim\! \mathrm{S}\Gamma\!\left(2,\frac{1}{5\sqrt{2}}\right)\!$} &  &  &  &  \\
\quad $n=200$ & 0.093 (0.081) & 0.077 (0.070) & \textbf{0.076 (0.071)} & 0.092 (0.092) \\
\quad $n=1000$ & 0.018 (0.017) & 0.017 (0.015) & \textbf{0.015 (0.014)} & 0.018 (0.018) \\
\quad $n=5000$ & 0.004 (0.004) & 0.004 (0.003) & \textbf{0.003 (0.003)} & 0.004 (0.004) \\
\midrule
\multicolumn{1}{l}{\rule[-.10ex]{0pt}{3.25ex}\!Non-smooth $F_Y$, $Z\!\sim\! N\!\left(0,\frac{1}{5}\right)\!$} &  &  &  &  \\
\quad $n=200$ & 0.361 (0.144) & 0.443 (0.092) & 0.373 (0.104) & \textbf{0.267 (0.109)} \\
\quad $n=1000$ & \textbf{0.162 (0.037)} & 0.300 (0.019) & 0.234 (0.020) & 0.171 (0.022) \\
\quad $n=5000$ & \textbf{0.113 (0.008)} & 0.232 (0.004) & 0.183 (0.004) & 0.131 (0.005) \\
\multicolumn{1}{l}{\rule[-.10ex]{0pt}{3.25ex}\!Non-smooth $F_Y$, $Z\!\sim\! \Gamma\!\left(2,\frac{1}{5\sqrt{2}}\right)\!$} &  &  &  &  \\
\quad $n=200$ & 0.325 (0.148) & 0.516 (0.082) & 0.376 (0.098) & \textbf{0.247 (0.100)} \\
\quad $n=1000$ & \textbf{0.091 (0.030)} & 0.271 (0.018) & 0.230 (0.019) & 0.144 (0.020) \\
\quad $n=5000$ & \textbf{0.036 (0.007)} & 0.151 (0.004) & 0.185 (0.004) & 0.099 (0.005) \\
\multicolumn{1}{l}{\rule[-.10ex]{0pt}{3.25ex}\!Non-smooth $F_Y$, $Z\!\sim\! \mathrm{S}\Gamma\!\left(2,\frac{1}{5\sqrt{2}}\right)\!$} &  &  &  &  \\
\quad $n=200$ & 0.342 (0.165) & 0.540 (0.095) & 0.399 (0.112) & \textbf{0.227 (0.123)} \\
\quad $n=1000$ & \textbf{0.087 (0.030)} & 0.269 (0.018) & 0.230 (0.020) & 0.107 (0.022) \\
\quad $n=5000$ & \textbf{0.034 (0.007)} & 0.150 (0.004) & 0.184 (0.003) & 0.073 (0.004) \\
\bottomrule
\end{tabular}%

}

    \tabnote{Boldface identifies the estimator with
        the smallest unrounded MSE in each row.
        Smooth $F_Y$ includes 
        $Y\sim N(0,1)$ and $Y\sim\Gamma(3,1/\sqrt{3})$; non-smooth $F_Y$ includes
        $Y\sim K_1$, $K_2,$ and $K_3$.
        Within each $Y$ design, 
        empirical MSE and variance are first averaged over the
        evaluation points; these design-level quantities are then averaged 
        across the $Y$ designs in each group and multiplied by 100 for readability.
        Full DGP-by-DGP results are reported in Tables \ref{tab:FYx_app_smth} and \ref{tab:FYx_app_nonsmth} in Appendix~\ref{app:Full_sim_results}.}
\end{table}
}

Table \ref{tab:FYx_main} reports empirical MSE and variance for the four estimators of $F_Y(x)$ across the smooth and
non-smooth $Y$ distribution designs. MSE decreases with the sample size for every
estimator, but the relative rankings differ markedly between the smooth and
non-smooth designs.

% Table \ref{tab:FYx_main}
% reports empirical MSE and variance across the smooth and
% non-smooth designs for the four estimators of $F_Y(x)$. The MSE of every estimator decreases with the sample size,
% although the relative ranking depends strongly on whether $F_Y$ is smooth at
% the evaluation points. 
% % The distinction between ordinary- and super-smooth
% % measurement errors is most apparent in the non-smooth designs, whereas error
% % asymmetry has comparatively little effect.

For the smooth designs, $\FDaRe(x)$ has the smallest grouped MSE
for every measurement-error distribution and sample size. However, its
advantage is generally modest, with $\FYhatAR(x)$ also performing well.

When estimating non-smooth $F_Y$ at kink points, the ranking reverses.
% The ranking changes substantially at the kink points.
Across all three error distributions,
% Under each error distribution, 
$\FHaLa(x)$ has the smallest grouped MSE when $n=200$, and 
$\FYhatCV(x)$ has the smallest grouped MSE when $n=1000$ and $n=5000$.
The advantage of $\FYhatCV(x)$ is most evident under
ordinary-smooth errors.
At $n=5000$, for example,
its MSE is approximately $14\%$, $64\%$, and $53\%$ smaller than that of 
% the next-best estimator 
$\FHaLa(x)$
under normal, Gamma, and symmetrized-Gamma errors,
respectively.
Table~\ref{tab:FYx_app_nonsmth} in Appendix~\ref{app:Full_sim_results} reports results for each $Y$ distribution separately,
% DGP-specific results 
which shows that this pattern is not driven by a single non-smooth distribution: $\FYhatCV(x)$ has the smallest MSE in all nine non-smooth DGP--error combinations when $n=5000$, and in seven of the nine combinations when $n=1000$.

Because MSE and variance are reported on the same scale, their difference
equals squared bias. 
Bias is relatively small in the smooth designs, and the lower-variance estimators tend to perform best. 
At the kink points, squared bias accounts for a much larger share of MSE. In these cases, $\FYhatCV(x)$ trades higher variance for a reduction in squared bias, resulting in lower MSE once the sample
size is sufficiently large.
Importantly, the variance of $\FYhatCV(x)$ is not uniformly larger across the simulation designs, so its performance
at the kink points does not simply reflect a generally more variable estimator.

%%%%%%%%%%%%%%%%%%%%%%%%%%%%%%%%%%
\subsection{Estimation of ${F}_Y(x,y)$}
%%%%%%%%%%%%%%%%%%%%%%%%%%%%%%%%%%
This subsection studies the finite-sample performance of the proposed estimator of the interval probability $F_Y(x,y)$ for $x<y$ and $x,y \in C(F_Y)$. 
% We use the same five distributions for $Y$, three measurement-error
% distributions, sample sizes $n\in\{200,1000,5000\}$, and number of Monte Carlo
% replications as in Section~\ref{sec:FYxSims}.
We use the same distributions for $Y$ and $Z$, sample sizes, and number of Monte Carlo replications as in Section \ref{sec:FYxSims}. 

For the two smooth $Y$ distributions, we estimate $F_Y(x,y)$ at the points 
$
x=F_Y^{-1}\!\left(0.5-{p}/{2}\right), 
$
and
$
y=F_Y^{-1}\!\left(0.5+{p}/{2}\right),
$
for $p\in\{0.05,0.20,0.80\}$,
so that $F_Y(x,y)=p$. 
% For the non-smooth distributions, we instead use intervals based on the kink points. For the seven-kink design $K_7$, we use $ (x,y)\in\left\{ \left(-{1}/{2},{1}/{2}\right), (-1,1), \left(-{3}/{2},{3}/{2}\right) \right\}, $ 
% which correspond to true probabilities $F_Y(x,y)$ of approximately 
% $
% % F_Y(x,y)\approx 
% 0.14, 0.69,$ and $ 0.76$, respectively.
% For the one-kink design $K_1$, we use  $(x,y)\in\left\{ (-1,0), (0,1), \left(-{3}/{2},{3}/{2}\right) \right\}$, 
% which correspond to true probabilities $F_Y(x,y)$ of approximately 
% $
% % F_Y(x,y)\approx 
% 0.60, 0.09,$ and $ 0.87$, respectively.
% 
 For the non-smooth distributions, we use the following
interval pairs:
    $
    (x,y)\in
    \left\{
        (-1,0),\ (0,1),\
        \left(-\frac32,\frac32\right)
    \right\}
    $
    for $Y\sim K_1$;
    $
    (x,y)\in
    \left\{
        \left(-1,\frac32\right),\
        \left(-\frac32,1\right),\
        \left(-\frac32,\frac32\right)
    \right\}
    $
    for $Y\sim K_2$;
    and
    $
    (x,y)\in
    \left\{
        \left(-\frac54,-\frac12\right),\
        \left(-\frac12,\frac14\right),\
        \left(\frac14,\frac98\right)
    \right\}
    $
    for $Y\sim K_3$.
% The corresponding interval probabilities are approximately
% \[
% \begin{aligned}
% K_1 &: \quad (0.512,\ 0.171,\ 0.866),\\
% K_2 &: \quad (0.289,\ 0.289,\ 0.579),\\
% K_3 &: \quad (0.100,\ 0.600,\ 0.100).
% \end{aligned}
% \]
Seven of these nine intervals have at least one kink point as an endpoint;
the remaining two are wider intervals that span the non-smooth region. The
design therefore examines interval-probability estimation both when an
endpoint is a point of non-differentiability and when the interval contains
non-smooth features in its interior.

We consider two feasible implementations of $\FYhat(x,y)$ which are analogous to those considered in Section \ref{sec:FYxSims} and use the same $H$ defined as in \eqref{eq:Hsimsdef}.
The first implementation, denoted
$\FYhatAR(x,y)$, uses the rate-based bandwidth
\[
h_n=
\begin{cases}
\dfrac{1}{5}n^{-5/28},
    & Z\sim\Gamma\left(2,\dfrac{1}{5\sqrt{2}}\right)
      \text{ or }
      Z\sim\mathrm{S}\Gamma\left(2,\dfrac{1}{5\sqrt{2}}\right),\\[1em]
\dfrac{0.8}{5}(\log n)^{-1/2},
    & Z\sim N\left(0,\dfrac{1}{5}\right).
\end{cases}
\]
The second, denoted $\widehat F_Y^{CV}(x,y)$, chooses $h$ by a similar reconvolution cross-validation procedure as before; each candidate $h$ is evaluated by reconvolving the corresponding estimate with the known distribution of $Z$ and comparing the resulting estimate of interval probabilities for $X$ with empirical interval probabilities from the observed $X$-sample.
The CV criterion includes a variance penalty term following from Theorem \ref{thm8a}, and
we fix the variance-penalty weight for $\FYhatCV(x,y)$ at $\cvvarpen=0.01$ across all designs. 
Table~\ref{tab:FYxy_CV_sensitivity} in Appendix~\ref{app:Full_sim_results} reports results for $\cvvarpen\in\{0.005,0.010,0.0125\}$, which indicate that overall performance is relatively stable across alternative $\cvvarpen$ choices. 
A detailed description of $\FYhatCV(x,y)$ is provided in Appendix~\ref{app:FYxy_CV}.

Although the estimators of \citet{Dattner2013}  and \citet{Hall2008} are developed for the pointwise distribution function $F_Y(x)$, they imply natural interval-probability estimators. 
For comparison, we therefore include
$$
\FDaRe(x,y) = \min\left\{1,\max\right\{0,\FDaRe(y)-\FDaRe(x)\left\}\right\}
, \quad \text{and} \quad
\FHaLa(x,y) = \min\left\{1,\max\right\{0,\FHaLa(y)-\FHaLa(x)\left\}\right\}
.
$$

%%%%%%%%%%%%%%%%%%%%%%%%%%%%%%%%%%
\subsubsection{Numerical results}
%%%%%%%%%%%%%%%%%%%%%%%%%%%%%%%%%%

{
\setlength{\tabcolsep}{9pt}
\begin{table}[h!t]
    \caption{
        Monte Carlo MSE$\times 100$ and variance$\times 100$ (in parentheses) for estimators of $F_Y(x,y)$.  
        }\label{tab:FYxy_main}
    \adjustbox{max width=\textwidth, center = \textwidth}{ 
    \small
    % Requires \usepackage{booktabs}
% Displayed MSE x 100; displayed variance x 100.
% The difference between MSE and variance is squared bias.
% Boldface marks the smallest unrounded MSE in each row.
\begin{tabular}{lcccc}
\toprule
&
$\FYhatCV(x,y)$ &
$\FYhatAR(x,y)$ &
$\FDaRe(x,y)$ &
$\FHaLa(x,y)$ \\
\midrule
\multicolumn{1}{l}{\rule[-.10ex]{0pt}{3.25ex}\!Smooth $F_Y$, $Z\!\sim\! N\!\left(0,\frac{1}{5}\right)\!$} &  &  &  &  \\
\quad $n=200$ & 0.055 (0.048) & 0.048 (0.043) & \textbf{0.043 (0.037)} & 0.062 (0.061) \\
\quad $n=1000$ & 0.013 (0.011) & 0.013 (0.010) & \textbf{0.010 (0.007)} & 0.014 (0.013) \\
\quad $n=5000$ & 0.004 (0.004) & 0.005 (0.003) & \textbf{0.003 (0.002)} & 0.004 (0.003) \\
\multicolumn{1}{l}{\rule[-.10ex]{0pt}{3.25ex}\!Smooth $F_Y$, $Z\!\sim\! \Gamma\!\left(2,\frac{1}{5\sqrt{2}}\right)\!$} &  &  &  &  \\
\quad $n=200$ & 0.054 (0.048) & 0.046 (0.040) & \textbf{0.042 (0.035)} & 0.057 (0.056) \\
\quad $n=1000$ & 0.012 (0.011) & 0.012 (0.010) & \textbf{0.010 (0.007)} & 0.013 (0.012) \\
\quad $n=5000$ & 0.003 (0.003) & 0.003 (0.003) & \textbf{0.003 (0.002)} & 0.003 (0.003) \\
\multicolumn{1}{l}{\rule[-.10ex]{0pt}{3.25ex}\!Smooth $F_Y$, $Z\!\sim\! \mathrm{S}\Gamma\!\left(2,\frac{1}{5\sqrt{2}}\right)\!$} &  &  &  &  \\
\quad $n=200$ & 0.054 (0.048) & 0.045 (0.040) & \textbf{0.040 (0.034)} & 0.069 (0.068) \\
\quad $n=1000$ & 0.013 (0.011) & 0.013 (0.011) & \textbf{0.010 (0.007)} & 0.015 (0.014) \\
\quad $n=5000$ & 0.003 (0.003) & 0.003 (0.003) & \textbf{0.003 (0.002)} & 0.004 (0.003) \\
\midrule
\multicolumn{1}{l}{\rule[-.10ex]{0pt}{3.25ex}\!Non-smooth $F_Y$, $Z\!\sim\! N\!\left(0,\frac{1}{5}\right)\!$} &  &  &  &  \\
\quad $n=200$ & 0.319 (0.116) & 0.393 (0.074) & 0.390 (0.085) & \textbf{0.274 (0.095)} \\
\quad $n=1000$ & \textbf{0.180 (0.033)} & 0.270 (0.017) & 0.234 (0.018) & 0.188 (0.021) \\
\quad $n=5000$ & \textbf{0.138 (0.008)} & 0.210 (0.004) & 0.176 (0.004) & 0.144 (0.005) \\
\multicolumn{1}{l}{\rule[-.10ex]{0pt}{3.25ex}\!Non-smooth $F_Y$, $Z\!\sim\! \Gamma\!\left(2,\frac{1}{5\sqrt{2}}\right)\!$} &  &  &  &  \\
\quad $n=200$ & 0.283 (0.116) & 0.414 (0.071) & 0.402 (0.084) & \textbf{0.261 (0.091)} \\
\quad $n=1000$ & \textbf{0.106 (0.027)} & 0.210 (0.017) & 0.232 (0.017) & 0.152 (0.018) \\
\quad $n=5000$ & \textbf{0.052 (0.007)} & 0.115 (0.004) & 0.177 (0.004) & 0.103 (0.004) \\
\multicolumn{1}{l}{\rule[-.10ex]{0pt}{3.25ex}\!Non-smooth $F_Y$, $Z\!\sim\! \mathrm{S}\Gamma\!\left(2,\frac{1}{5\sqrt{2}}\right)\!$} &  &  &  &  \\
\quad $n=200$ & 0.289 (0.128) & 0.424 (0.077) & 0.414 (0.090) & \textbf{0.230 (0.111)} \\
\quad $n=1000$ & \textbf{0.101 (0.031)} & 0.210 (0.018) & 0.233 (0.018) & 0.116 (0.022) \\
\quad $n=5000$ & \textbf{0.045 (0.008)} & 0.115 (0.004) & 0.176 (0.003) & 0.077 (0.005) \\
\bottomrule
\end{tabular}%

    }

    \tabnote{Boldface identifies the estimator with
        the smallest unrounded empirical MSE in each row.
        Smooth $F_Y$ includes 
        $Y\sim N(0,1)$ and $Y\sim\Gamma(3,1/\sqrt{3})$; non-smooth $F_Y$ includes
        $Y\sim K_1$, $K_2,$ and $K_3$.
        Within each $Y$ design, 
        empirical MSE and variance are first averaged over the
        evaluation points; these design-level quantities are then averaged 
        across the $Y$ designs in each group and multiplied by 100 for readability.
        Full DGP-by-DGP results are reported in Tables \ref{tab:FYxy_app_smth} and \ref{tab:FYxy_app_nonsmth} in Appendix~\ref{app:Full_sim_results}.}
\end{table}
}

Table~\ref{tab:FYxy_main}
reports empirical MSE and variance for the four estimators of $F_Y(x,y)$.
The patterns are broadly similar to those
for pointwise estimation of $F_Y(x)$.

For smooth designs,
$\FDaRe(x,y)$ has the smallest grouped MSE throughout, and $\FYhatAR(x,y)$ performs competitively. 
However, the DGP-level results from Table~\ref{tab:FYxy_app_smth} in Appendix~\ref{app:Full_sim_results} show that the advantage of $\FDaRe(x,y)$ is concentrated in the $Y\sim \Gamma\big(3,1/\sqrt{3}\big)$ designs, and $\FYhatAR(x,y)$ generally has a small advantage in the $Y\sim N(0,1)$ designs.

For non-smooth $F_Y$, $\FHaLa(x,y)$ has the smallest grouped MSE when $n=200$, whereas $\FYhatCV(x,y)$ performs best under all three error distributions when $n=1000$ and $n=5000$. 
As with the $F_Y(x)$ results, the advantage of $\FYhatCV(x,y)$ is most prominent under ordinary-smooth errors. 
For example, at $n=5000$, $\FYhatCV(x,y)$ reduces MSE  relative to $\FHaLa(x,y)$ by approximately $4\%$, $50\%$, and $42\%$ under normal, Gamma, and symmetrized-Gamma errors, respectively.
Table~\ref{tab:FYxy_app_nonsmth} in Appendix~\ref{app:Full_sim_results} shows that
$\FYhatCV(x,y)$ has the smallest MSE in all nine non-smooth combinations when
$n=5000$, and has the smallest MSE in four of the nine non-smooth designs when $n=1000$. 

The corresponding bias--variance pattern is similar to the
$F_Y(x)$ estimation results. 
In the non-smooth designs, the higher variance of $\FYhatCV(x,y)$ is offset by a larger reduction in squared
bias when sample size is large, whereas in smooth designs the variance of $\FYhatCV(x,y)$ is comparable to
that of the other estimators.

%%%%%%%%%%%%%%%%%%%%%%%%%%%%%%%%%%
\subsection{Estimation of ${p}_x$}
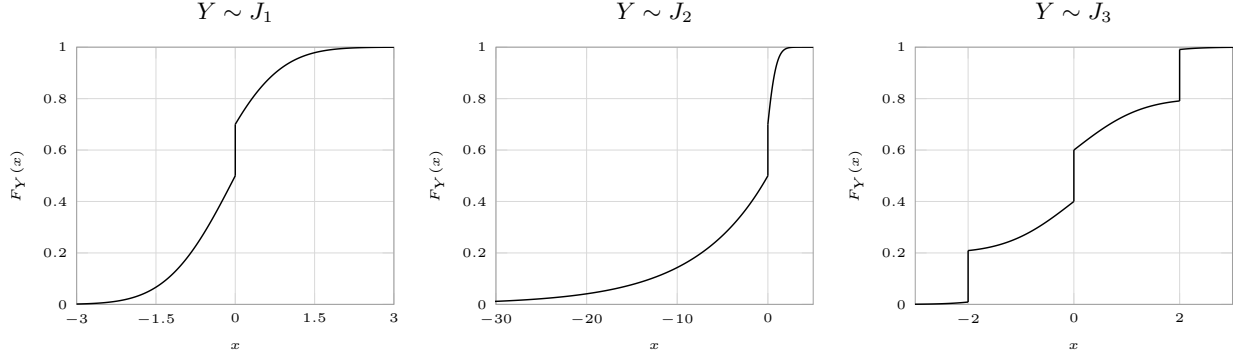
\begin{figure}[h!]
\centering
\makebox[\textwidth][c]{%
\begin{tikzpicture}
\def\pzero{0.20}
\def\mujump{-0.5244005}

\begin{groupplot}[
    group style={group size=3 by 1, horizontal sep=1.35cm},
    width=0.35\textwidth,
    % width=0.33\textwidth,
    % height=0.24\textwidth,
    ymin=0, ymax=1,
    samples=200,
    axis lines=box,
    grid=major,
    grid style      =   {gray!30, line width=0.15pt},
    tick style      =   {gray!30, line width=0.20pt},
    axis line style =   {gray!75, line width=0.20pt},
    every axis plot/.append style={
        black,
        line width=0.55pt,
        mark=none,
        smooth
    },
    xlabel={$x$},
    ylabel={$F_Y(x)$},
    title style={font=\small},
    label style={font=\tiny},
    ticklabel style={font=\tiny},
    ytick={0,0.2,...,1},
    scaled ticks=false,
    enlargelimits=false,
    clip=true
]

% ------------------------------------------------------------
% J1
% ------------------------------------------------------------
\nextgroupplot[
    title={$Y\sim J_1$},
    xmin=-3, xmax=3,
    xtick={-3,-1.5,0,1.5,3},
]

\addplot[domain=-3:0] {\Phistd{x}};
\addplot[domain=0:3] {\Phistd{x-(\mujump)}};

\addplot[sharp plot] coordinates {
    (0,0.5)
    (0,{0.5+\pzero})
};

% ------------------------------------------------------------
% J2
% ------------------------------------------------------------
\nextgroupplot[
    title={$Y\sim J_2$},
    xmin=-30, xmax=5,
    xtick={-40,-30,-20,-10,0},
]

\addplot[domain=-40:0] {0.5*exp(x/8)};
\addplot[domain=0:5] {\Phistd{x-(\mujump)}};

\addplot[sharp plot] coordinates {
    (0,0.5)
    (0,{0.5+\pzero})
};

% ------------------------------------------------------------
% J3
% ------------------------------------------------------------
\nextgroupplot[
    title={$Y\sim J_3$},
    xmin=-3, xmax=3,
    xtick={-4,-2,0,2,4},
    % xtick={-3,-2,0,2,3},
    % ylabel={}
    % ylabel={\tiny $F_Y^{(J_3)}(x)$},
    % label style={font=\footnotesize},
]

\addplot[domain=-4:-2] {(1-3*\pzero)*\Phistd{x}};
\addplot[domain=-2:0] {\pzero + (1-3*\pzero)*\Phistd{x}};
\addplot[domain=0:2] {2*\pzero + (1-3*\pzero)*\Phistd{x}};
\addplot[domain=2:4] {3*\pzero + (1-3*\pzero)*\Phistd{x}};

\addplot[sharp plot] coordinates {
    (-2,{(1-3*\pzero)*\Phistd{-2}})
    (-2,{\pzero+(1-3*\pzero)*\Phistd{-2}})
};

\addplot[sharp plot] coordinates {
    (0,{\pzero+(1-3*\pzero)*\Phistd{0}})
    (0,{2*\pzero+(1-3*\pzero)*\Phistd{0}})
};

\addplot[sharp plot] coordinates {
    (2,{2*\pzero+(1-3*\pzero)*\Phistd{2}})
    (2,{3*\pzero+(1-3*\pzero)*\Phistd{2}})
};

\end{groupplot}
\end{tikzpicture}
}
\caption{Jump distribution functions used in the simulation study, illustrated for $p_x=0.20$.}
\label{fig:jump-dgps}
\end{figure}

%%%%%%%%%%%%%%%%%%%%%%%%%%%%%%%%%%
This subsection studies the finite-sample performance of the proposed estimator of a jump
$p_x=F_Y(x)-\underset{\varepsilon \downarrow 0}{\lim}\,F_Y(x-\varepsilon)$
% $p_x = F_Y(x)-F_Y(x-0)$ 
at the point $x\in J(F_Y)$.
We consider three types of jump distributions for $F_Y$. 
Let $\Phi$ denote the standard normal distribution function.
The first two,
defined as in \citet{Mynbaev2022},
have a single jump of size $p$ at $x=0$ 
    $$
    F_Y^{(J_1)}(x;p) 
    \;=\;
    \begin{cases}
            \Phi(x), 
        & 
            x <0, 
        \\
            \frac{1}{2} + p, 
        & 
            x = 0, 
        \\
            \int_{-\infty}^x 
            \frac{1}{\sqrt{2\pi}}e^{-\frac{1}{2}(t-\mu_R)^2}
            \,dt
        & 
            x >0, 
    \end{cases}
    \qquad
    F_Y^{(J_2)}(x;p)
    \;=\;
    \begin{cases}
            \frac{1}{2}e^{x/8}, 
        & 
            x <0, 
        \\%[1.2em]
            \frac{1}{2} + p, 
        & 
            x = 0, 
        \\%[1.2em]
            \int_{-\infty}^x 
            \frac{1}{\sqrt{2\pi}}e^{-\frac{1}{2}(t-\mu_R)^2}
            \,dt
        & 
            x >0, 
    \end{cases}
    $$
where $\mu_R=-\Phi^{-1}(1/2+p)$.
The third distribution has three jumps, each of
size $p$, at $x\in\{-2,0,2\}$,
    $$
    F_Y^{(J_3)}(x;p) \;=\;  
        p \cdot \chi_{\{x \geq -2\}}
        \;+\; p \cdot \chi_{\{x \geq 0\}}
        \;+\; p \cdot \chi_{\{x \geq 2\}}
        \;+\; (1-3p) \cdot 
        \Phi\!\left(x\right)
        % \Phi\!\left(x\right)
        .
    $$
We write $Y\sim J_1$, $J_2$, or $J_3$ when $F_Y$ is given by $F_Y^{(J_1)}$, $F_Y^{(J_2)}$, or $F_Y^{(J_3)}$, respectively.
Figure~\ref{fig:jump-dgps} illustrates these three distribution functions for $p_x=0.20$.

For each $p\in\{ 0.01, 0.10, 0.20\}$ we estimate the jump at $x=0$ for $F_Y^{(J_1)}$ and $F_Y^{(J_2)}$, and the jumps at $x\in\{-2,0,2\}$ for  $F_Y^{(J_3)}$.
We consider the same error distributions and sample sizes as before,
$
    n\in\{200,1000,5000\},
$
and generate 500 independent Monte Carlo samples for each design.

We report two feasible implementations of the proposed jump estimator. Both use an Epanechnikov kernel,
\[
    H(t)=\frac{3}{4}(1-t^2)\chi_{\{|t|\leq 1\}}.
\]

The first implementation, denoted $\pxhatAR$, chooses $h$ according to an asymptotic rule motivated by Corollary~\ref{coro2}. Applying the corollary with $k=m=2$ and $s=1$ gives 
\[
h_n=
\begin{cases}
Cn^{-3/16},
    & 
    % \text{under ordinary-smooth errors},
    \text{if } 
    Z\sim\Gamma\left(2,\frac{1}{5\sqrt{2}}\right)  
    \text{ or }
    Z\sim\mathrm{S}\Gamma\left(2,\frac{1}{5\sqrt{2}}\right)\!,
    \\[0.8em]
1.01\frac{\sqrt{2}}{5}
% 1.01\sqrt{4\tau}\,
(\log n)^{-1/2},
    & 
    \text{if } Z\sim N\left(0,\frac{1}{5}\right)\!
    % \text{under super-smooth errors}
    ,
\end{cases}
\]
where
$C$
is calibrated so that
% $h_{1000}=0.01461$,
$h_{1000} \approx 0.015$ for ordinary-smooth errors,
and
the factor
$1.01$ provides a margin above the lower bound on the
super-smooth constant in Corollary~\ref{coro2}.

The second implementation, denoted $\pxhatGL$, chooses $h$ using a feasible Goldenshluger--Lepski style bandwidth selector, motivated by the procedure developed in \citet{goldenshluger_universal_2008}.
For each candidate $h$ in a finite grid $\mathcal H_{GL}$, we compute $\widehat p_{x,h}$ and the variance bound 
$
    V_h
    =
    % \frac{u_h^2}{n}
    u_h^2/n
$
defined as in Theorem \ref{thm12}.
With this, we construct a proxy MSE criterion
\[
    Q_{\kappa_A,\kappa_V}(h)
    =
    A_{\kappa_A}(h)+\kappa_V V_h
,\quad \text{where}\quad
    A_{\kappa_A}(h)
    =
    \max_{h'\in\mathcal H_{GL}: h'\leq h}
    \left[
        \left(
            \widehat p_{x,h}-\widehat p_{x,h'}
        \right)^2
        -
        \kappa_A\left(V_h+V_{h'}\right)
    \right]_+,
\]
and $[a]_+=\max\{a,0\}$.
Then, the corresponding estimator is
\[
    \pxhatGL
=
    \widehat p_{x,h_{GL}(x)} 
,\quad \text{where} \quad
    h_{GL}(x)
=
    \arg\min_{h\in\mathcal H_{GL}}
    Q_{\kappa_A,\kappa_V}(h)
    .
\]
The idea is that 
$A_{\kappa_A}(h)$ acts as a feasible proxy for squared smoothing bias: pairwise differences between $\widehat p_{x,h}$ and finer-bandwidth estimators are interpreted as evidence of bias, not just random noise, only when they exceed the threshold $\kappa_A\left(V_h+V_{h'}\right)$.
In our simulations, the candidate grid $\mathcal H_{GL}$ is constructed around the asymptotic-rate
bandwidth $h_{n}$.
For the tuning parameters we use $\kappa_A=0.05,$ and $\kappa_V=0.25.$

%%%%%%%%%%%%%%%%%%%%%%%%%%%%%%%%%%
\subsubsection{Numerical results}
%%%%%%%%%%%%%%%%%%%%%%%%%%%%%%%%%%

{
\setlength{\tabcolsep}{5pt}
\begin{table}[h!t]
    \caption{
        Monte Carlo MSE$\times 100$ and variance$\times 100$ (in parentheses) for estimators of $p_x$.
        % Empirical root MSE and empirical variance$\times 10$ (in parenthesis) for estimating $p_x$. 
        }\label{tab:jump_main}
    \adjustbox{max width=\textwidth, center = \textwidth}{ 
    \small
    % Requires \usepackage{booktabs}
% Displayed MSE x 100; displayed variance x 100.
\begin{tabular}{lcccccc}
\toprule
& \multicolumn{2}{c}{$p_x = 0.01$} & \multicolumn{2}{c}{$p_x = 0.10$} & \multicolumn{2}{c}{$p_x = 0.20$} \\
\cmidrule(lr){2-3}
\cmidrule(lr){4-5}
\cmidrule(lr){6-7}
& $\pxhatGL$ & $\pxhatAR$ & $\pxhatGL$ & $\pxhatAR$ & $\pxhatGL$ & $\pxhatAR$ \\
\midrule
\multicolumn{1}{l}{\rule[-1.6ex]{0pt}{6.0ex}\!$Z\!\sim\! N\!\left(0,\frac{1}{5}\right)\!$} &  &  &  &  &  &  \\
\quad $n=200$ & \textbf{2.626 (0.127)} & 3.040 (0.088) & \textbf{2.358 (0.155)} & 2.613 (0.113) & \textbf{1.944 (0.189)} & 2.086 (0.136) \\
\quad $n=1000$ & \textbf{1.737 (0.038)} & 2.299 (0.018) & \textbf{1.525 (0.051)} & 1.956 (0.025) & \textbf{1.250 (0.059)} & 1.556 (0.030) \\
\quad $n=5000$ & \textbf{1.235 (0.019)} & 1.828 (0.005) & \textbf{1.089 (0.022)} & 1.571 (0.006) & \textbf{0.896 (0.023)} & 1.251 (0.007) \\
\multicolumn{1}{l}{\rule[-1.6ex]{0pt}{6.0ex}\!$Z\!\sim\! \Gamma\!\left(2,\frac{1}{5\sqrt{2}}\right)\!$} &  &  &  &  &  &  \\
\quad $n=200$ & 0.922 (0.199) & \textbf{0.823 (0.544)} & \textbf{0.871 (0.300)} & 1.349 (1.207) & \textbf{0.852 (0.440)} & 1.854 (1.775) \\
\quad $n=1000$ & 0.479 (0.068) & \textbf{0.411 (0.270)} & \textbf{0.425 (0.095)} & 0.614 (0.567) & \textbf{0.371 (0.131)} & 0.816 (0.785) \\
\quad $n=5000$ & 0.274 (0.032) & \textbf{0.211 (0.142)} & \textbf{0.239 (0.041)} & 0.316 (0.289) & \textbf{0.202 (0.047)} & 0.371 (0.355) \\
\multicolumn{1}{l}{\rule[-1.6ex]{0pt}{6.0ex}\!$Z\!\sim\! \mathrm{S}\Gamma\!\left(2,\frac{1}{5\sqrt{2}}\right)\!$} &  &  &  &  &  &  \\
\quad $n=200$ & \textbf{0.830 (0.225)} & 1.034 (0.697) & \textbf{0.967 (0.484)} & 1.950 (1.785) & \textbf{1.132 (0.793)} & 3.163 (3.029) \\
\quad $n=1000$ & \textbf{0.482 (0.101)} & 0.592 (0.390) & \textbf{0.472 (0.190)} & 1.003 (0.919) & \textbf{0.544 (0.353)} & 1.623 (1.575) \\
\quad $n=5000$ & \textbf{0.264 (0.037)} & 0.268 (0.183) & \textbf{0.261 (0.089)} & 0.611 (0.566) & \textbf{0.279 (0.174)} & 0.982 (0.968) \\
\bottomrule
\end{tabular}%

    }
    
    \tabnote{Boldface identifies the estimator with
        the smallest unrounded empirical MSE in each row.
    For each measurement-error distribution,
    jump size, and sample size, entries are averaged over the three jump
    designs $J_1,J_2,J_3$, with the $J_3$ entries first averaged over the jump
    locations $x=-2,0,2$. 
    % and the square root is then taken. Variance entries are
    % averaged over the same designs and jump locations. 
    Full DGP-by-DGP results
    are reported in Table \ref{tab:jump_appendix} in Appendix~\ref{app:Full_sim_results}.
    }
\end{table}
}

% {
% \setlength{\tabcolsep}{7pt}
% % 
% \begin{table}[h!t]
%     \caption{
%         Empirical root MSE and empirical variance$\times 10$ (in parenthesis) for estimating $p_x$. 
%         }\label{tab:jump_mainOLD}
%     \adjustbox{max width=\textwidth, center = \textwidth}{ 
%     \def\arraystretch{1.3}
%     \input{Tables/Table_jump_byZ_main_OLD}
%     }
    
%     \tabnote{
%     For each measurement-error distribution,
%     jump size, and sample size, MSE entries are averaged over the three jump
%     designs $J_1,J_2,J_3$, with the $J_3$ entries first averaged over the jump
%     locations $x=-2,0,2$, and the square root is then taken. Variance entries are
%     averaged over the same designs and jump locations. Full DGP-by-DGP results
%     are reported in Table \ref{tab:jump_appendix} in Appendix~\ref{app:Full_sim_results}.
%     }
% \end{table}
% }

Table~\ref{tab:jump_main} reports the simulation results for estimating the jump
size $p_x$. MSE decreases with $n$ for both estimators in every grouped design.
Performance nevertheless depends strongly on the measurement-error
distribution. The normal-error designs produce the largest MSEs despite having
the smallest empirical variances, indicating that estimation under super-smooth
errors is primarily limited by regularization bias. MSE is substantially
smaller under the two ordinary-smooth error distributions.

Under normal measurement error, $\pxhatGL$ has lower MSE than
$\pxhatAR$ for every jump size and sample size, although $\pxhatAR$ has lower
variance. Thus, the improvement from GL selection does not arise from choosing
a uniformly less variable estimator; instead, its higher variance is more than
offset by a reduction in squared bias. The relative MSE advantage of
$\pxhatGL$ also generally increases with the sample size.

Under Gamma measurement error, the ranking depends on the jump size. At each
sample size, $\pxhatAR$ has lower MSE when $p_x=0.01$, whereas $\pxhatGL$
performs better when $p_x=0.10$ or $p_x=0.20$. Under symmetrized-Gamma
errors, $\pxhatGL$ has lower MSE throughout. In the ordinary-smooth designs,
the asymptotic-rate estimator often has substantially greater variance,
particularly for the larger jumps.

The DGP-specific results in Table~\ref{tab:jump_appendix} in Appendix~\ref{app:Full_sim_results} reveal some
heterogeneity behind these grouped comparisons, particularly at the smallest
sample size. The grouped rankings should therefore not be interpreted as
uniform across every latent distribution. Because the baseline GL estimator
is bias-dominated in several designs, Appendix
Table~\ref{tab:jump_GL_sensitivity} also reports sensitivity to reducing either
of the GL tuning coefficients $\kappa_A$ and $\kappa_V$. These less
conservative specifications tend to improve performance under normal errors
and for smaller jumps, but can perform worse for larger jumps under
symmetrized-Gamma errors. Overall, the results are similar across the three GL
specifications, and no alternative uniformly dominates.
% ; the main qualitative
% conclusions are therefore not driven by a single choice of the tuning
% coefficients.

% \newpage
% \input{Explicit_rate_corollary_PROB}

%%%%%%%%%%%%%%%%%%%%%%%%%%%%%%%%%%%%%
\section{Conclusion}\label{sec:conclusion}
%%%%%%%%%%%%%%%%%%%%%%%%%%%%%%%%%%%%%
In this paper we proposed three nonparametric deconvolution estimators under the classical additive error-in-measurement model: an estimator of the distribution function at its continuity points, an estimator of interval probabilities, and a new estimator of the sizes of jump discontinuities. 
% While previous work has primarily focused on estimating distribution functions at continuity points, to the best of our knowledge the estimator proposed here for jump sizes is the first developed for the classical additive measurement-error model.
% While previous work has primarily focused on estimating distribution functions at continuity points, to the best of our knowledge the estimator proposed here for jump sizes is the first developed for a general class of latent distributions in the additive measurement-error model.
These estimators extend the scope of existing work by accommodating much broader classes of latent distributions.
In particular, whereas existing jump-size estimators under additive measurement error impose discrete-continuous mixture structures and smoothness conditions, to the best of our knowledge the estimator proposed here is the first to accommodate an otherwise unrestricted latent distribution.

The proposed estimators are motivated by a generalized Fourier inversion theorem established in \cite{Mynbaev2022}, which reveals a direct connection between generalized Fourier inversion formulas and a broad class of kernel-based estimators. This connection provides a unified framework for constructing deconvolution estimators of several distributional functionals. Unlike existing deconvolution estimators based on Sobolev smoothness assumptions, the non-asymptotic bias bounds obtained here require neither the existence of a density nor global smoothness of the latent distribution. Instead, they depend primarily on local properties of the distribution function together with regularity assumptions on the measurement-error distribution and the regularization kernel.

For each estimator we derived explicit non-asymptotic bounds on the bias, variance, and mean squared error, and established asymptotic unbiasedness and consistency under both ordinary-smooth and super-smooth measurement-error distributions. Under additional local regularity assumptions we also obtained explicit convergence rates for the mean squared error. In particular, the pointwise and interval estimators remain valid for distributions that need not possess densities, while the jump estimator provides a direct method for recovering point masses obscured by measurement error.

The simulation study demonstrates that the proposed procedures perform well across a wide range of latent distributions and measurement-error models. The cross-validation implementation of the pointwise and interval estimators provides stable performance across smooth and nonsmooth designs, while the Goldenshluger–Lepski bandwidth selection procedure substantially improves the finite-sample performance of the jump estimator in many settings.

Several directions for future research remain. An important extension would be to establish minimax optimality of the proposed estimators over suitable classes of distributions. Another natural direction is the development of adaptive bandwidth selection procedures supported by asymptotic optimality theory, particularly for the jump estimator. It would also be of interest to extend the methodology to settings with unknown measurement-error distributions, repeated measurements, Berkson-type measurement errors, or multivariate latent variables. Finally, the development of confidence intervals and confidence bands for the proposed estimators remains an open problem.

%%%%%%%%%%%%%%%%%%%%%%%%%%%%%%%%%%%%%%%%%%%%%%%%%%
%------------------------------------------------%
                % APPENDIX %
%------------------------------------------------%
%%%%%%%%%%%%%%%%%%%%%%%%%%%%%%%%%%%%%%%%%%%%%%%%%%
% \pagebreak
% \addappheadtotoc 
% \appendix
% \appendixpage
\begin{appendices}

\clearpage
%%%%%%%%%%%%%%%%%%%%%%%%%%%%%%%%%%%%%%%%%
% Appendix
%%%%%%%%%%%%%%%%%%%%%%%%%%%%%%%%
% \section*{Appendix}
\section{Mathematical proofs}\label{app:proofs}
This appendix contains the proofs for all lemmas and theorems.

%%%%%%%%%%%%%%%%%%%%%%%%%%%%%%%%%%%%%%%%%%%%%
 %Proof of Lemma 1
%%%%%%%%%%%%%%%%%%%%%%%%%%%%%%%%%%%%%%%%%%%%%
\subsection{Proofs for Section \ref{sec:FYx}}
\noindent \bf Lemma \ref{lem1}: \it Proof. \rm 
Let $G=G_H$ and define 
\begin{equation}\label{35}
A_{x}(h,\lambda ):=\frac{1}{2\pi }\int_{\R}\frac{e^{-it\left(x-\frac{1}{\lambda} \right)}-e^{-itx}}{it} H(ht)(\mathcal{F}F) (t)dt, \quad \mbox{ for $h,\lambda >0$ and $x\in \R$.}
\end{equation}
Let $g_{\lambda ,x}(s)=\chi _{(x-1/\lambda,x)}(-s)$ and note that
\[
(\mathcal{F}g_{\lambda ,x})(t)=\int_{\mathbb{R}}e^{its}g_{\lambda,x}(s)\,ds=\int_{-x}^{-x+1/\lambda }e^{its}\,ds=\frac{e^{-it(x-\frac{1}{\lambda})}-e^{-itx}}{it}.
\]
The inverse Fourier transform of $H(h\,\cdot )(\mathcal{F}g_{\lambda ,x})(\mathcal{F}F)$ is given by 
\[
\mathcal{F}^{-1}\left[ H(h\cdot )(\mathcal{F}g_{\lambda ,x})(\mathcal{F}F) \right] (s)=\frac{1}{2\pi }\int_{\R}e^{-its}H(ht)\left( \frac{e^{-it(x-\frac{1}{\lambda})}-e^{-itx}}{it}\right) (\mathcal{F}F) (t)\,dt,
\]
and
\[
\mathcal{F}^{-1}\left[ H(h\,\cdot )(\mathcal{F}g_{\lambda ,x})(\mathcal{F}F) \right] (0)=\frac{1}{2\pi }\int_{\R}H(ht)\left( \frac{e^{-it(x-\frac{1}{\lambda})}-e^{-itx}}{it}\right)(\mathcal{F}F) (t)\,dt=A_{x}(h,\lambda ).
\]
Since $(\mathcal{F}F) (t)=\int_{\R}e^{iut}\,dF(u)$, we have
\[
A_{x}(h,\lambda )=\frac{1}{2\pi }\int_{\R}H(ht)\left( \frac{e^{-it(x-\frac{1}{\lambda})}-e^{-itx}}{it}\right)\left( \int_{\R}e^{iut}\,dF(u)\right) dt.
\]
Given integrability of $H(h\cdot )$ and $\big|\frac{e^{-it(x-1/\lambda)}-e^{-itx}}{it}\big|\le\min\{1/\lambda,2/|t|\}$, by Fubini's theorem we have
\begin{equation}\label{63.1}
A_{x}(h,\lambda )=\frac{1}{2\pi }\int_{\R}\left( \int_{\R}H(ht)\left( \frac{e^{-it(x-\frac{1}{\lambda})}-e^{-itx}}{it}\right)e^{iut}\,dt\right) dF(u).
\end{equation}
For the inner integral $\int_{\R}H(ht)\left( \frac{e^{-it(x-\frac{1}{\lambda})}-e^{-itx}}{it}\right)e^{iut}dt$, let $v=ht$ and write
\[
\int_{\R}H(v)e^{i(v/h)(u-x)}\frac{e^{i(v/h)/\lambda
}-1}{i(v/h)}\frac{dv}{h}=\int_{\R}H(v)e^{iv(u-x)/h}\frac{e^{iv/(h\lambda )}-1}{iv}\,dv.
\]
Since $\frac{e^{iv/(h\lambda )}-1}{iv}=\int_{0}^{1/(h\lambda )}e^{ivs}\,ds$, the inner integral is
\[
\int_{\R}H(v)e^{iv(u-x)/h}\left(
\int_{0}^{1/(h\lambda )}e^{ivs}\,ds\right) dv=\int_{0}^{1/(h\lambda )}\left( \int_{\R}H(v)e^{iv((u-x)/h+s)}\,dv\right) ds.
\]
Since $\mathcal{F}H(u)=\int_{\R}H(v)e^{ivu}\,dv$, we have
$
\int_{\R}H(v)e^{iv((u-x)/h+s)}\,dv=(\mathcal{F}H)\left( \frac{u-x}{h}%
+s\right) .
$  Thus, letting $w=\frac{u-x}{h}+s$ we obtain
\begin{align}
\frac{1}{2\pi }\int_{\R}H(ht)\left( \frac{e^{-it(x-\frac{1}{\lambda})}-e^{-itx}}{it}\right)e^{iut}dt&=\frac{1}{2\pi }\int_{0}^{1/(h\lambda )}(\mathcal{F}H)\left( \frac{u-x}{h}+s\right) ds \label{64.00}\\
& =\frac{1}{2\pi }\int_{(u-x)/h}^{(u-x)/h+1/(h\lambda )}(\mathcal{F}H)(w)\,dw \notag\\
&=G\left( \left[ \frac{u-x}{h},\frac{u-(x-1/\lambda) }{h}\right]\right).\label{64.0}
\end{align}
Equations \eqref{63.1} and \eqref{64.0} give $A_{x}(h,\lambda )=\int_{\R}G\left( \left[ \frac{u-x}{h},\frac{u-(x-1/\lambda) }{h}\right] \right) dF(u)$.

For the second part, 
we derive the integral representation for $E\left( \FYhat(x) \right)$. Note that
\begin{align*}
E\left( K_{h,\lambda }\left( X-x\right) |Y\right)  &=\int_{\R}K_{h,\lambda
}\left( Y+z-x\right) dF_{Z}(z) =\int_{\R}\left[ \frac{1}{2\pi }\int_{\R}e^{i(Y+z-x)s}\alpha _{h,\lambda
}\left( s\right) ds\right] dF_{Z}(z) \\
&=\frac{1}{2\pi }\int_{\R}\left[ \int_{\R}e^{izs}dF_{Z}(z)\right]
e^{i(Y-x)s}\alpha _{h,\lambda }\left( s\right) ds =\frac{1}{2\pi }\int_{\R} e^{i(Y-x)s}\frac{e^{is/\lambda }-1}{is}H(hs)ds \\
&=\frac{1}{2\pi }\int_{\R}e^{i(Y-x)s}\frac{e^{is/\lambda }-1}{is}H(hs)ds=G\left( \left[ \frac{Y-x}{h},\frac{Y-(x-\frac{1}{\lambda})}{h}\right]\right),
\end{align*}
where the last equality follows from equation \eqref{64.0}.
% in the proof of Lemma \ref{lem1}.  
Hence, by the law of iterated expectations
\begin{equation}\label{mean}
E\left( K_{h,\lambda }\left( X-x\right) \right) =\int_{\R}G\left( \left[
\frac{y-x}{h},\frac{y-(x-\frac{1}{\lambda}) }{h}\right] \right) dF_{Y}(y).  
\end{equation}
Then, $E \left(\FYhat(x)\right)=E\left(K_{h,\lambda }\left(X-x\right) \right)$ and by the first part of this lemma, $
E \left(\FYhat(x)\right)=A_{x}(h,\lambda )$.
\hfill$\square$

%%%%%%%%%%%%%%%%%%%%%%%%%%%%%%%%%%%%%%%%%%%%%
% Proof of lemma 1.5
%%%%%%%%%%%%%%%%%%%%%%%%%%%%%%%%%%%%%%%%%%%%%
% \noindent \bf Lemma \ref{lem1.5}: 
\begin{applemma}\label{lem1.5}
Let $F$ be a distribution function and $H \in L(\R)$. Define $G([a,b]):=\frac{1}{2\pi}\int_a^b(\mathcal{F}H)(v)dv$ for $a<b$, $a,b\in \R$.
Let $A_x(h,\lambda)$ be defined as in \eqref{35}.
% and $G$ be defined as in Lemma \ref{lem1}.  
In addition, assume that $G$ is a continuous function of intervals $[a,b]$ which satisfies
\begin{equation} \label{Gcond1}
\lim_{a\rightarrow -\infty ,b\rightarrow \infty}
G([a,b])=1,
\quad
\lim_{a\rightarrow \infty }G([a,b])=0,
\quad
\lim_{b\rightarrow -\infty}G([a,b])=0,
\quad
\left\Vert G\right\Vert _{\infty} < \infty.
% \mbox{ $G$ is bounded,} 
\end{equation}
% and a continuous function of intervals $[a,b]$.  
Then, for any distribution function $F$, and parameters $h \in (0,1]$, $\delta \in (0,1)$ and $\lambda >0$ such that $\lambda h^\delta \le 1/2$ we have
\begin{equation}\label{lem2eq2}
\left\vert A_{x}(h,\lambda )-F(x)\right\vert \leq \phi _{G}(h^{\delta -1})+\left( 1+\left\Vert G\right\Vert _{\infty}\right)
\omega_F (x,h^{\delta })+\left( 2+\left\Vert G\right\Vert _{\infty}\right) F\left(
x-\frac{1}{\lambda }+1\right),
\end{equation}
where $\phi _{G}$ is defined as in \eqref{phiGdef}.
% where $\phi _{G}(N):=\max \left\{ \underset{b<-N}{\sup} |G([a,b])|,\ \underset{a>N}{\sup} |G([a,b])|,\ \underset{a<-N,\ b>N}{\sup}\,|G([a,b])-1|\right\}$ for $N>0$.
\end{applemma}
\noindent
\it Proof. \rm Lemma 2 in \cite{Mynbaev2022} states that if $G$ satisfies \eqref{Gcond1} and $\delta \in (0,1)$, then for all $F$, $h>0$ such that $x<y$ and $h^\delta\leq(y-x)/2$, 
\begin{equation}\label{42}
\left\vert \int_{\R}G\left( \left[ \frac{t-y}{h},\frac{t-x}{h}\right]
\right) dF(t)-F(x,y)\right\vert \leq \phi _{G}(h^{\delta -1})\notag+\left( 1+\left\Vert G\right\Vert _{\infty}\right)\left[ \omega_F (x,h^{\delta })+\omega_F (y,h^{\delta })\right],  
\end{equation}
where $F(x,y):=F(y)-F(x)$. Since $\lambda h^\delta \le 1/2$, we replace $(x,y)$ with the pair $(x-\frac{1}{\lambda} ,x)$ and use Lemma \ref{lem1} to obtain $\left\vert A_{x}(h,\lambda )-F(x-\frac{1}{\lambda},x)\right\vert \leq \phi
_{G}(h^{\delta -1})+\left( 1+\left\Vert G\right\Vert _{\infty}\right) \left[
\omega_F (x-1/\lambda ,h^{\delta })+\omega_F (x,h^{\delta })\right]$.  Then, 
\begin{align*}
\left\vert A_{x}(h,\lambda )-F(x)\right\vert &\leq \left\vert A_{x}(h,\lambda )-\left[ F(x)-F\left( x-\frac{1}{\lambda }\right) \right]\right\vert +F\left( x-\frac{1}{\lambda }\right) \\
&\leq \phi _{G}(h^{\delta -1})+\left( 1+\left\Vert G\right\Vert _{\infty}\right)\left[ F\left( x-\frac{1}{\lambda }+h^{\delta}\right)-F\left( x-\frac{1}{\lambda }-h^{\delta }\right)+\omega_F (x,h^{\delta }) \right]+F\left( x-\frac{1}{\lambda }\right)\\ 
&\leq \phi _{G}(h^{\delta -1})+\left( 1+\left\Vert G\right\Vert _{\infty}\right)\omega_F (x,h^{\delta })+\left( 2+\left\Vert G\right\Vert _{\infty}\right) F\left(x-\frac{1}{\lambda }+h^\delta\right)\\
&\leq \phi _{G}(h^{\delta -1})+\left( 1+\left\Vert G\right\Vert _{\infty}\right)\omega_F (x,h^{\delta })+\left( 2+\left\Vert G\right\Vert _{\infty}\right) F\left(x-\frac{1}{\lambda }+1\right), \mbox{ since $h \in (0,1]$. \hfill$\square$}
\end{align*}
%%%%%%%%%%%%%%%%%%%%%%%%%%%%%%%%%%%%%%%%%%%%%

%%%%%%%%%%%%%%%%%%%%%%%%%%%%%%%%%%%%%%%%%%%%%%Proof of Theorem thm7
%%%%%%%%%%%%%%%%%%%%%%%%%%%%%%%%%%%%%%%%%%%%%
\noindent \bf Theorem \ref{thm7}: \it Proof. \rm
Let $A_x(h,\lambda)$ be defined as in \eqref{35} with $G=G_H$ and $F=F_Y$.
% Set $G=G_H$ and let $A_x(h,\lambda)$ denote the Fourier integral
% on the left-hand side of Lemma~1 with $F=F_Y$. 
By Lemma~\ref{lem1},
$
    \mathbb E[\widehat F_Y(x)]=A_x(h,\lambda).
$
Since $H\in L(\mathbb R)$, $\mathcal FH\in C_b(\mathbb R)$, so
$G_H$ is continuous in its interval endpoints. Assumption~\ref{AG}
therefore verifies the conditions of Lemma~\ref{lem1.5}, which implies \eqref{gen_bound2}.

Fix any $\delta\in(0,1)$. For all $h,\lambda$ sufficiently small $\lambda h^\delta\leq1/2$, so \eqref{gen_bound2} holds. As $h,\lambda\to0$,
\[
\varphi_{G_H}(h^{\delta-1})\to0,\qquad
\omega_{F_Y}(x,h^\delta)\to0,
\qquad
F_Y(x-\lambda^{-1}+1)\to0,
\]
where the second convergence follows from $x\in C(F_Y)$.
% Moreover, $\lambda h^\delta\leq1/2$ eventually. 
This proves
\eqref{unqua2}.
\hfill$\square$

\noindent \bf Theorem \ref{thm8}: \it Proof. \rm We first derive an upper bound for variance $V\left( \FYhat(x)\right)$.  Since $V\left( \FYhat(x)\right) \leq \frac{1}{n}E\left(K_{h,\lambda }^{2}\left(X-x\right)\right)$, we use the algebraic structure of the estimator to obtain
\begin{align}
E\left(K_{h,\lambda }^{2}\left( X-x\right) \right)&=\int_{\R} \int_{\R} K_{h,\lambda }^{2}\left(
y+z-x\right) dF_{Y}(y)dF_{Z}(z)  \notag \\
&=(2\pi )^{-2}\int_{\R} \int_{\R} \left[ \int_{\R}e^{i(y+z-x)s}\alpha _{h,\lambda
}\left( s\right) ds\right] \left[ \int_{\R}e^{i(y+z-x)t}\alpha _{h,\lambda
}\left( t\right) dt\right] dF_{Y}(y)dF_{Z}(z)  \notag \\
&=(2\pi )^{-2}\int_{\R}\int_{\R}e^{-ix(s+t)}\alpha _{h,\lambda }\left(
s\right) \alpha _{h,\lambda }\left( t\right) \left( \int_{\R}
e^{iy(s+t)}dF_{Y}(y)\int_{\R} e^{iz(s+t)}dF_{Z}(z)\right) ds dt  \notag \\
&=(2\pi )^{-2}\int_{\R}\int_{\R}\int_{\R}e^{-ix(s+t)}\alpha _{h,\lambda
}\left( s\right) \alpha _{h,\lambda }\left( t\right) \Phi _{Y}\left(s+t\right) \Phi _{Z}\left( s+t\right) dsdt.  \label{cons}
\end{align}
This expectation is bounded as
\begin{equation*}
E\left(K_{h,\lambda }^{2}\left( X-x\right)\right) \leq (2\pi
)^{-2}\int_{\R}\int_{\R}\left\vert \alpha _{h,\lambda }\left( s\right) \alpha
_{h,\lambda }\left( t\right) \right\vert dsdt=(2\pi )^{-2}\left(
\int_{\R}\left\vert \alpha _{h,\lambda }\left( s\right) \right\vert ds\right)
^{2}.
\end{equation*}
Hence, $V\left( \FYhat(x)\right) \leq \frac{1}{(2\pi )^{2}}\left( \frac{v_{h,\lambda}}{\sqrt{n}}\right) ^{2}\rightarrow 0$ since we assume $v_{h,\lambda }/\sqrt{n}\rightarrow 0$ as $n\rightarrow \infty$.  By Chebyshev's inequality and Theorem \ref{thm7} we obtain $ \FYhat(x) \overset{p}{\rightarrow} F_Y(x)$. \hfill$\square$
%%%%%%%%%%%%%%%%%%%%%%%%%%%%%%%%%%%%%%%%%%%%

%%%%%%%%%%%%%%%%%%%%%%%%%%%%%%%%%%%
% Lemma 6
%%%%%%%%%%%%%%%%%%%%%%%%%%%%%%%%%%
\noindent \bf Lemma \ref{lem6a}: \it Proof. \rm 
% a) 
Let $C_H:=\underset{t \in [-\gamma,\gamma]}{\sup}|H(t)|$.
First, note that $|e^{it}-1|^2=4\sin^2\left(\frac{t}{2}\right)$ for all $t\in \R$.  Hence, letting $\phi_\lambda(t):=\left|\frac{e^{it/\lambda}-1}{it}\right|$ for $\lambda \in (0,1)$ and $t \neq 0$ we have 
$
\phi_\lambda(t)=\frac{2 \left| \sin \left( \frac{t}{2 \lambda}\right) \right|}{|t|}=\frac{1}{\lambda}\frac{\left| \ \sin \left(\frac{t}{2 \lambda}\right)\right|}{\left|\frac{t}{2 \lambda}\right|}
$ and  
\begin{equation}\label{eq75}
 \phi_{\lambda}(t) \le \min\left\{\frac{1}{\lambda},\frac{2}{|t|} \right\}
\end{equation}
Since $\Phi_Z(0)=1$, continuity of $\Phi_Z$ at $t=0$ implies that for every $\epsilon \in (0,1)$ there exists $\beta>0$ such that whenever $|t| \le \beta$ we have $|\Phi_Z(t)| \ge 1- \epsilon$. Since $|\Phi_Z(t)| \neq 0$, we have $\frac{1}{|\Phi_Z(t)|}\le \frac{1}{1- \epsilon}$ whenever $|t| \le \beta$.
For the assumed ranges of $\lambda$ and $h$, we have $\pi\lambda < \beta < \gamma/h$.
% For any $\gamma>0$ there exist $\lambda, h>0$ such that $\pi \lambda \le \gamma/h$.  
Then, given the assumptions on $H$ and the two bounds on $\phi_{\lambda}$ we have
\begin{align*}
v_{h,\lambda} &:= \int_{\R}|\alpha_{h,\lambda}(t)|dt = \underset{|t| \le \gamma/h}{\int}|\alpha_{h,\lambda}(t)|dt  = \underset{|t| \le \gamma/h}{\int}\phi_\lambda (t) \frac{|H(ht)|}{|\Phi_Z(t)|}d t \\
&\le \frac{C_H}{\lambda}\int_{-\pi \lambda}^{\pi \lambda}\frac{1}{|\Phi_Z(t)|}dt+4 C_H \int_{\pi \lambda}^{\gamma/h}\frac{1}{t}\frac{1}{|\Phi_Z(t)|}dt.
\end{align*} 
For $\lambda < \beta/ \pi$, $\int_{-\pi \lambda}^{\pi \lambda}\frac{1}{|\Phi_Z(t)|}dt \le \frac{2 \pi \lambda}{1- \epsilon}$.  In addition, for $\beta <\gamma/h$ we have
\begin{align*}
\int_{\pi \lambda}^{\gamma/h}\frac{1}{t}\frac{1}{|\Phi_Z(t)|}dt&= \int_{\pi \lambda}^{\beta}\frac{1}{t}\frac{1}{|\Phi_Z(t)|}dt + \int_{\beta}^{\gamma/h}\frac{1}{t}\frac{1}{|\Phi_Z(t)|}dt \le \frac{1}{1- \epsilon}  \int_{\pi \lambda}^{\beta}\frac{1}{t}dt+\int_{\beta}^{\gamma/h}\frac{1}{t}\frac{1}{|\Phi_Z(t)|}dt\\
&= \frac{1}{1- \epsilon} \left(\log \ \beta - \log \pi + \log \lambda^{-1} \right)+\int_{\beta}^{\gamma/h}\frac{1}{t}\frac{1}{|\Phi_Z(t)|}dt.
\end{align*}
 Hence,
 $$
 v_{h,\lambda} \le \frac{2\pi C_H}{1- \epsilon}+  \frac{4 C_H}{1- \epsilon} (\log \,\beta - \log \, \pi  + \log \lambda^{-1})+ 4C_H \int_{\beta}^{\gamma/h}\frac{1}{t}\frac{1}{|\Phi_Z(t)|}dt
 .
 $$ 
 Clearly, the above bound holds when $\beta$ is replaced with any $\beta'\in (0,\beta)$, so we may always choose $\beta \in(0,\pi/2)$.
 Consequently, $v_{h,\lambda} =O\left( \log \lambda^{-1}+  \underset{ \beta < t\le\gamma/h}{\int} \frac{1}{t|\Phi_Z(t)|}dt\right)$.
 \hfill$\square$
 %%%%%%%%%%%%%%%%%%%%%%%%%%%%%%%%%%%%%%%%%%%

%%%%%%%%%%%%%%%%%%%%%%%%%%%%%%%%%%%%
% LEMMA 7
%%%%%%%%%%%%%%%%%%%%%%%%%%%%%%%%%%%
\begin{applemma}\label{lem7a}
Let $0<\beta<\xi_n$ for $n \in \mathds{N}$ with $\xi_n \to \infty$ as $n \to \infty$ and 
$I(\xi_n):=\int_{\beta <t \le {\xi_n}}\frac{1}{t|\Phi_Z(t)|}dt$.  
% $\underset{ \beta <t \le {\xi_n}}{\int}\frac{1}{t|\Phi_Z(t)|}dt$.
\begin{enumerate}
\item[(a)]
% a)  
If there exist $\tau, \rho>0$ such that 
\eqref{eq:error_smoothness_super} holds,
% $|\Phi_Z(t)| \asymp \exp (-\tau |t|^{\rho})$ as $|t| \to \infty$,  
then $I(\xi_n)=O\left( \frac{ \exp(\tau \xi_n^\rho)}{\tau \xi_n^\rho} \right)$ and if $\xi_n \le \left( \frac{\log n}{2 \tau} \right)^{1/\rho}$, $I(\xi_n)=o(n^{1/2})$. 
% b)
\item[(b)]
If there exists $\rho>0$ such that 
% $|\Phi_Z(t)| \asymp |t|^{-\rho}$ as $|t| \to \infty$, 
\eqref{eq:error_smoothness_ordinary} holds,
then $I(\xi_n) = O\left( \xi_n^\rho  \right)$ and if $\xi_n= o(n^{1/2\rho})$, $I(\xi_n)=o(n^{1/2})$. 
 \end{enumerate}
\end{applemma}
\noindent 
% \bf Lemma \ref{lem7a}: 
\it Proof. \rm 
\textbf{(a)}
By Assumption \ref{A1}.1 $|\Phi_Z(t)|>0$ and given that $|\Phi_Z(t)| \asymp \exp (-\tau |t|^{\rho})$ as $|t| \to \infty$, there exists a constant $C>0$ such that 
\begin{align*}
I(\xi_n) \le C \int_\beta^{\xi_n}\frac{\exp(\tau t^\rho)}{t}dt=\frac{C}{\rho}\int_{\tau\beta^\rho}^{\tau \xi_n^\rho} \frac{\exp (u)}{u}du.
\end{align*}
Letting $g(x):=\int_{\tau\beta^\rho}^{x} \frac{\exp (u)}{u}du$ and using L'H\^opital's rule we obtain
$
\underset{x \to \infty}{\lim}\frac{g(x)}{\exp(x)/x}=1
$.  Hence,  $I(\xi_n) =O\left( \frac{\exp(\tau \xi_n^\rho)}{\tau\xi_n^\rho} \right)$.  If $\xi_n \le \left( \frac{\log n}{2 \tau} \right)^{1/\rho}$ for all $n$ sufficiently large, then $\exp(\tau \xi_n^\rho) \le n^{1/2}$ and 
$$
\frac{1}{\sqrt{n}}I(\xi_n) \le \frac{C}{\sqrt{n}}\frac{\exp(\tau \xi_n^\rho)}{\tau \xi_n^\rho}\le \frac{C}{\tau  \xi_n^\rho}=o(1).
$$

\medskip
\noindent
\textbf{(b)}
Given that $|\Phi_Z(t)| \asymp |t|^{-\rho}$ as $|t| \to \infty$, there exists a constant $C>0$ such that 
$
I(\xi_n) \le C \int_\beta^{\xi_n}t^{\rho-1}dt=\frac{C}{\rho}\left( \xi_n^\rho - \beta^\rho\right)
$.  Hence, $I(\xi_n)=O\left( \xi_n^\rho  \right)$ and it follows immediately that if $\xi_n = o(n^{1/2\rho})$, $\frac{1}{\sqrt{n}}I(\xi_n)=o(1)$.
\hfill$\square$
%%%%%%%%%%%%%%%%%%%%%%%%%%%%%%%%%%%%%%%%%%%%

%%%%%%%%%%%%%%%%%%%%%%%%%%%%%%%%%
% Proof Theorem thm10a
%%%%%%%%%%%%%%%%%%%%%%%%%%%%%%%%
\noindent \bf Theorem \ref{thm10a}: \it Proof. \rm
Suppose Assumption \ref{A1}.1 holds.
Let 
% $C_H:=\underset{t \in [-\gamma,\gamma]}{\sup}|H(t)|$ and 
$I$ be defined as in Lemma \ref{lem7a}. 

\noindent
\textbf{(a)}
It follows from part (a) of Lemma \ref{lem6a} that 
\begin{align}\label{eq76a}
 \frac{1}{\sqrt{n}} v_{h,\lambda} &\le C \left( \frac{1}{\sqrt{n}} \log \lambda^{-1}+ \frac{1}{\sqrt{n}}  \underset{ \beta < t\le\gamma/h}{\int} \frac{1}{t|\Phi_Z(t)|}dt\right) 
 % \notag
 % \\
 % &
 =C \left( \frac{1}{\sqrt{n}} \log \lambda^{-1}+ \frac{1}{\sqrt{n}} I(\gamma/h)\right).
\end{align}
% where $I$ is defined in Lemma \ref{lem7a}.
It follows immediately that if $\lambda=\exp(-o(\sqrt{n}))$, the first term on the right-hand side is $o(1)$.  For the second term, under the conditions on $h$ we have $\gamma /h \le (\log \,n/2\tau)^{1/\rho}$ for all large $n$, so by part (a) of Lemma \ref{lem7a}, $I(\gamma/h) = o(\sqrt {n})$ and hence
$ (1/\sqrt {n}) I(\gamma/h) = o(1)$.  Therefore $(1/\sqrt {n}) v_{h,\lambda} = o(1)$.  

\medskip
\noindent
\textbf{(b)}
By part (b) of Lemma \ref{lem7a}, $\frac{1}{\sqrt{n}}I(\gamma/h)=o(1)$ provided $\gamma/h=o(n^{1/2\rho})$, or equivalently, $nh^{2\rho}\to \infty$.  Hence, using equation \eqref{eq76a}, if $\lambda=\exp(-o(\sqrt{n}))$,  $\frac{1}{\sqrt{n}}v_{h,\lambda}=o(1)$. \hfill$\square$
%%%%%%%%%%%%%%%%%%%%%%%%%%%%%%%%%%%%%%%%%%%%

%%%%%%%%%%%%%%%%%%%%%%%%%%%%%%%%%
% Proof Theorem thm:FYx_MSEbnd
%%%%%%%%%%%%%%%%%%%%%%%%%%%%%%%%
\noindent \bf Theorem \ref{thm:FYx_MSEbnd}: \it Proof. \rm 
If the bound in Lemma~\ref{lem6a} holds for $\beta>0$, then it also holds for any $\beta'\in(0, \beta]$.
Hence, we may always apply Lemma~\ref{lem6a} with $\beta$ so such that $0<\beta<\min\{\gamma,\pi/2\}$.
For such a $\beta$, if $\delta \in (0,1)$, $0<h\leq 1$, and $0<\lambda<\beta/\pi$, then 
$h<\gamma/\beta$ and $\lambda h^\delta\leq 1/2$. 
Hence, Theorems \ref{thm7} and Lemma \ref{lem6a} can be applied. This, with Theorem \ref{thm8}, completes the proof.
\hfill$\square$
%%%%%%%%%%%%%%%%%%%%%%%%%%%%%%%%%%%%%%%%%%%%

%%%%%%%%%%%%%%%%%%%%%%%%%%%%%%%%%%%%%%%%%%%
% commented out proof of b)
\iffalse
 \noindent b) Let $t_0<t(h):=h^{-\delta}$ and note that $t(h) \to \infty$ as $h \to 0$.  Now, given equation \eqref{eq77b}, there exists $C>0$ such that
$
\frac{|H(ht)|}{t|\Phi_Z(t)|} \le C |t|^\rho 
$
and consequently 
$$
\int_{t_0<t \le t(h)}\frac{|H(ht)|}{t|\Phi_Z(t)|}dt \le C \int_{t_0<t \le t(h)}t^\rho dt=\frac{1}{\rho+1}\left(t(h)^{\rho+1}-t_0^{\rho+1}\right) \to \infty \mbox{ as $h \to 0$.}
$$
Now, given that $H(t)=O\left( |t|^{ \frac{\delta}{\delta-1} (\rho+1)} \right)$
$$
\int_{t > t(h)}\frac{|H(ht)|}{t|\Phi_Z(t)|}dt \le \frac{C}{h^{\rho+1}}\int_{t>h^{1-\delta}}t^\rho|H(t)|dt \le \frac{C}{h^{\rho+1}} \int_{t>h^{1-\delta}}t^{- \frac{\rho+\delta}{\delta-1}}dt=\frac{\delta-1}{\rho+1}<\infty.
$$
Hence, as in part (a) $\underset{t> t(h)}{\int} \frac{|H(ht)|}{t|\Phi_Z(t)|}dt=o\left(\underset{\beta < t \le t(h)}{\int} \frac{|H(ht)|}{t|\Phi_Z(t)|}dt\right)$.  Again, by part (b) of Lemma \ref{lem6a} we have that equation \eqref{eq74n} holds if $\lambda=\exp(-o(\sqrt{n}))$ and $t(h) =o(n^{1/2\rho}) $ by part (a) of Lemma \ref{lem7a}.  For the last condition it suffices to have $nh^{2\rho\delta}\to \infty$.
\fi
%%%%%%%%%%%%%%%%%%%%%%%%%%%%%%%%%%%%%%%%%%%

%%%%%%
% Proof of Corollary 1
%%%%%%%%%

\noindent \bf Corollary \ref{coro1}: \it Proof. \rm By the stated assumptions,
\[
\phi_G(h^{\delta-1})
=
O(h^{k(1-\delta)}),
\qquad
\omega_{F_Y}(x,h^\delta)
=
O(h^{s\delta}),
\qquad
F_Y(x+1-\lambda^{-1})
=
O(\lambda^q).
\]
Let
$
\delta^*=\frac{k}{k+s},
\,
r=\frac{ks}{k+s}.
$
Then
$
k(1-\delta^*)
=
s\delta^*
=
r,
$
so that
$
\phi_G(h^{\delta^*-1})
+
\omega_{F_Y}(x,h^{\delta^*})
=
O(h^r)$.
Choose
$
\lambda=h^{r/q},
$
so that
$
F_Y(x+1-\lambda^{-1})
=
O(\lambda^q)
=
O(h^r)$.  Therefore, by Theorem \ref{thm:FYx_MSEbnd} a),
\[
\operatorname{MSE}(\widehat F_Y(x))
\le
\frac{C}{n}
\left(
\log h^{-1}
+
\int_{\beta<t\le\gamma/h}
\frac{dt}{t|\Phi_Z(t)|}
\right)^2
+
Ch^{2r}.
\]

\medskip
\noindent
\textbf{(a)}
Since
$
|\Phi_Z(t)|
\asymp
\exp(-\tau |t|^\rho),
$
we have
$
\int_{\beta<t\le\gamma/h}
\frac{dt}{t|\Phi_Z(t)|}
=
O\!\left(
\frac{\exp\{\tau(\gamma/h)^\rho\}}
{(\gamma/h)^\rho}
\right)
$.  Furthermore,
$
h_n
=
A(\log n)^{-1/\rho}
$
implies
$
\exp\{2\tau(\gamma/h_n)^\rho\}
=
\exp\!\left(
2\tau(\gamma/A)^\rho\log n
\right)
=
n^{2\tau(\gamma/A)^\rho}$.  Also,
$
\log h_n^{-1}
=
O(\log\log n)
$.  Hence
\[
\operatorname{MSE}(\widehat F_Y(x))
\lesssim
n^{-1}(\log\log n)^2
+
n^{2\tau(\gamma/A)^\rho-1}
(\log n)^{-2}
+
h_n^{2r}.
\]
Since
$
h_n^{2r}
=
A^{2r}(\log n)^{-2r/\rho}
=
O\!\left(
(\log n)^{-2r/\rho}
\right)
$,
and
$
A>\gamma(2\tau)^{1/\rho}
$
implies
$
2\tau(\gamma/A)^\rho<1,
$
we also have
$
n^{2\tau(\gamma/A)^\rho-1}
(\log n)^{-2}
=
O\!\left(
(\log n)^{-2r/\rho}
\right).
$
Therefore,
\[
\operatorname{MSE}(\widehat F_Y(x))
=
O\!\left(
(\log n)^{-2r/\rho}
\right).
\]
% as required.

\medskip
\noindent
\textbf{(b)}
Since
$
|\Phi_Z(t)|
\asymp
|t|^{-\rho},
$ we have
$
\int_{\beta<t\le\gamma/h}
\frac{dt}{t|\Phi_Z(t)|}
=
O(h^{-\rho})
$.
Moreover,
$
\log h^{-1}
=
o(h^{-\rho}),
$
and with
$
h_n
\asymp
n^{-1/\{2(\rho+r)\}},
$
we obtain
\[
\operatorname{MSE}\{\widehat F_Y(x)\}
\le
Cn^{-1}h^{-2\rho}
+
Ch^{2r}
\asymp
n^{-r/(\rho+r)}.
\]
\hfill$\square$

\subsection{Proofs for Section \ref{sec:FYxy}}

%%%%%%%%%%%%%%%%%%%%%%%%
% Proof of Lemma 6ab
%%%%%%%%%%%%%%%%%%%%%%
\noindent \bf Lemma \ref{lem6ab}: \it Proof. \rm 
Let $C_H:=\underset{t \in [-\gamma,\gamma]}{\sup}|H(t)|$.
As in the proof of Lemma \ref{lem6a}, for every $\epsilon \in (0,1)$ there exists $\beta>0$ such that whenever $|t| \le \beta$ we have $\frac{1}{|\Phi_Z(t)|}\le \frac{1}{1- \epsilon}$.  Furthermore, letting $\phi_{y-x}(t)=\frac{1-e^{-i(y-x)t}}{it}$ gives 
$$
|\phi_{y-x}(t)|=\frac{2\left|\sin\left(\frac{(y-x)t}{2}\right)\right|}{|t|} \mbox{ for all } t \in \R.
$$
Then, given the assumptions on $H$ we have
\begin{align*}
v_{h}(x,y) 
&:= 
    \int_{\R}|\beta_{x,y,h}(t)|dt 
= 
    \int_{-\gamma/h}^{\gamma/h}|\phi_{y-x}(t)|\frac{|H(ht)|}{|\Phi_Z(t)|}dt
=
    4\int_{0}^{\gamma/h}\frac{\left|\sin\left(\frac{(y-x)t}{2}\right)\right|}{|t|} \frac{|H(ht)|}{|\Phi_Z(t)|}dt
\\&=
    4 \int_{0}^{\beta}\frac{\left|\sin\left(\frac{(y-x)t}{2}\right)\right|}{|t|} \frac{|H(ht)|}{|\Phi_Z(t)|}dt 
    + 
    4 \int_{\beta \le t <\gamma/h}
    \frac{\left|\sin\left(\frac{(y-x)t}{2}\right)\right|}{|t|} \frac{|H(ht)|}{|\Phi_Z(t)|}dt 
\\&
\le 
    \frac{4C_H}{1-\epsilon}
    \int_{0}^{\beta}
    % \frac{\left|\sin\left(\frac{(y-x)t}{2}\right)\right|}{|t|} 
    \frac{\min\{(y-x)t/2,1\}}{|t|} 
    dt 
    +
    4C_H \int_{\beta \le t <\gamma/h}\frac{1}{t|\Phi_Z(t)|}dt.
\end{align*} 
For the first term,
\begin{equation*}
    \int_{0}^{\beta}
    \frac{\min\{(y-x)t/2,1\}}{|t|} 
    dt 
=
    \begin{cases}
    (y-x)\beta/2, & (y-x)\beta\le2,\\[3pt]
    1+\log((y-x)\beta/2), & (y-x)\beta>2.
    \end{cases}
\end{equation*}
so for some $C_\beta>0$ depending only on $\beta$,
    $$
    \int_{0}^{\beta}
    \frac{\min\{(y-x)t/2,1\}}{|t|} 
    dt 
    \leq 
    2\log(1+(y-x)\beta/2)
    \leq 
    C_\beta
    \log(1+y-x)
    ,
    $$
since for $a\in(0,1)$ we have $a\leq 2\log(1+a)$ and for $a>1$ we have $1+\log(a)\leq 2\log(1+a)$.
The desired result follows applying this to the above bound on $v_{h}(x,y) $.
% Then, given the assumptions on $H$ we have
% \begin{align*}
% v_{h}(x,y) &:= \int_{\R}|\beta_{x,y,h}(t)|dt = \int_{-\gamma/h}^{\gamma/h}|\phi_{y-x}(t)|\frac{|H(ht)|}{|\Phi_Z(t)|}dt=2\int_{-\gamma/h}^{\gamma/h}\frac{\left|\sin\left(\frac{(y-x)t}{2}\right)\right|}{|t|} \frac{|H(ht)|}{|\Phi_Z(t)|}dt\\
% &=2 \int_{-\beta}^{\beta}\frac{\left|\sin\left(\frac{(y-x)t}{2}\right)\right|}{|t|} \frac{|H(ht)|}{|\Phi_Z(t)|}dt + 2 \int_{\beta \le |t| <\gamma/h}\frac{\left|\sin\left(\frac{(y-x)t}{2}\right)\right|}{|t|} \frac{|H(ht)|}{|\Phi_Z(t)|}dt \\
% & \le \frac{2C_H(y-x)\beta}{1-\epsilon}+4C_H \int_{\beta \le t <\gamma/h}\frac{1}{t|\Phi_Z(t)|}dt,
% \end{align*} 
% where the bound for the first integral is obtained using the fact that $|\sin(x)|\le |x|$, which gives
% $$
% \frac{2\left|\sin\left(\frac{(y-x)t}{2}\right)\right|}{|t|} \le (y-x),
% $$
% and the bound for the second integral follows from $|\sin(x)|\le 1$.  Hence, $v_{h}(x,y) =O\left( \underset{ \beta < t\le\gamma/h}{\int} \frac{1}{t|\Phi_Z(t)|}dt\right)$.
\hfill$\square$
%%%%%%%%%%%%%%%%%%%%%%%%%%%%%%%%%%%%%%%%%

%%%%%%%%%%%%%%%%%%%%%%%%%%%%%%%%%%%%
% Proof of Theorem Bias of \hat{F}(x,y)
%%%%%%%%%%%%%%%%%%%%%%%%%%%%%%%
\noindent \bf Theorem \ref{thm13n}: \it Proof. \rm We start by deriving an integral representation for $E\left(\FYhat(x,y)\right)$.
\begin{eqnarray*}
E\left( K_{x,y,h}\left( X-x\right) |Y\right)  &=&\int_{\R} K_{x,y,h}\left( Y+z-x\right) dF_{Z}(z)=\int_{\R} \left[ \frac{1}{2\pi }\int_{\R}e^{i(Y+z-x)s}\beta_{x,y,h} \left( s\right) ds\right] dF_{Z}(z) \\
&=&\frac{1}{2\pi }\int_{\R} \left[ \int_{\R}e^{izs}dF_{Z}(z)\right] e^{i(Y-x)s}\beta_{x,y,h}\left( s\right) ds \\
&=&\frac{1}{2\pi }\int_{\R} \Phi _{Z}\left( s\right) e^{i(Y-x)s}\frac{1-e^{-i(y-x)s}}{is}\frac{H(hs)}{\Phi _{Z}(s)}ds \\
&=&\frac{1}{2\pi }\int_{\R} \frac{e^{-ixs}-e^{-iys}}{is}H(hs)e^{iYs}ds.
\end{eqnarray*}
Now, for any $a<b$ the left-hand side of equation \eqref{64.00} in the proof of Lemma \ref{lem1}, can be written as 
\begin{equation*}
\frac{1}{2\pi }\int_{\R}H(ht)\left( \frac{e^{-ita}-e^{-itb}}{it}\right)e^{iut}dt.    
\end{equation*}
Take $b=y$ and $a=x$.  Then, 
$$
E\left( K_{x,y,h}\left( X-x\right) |Y\right)  =G\left( \left[ \frac{Y-y}{h},\frac{Y-x}{h}\right] \right)
$$
and
$$
E\left(\FYhat(x,y)\right)=E\left( K_{x,y,h}\left( X-x\right) \right)  =\int_{\R}G\left( \left[ \frac{t-y}{h},\frac{t-x}{h}\right] \right)dF_Y(t).
$$
Direct application of Lemma 2 in \cite{Mynbaev2022} gives \eqref{qual1}.  Since $x,y \in C(F_Y)$, as $h \to 0$ we obtain \eqref{unqual}. \hfill$\square$
%%%%%%%%%%%%%%%%%%%%%%%%%%%%%%%%%%%%%%%%%%%

\noindent \bf Theorem \ref{thm8a}: \it Proof. \rm We first derive an upper bound for $V\left(\FYhat(x,y) \right)$.  Since 
$$
V\left( \FYhat(x,y)\right) \leq \frac{1}{n}E \left( K_{x,y,h}^{2}\left( X-x\right) \right),
$$
we use the algebraic structure of the estimator to obtain
\begin{equation*}
E\left( K_{x,y,h}^{2}\left( X-x\right) \right)=\frac{1}{4\pi^2}\int_{\R}\int_{\R}e^{-ix(s+t)}\beta_{x,y,h}\left( s\right) \beta_{x,y,h} \left( t\right) \Phi _{Y}\left( s+t\right) \Phi
_{Z}\left( s+t\right) dsdt.
\end{equation*}
This expectation is bounded as
\begin{equation*}
E\left( K_{x,y,h}^{2}\left( X-x\right) \right) \leq \frac{1}{4\pi^2}\int_{\R}\int_{\R}\left\vert \beta_{x,y,h}
\left( s\right) \beta_{x,y,h} \left( t\right) \right\vert dsdt=\frac{1}{4\pi^2}\left(
\int_{\R}\left\vert \beta_{x,y,h} \left( s\right) \right\vert ds\right) ^{2}.
\end{equation*}
Hence, $V\left( \FYhat(x,y)\right) \leq \frac{1}{4\pi^{2}}\left( \frac{v_{h}(x,y)}{\sqrt{n}}\right) ^{2}\rightarrow 0$ since we assume $v_{h}(x,y)/\sqrt{n}\rightarrow 0$ as $n\rightarrow \infty$.  By Chebyshev's inequality and Theorem \ref{thm13n} we obtain $ \FYhat(x,y) \overset{p}{\rightarrow} F_Y(x,y)$. \hfill$\square$

%%%%%%%%%%%%%%%%%%%%%%%%%%%%%%%%%
% Proof Theorem thm10ab
%%%%%%%%%%%%%%%%%%%%%%%%%%%%%%%%
\noindent \bf Theorem \ref{thm10ab}: \it Proof. \rm 
% \medskip
% \noindent
\textbf{(a)}
% a) 
It follows from part (a) of Lemma \ref{lem6ab} that 
\begin{align}\label{eq76ab}
 \frac{1}{\sqrt{n}} v_{h}(x,y) &\le C \left(  \frac{1}{\sqrt{n}}  \underset{ \beta < t\le\gamma/h}{\int} \frac{1}{t|\Phi_Z(t)|}dt\right)=C \frac{1}{\sqrt{n}} I(\gamma/h), \mbox{ where $I$ is defined in Lemma \ref{lem7a}.}
\end{align}
By part (a) of Lemma \ref{lem7a}, if $\frac{\gamma}{h} \le \left( \frac{\log n}{2 \tau}\right)^{1/\rho}$, $\frac{1}{\sqrt{n}} I(\gamma/h)=o(1)$.  Under the conditions on $h$, $(1/\sqrt {n}) I(\gamma/h) = o(1)$.  Then, $\frac{1}{\sqrt{n}}v_{h}(x,y)=o(1)$.  

\medskip
\noindent
\textbf{(b)}
% \noindent b) 
By part (b) of Lemma \ref{lem7a}, $\frac{1}{\sqrt{n}}I(\gamma/h)=o(1)$ provided $\gamma/h=o(n^{1/2\rho})$, or equivalently, $nh^{2\rho}\to \infty$.  Hence, $\frac{1}{\sqrt{n}}v_{h}(x,y)=o(1)$. \hfill$\square$
%%%%%%%%%%%%%%%%%%%%%%%%%%%%%%%%%%%%%%%%%%%%

\subsection{Proofs for Section \ref{sec:px}}

%%%%%%%%%%%%%%%%%%%%%%%%%%%%%%%%%
% Proof Theorem thm11
%%%%%%%%%%%%%%%%%%%%%%%%%%%%%%%%
\noindent \bf Theorem \ref{thm11}: \it Proof. \rm 
Let \(W:=\mathcal{F}H\). We first derive a representation for the expectation of \(\widehat p_x\). By Assumption~\ref{A2} and Fubini's theorem,
\begin{align*}
E\left( K_{h}\left( X-x\right) |Y\right)  &=\int_{\R}\left[
\int_{\R}e^{i(Y+z-x)s}\frac{hH(hs)}{\Phi _{Z}(s)}ds\right] dF_{Z}(z) =\int_{\R}\left[ \int_{\R}e^{izs}dF_{Z}(z)\right] e^{i(Y-x)s}\frac{hH(hs)}{\Phi _{Z}(s)}ds \\
&=\int_{\R}\Phi _{Z}\left( s\right) e^{i(Y-x)s}\frac{hH(hs)}{\Phi _{Z}(s)}
ds=\int_{\R}e^{i(Y-x)s}hH(hs)ds
=
(\mathcal{F}H)\left(\frac{Y-x}{h}\right)
.
\end{align*}
where the last equality follows by the change of variables \(t=hs\).
Thus,
\begin{equation}
E(\hat{p}_{x})
=
E(E\left( K_{h}\left( X-x\right) |Y\right) )
=
\int_{\mathbb R}
W\left(\frac{u-x}{h}\right)dF_Y(u).
\label{pxexpec}
\end{equation}

$(1\Rightarrow 3)$.
Statement 1 implies $W(0)=(\mathcal FH)(0)=\int_{\R}H(t)dt=1$ since $W:=\mathcal{F}H$. In addition, $\mathcal FH$ is continuous and vanishes at infinity by the Riemann-Lebesgue lemma. Thus,
 $\mathcal{F}H$ satisfies equation \eqref{A3}.

 $(3\Rightarrow 1)$. This follows since $W(0)=1$ by statement 3, and $W(0)=(\mathcal FH)(0)=\int_{\R}H(t)dt$ since $W:=\mathcal{F}H$.

% $(1\Rightarrow 2)$. We have just shown Statement 1 implies that $W:=\mathcal{F}H$ satisfies equation \eqref{A3}. Then, statement 2 follows from \eqref{jumpmot}. 

$(3\Rightarrow 2)$.
Equation \eqref{A3} implies 
$W$ is bounded and
$\underset{h \to 0}{\lim}W\left(\frac{u-x}{h}\right)= \chi_{{x}}(u)$,
so bounded convergence gives 
\begin{equation*}%\label{jumpmot}
    \lim_{h\to 0}
    \int_{\R}W\left(\frac{u-x}{h}\right)dF_Y(u)
    % \to 
    =
    p_x
    % \quad
    % \text{ as }
    % h\downarrow0
    .
\end{equation*}
Then equation \eqref{pxexpec} implies statement 2 holds.

$(2\Rightarrow 1)$. 
Fix $x\in\R$ and take $F_Y$ to be the distribution function of a point mass at $x$, so $p_x=1$. 
By equation \eqref{pxexpec}
$$
% \lim_{h\to0}
E(\hat{p}_{x})
=
% \lim_{h\to0}
\int_{\mathbb R}
W\left(\frac{u-x}{h}\right)dF_Y(u)
=
W(0)
,
$$
so $W(0)=1$ since $p_x=1$ by construction and statement 2 implies
$\displaystyle\lim_{h\to0}
E(\hat{p}_{x})=p_x$. 
Thus, statement 1 holds since $W(0)=(\mathcal FH)(0)=\int_{\R}H(t)dt$. 
% 
% Because $H\in L(\mathbb R)$, the Riemann-Lebesgue lemma implies that $W$ is bounded and continuous and that $W(v)\to0$ as $|v|\to\infty$.
% Hence, for every \(u\in\mathbb R\),
% $$
% W\left(\frac{u-x}{h}\right)
% \rightarrow
% W(0)\chi_x(u),
% \qquad\text{as }h\to0.
% $$
% Bounded convergence applied to equation \eqref{pxexpec} gives
% $$
% \lim_{h\to0}
% E(\hat{p}_{x})
% =
% \lim_{h\to0}
% \int_{\mathbb R}
% W\left(\frac{u-x}{h}\right)dF_Y(u)
% =
% W(0)
% ,
% $$
% so $W(0)=1$ since $p_x=1$ by construction and statement 2 implies
% $\lim_{h\to0}
% E(\hat{p}_{x})=p_x$. 
% Thus, statement 1 holds since $W(0)=(\mathcal FH)(0)=\int_{\R}H(t)dt$. 

Finally for the bias bound, for all $h,\varepsilon_1 \in (0,1)$ and $\varepsilon_2>\varepsilon_1$, we obtain from equations \eqref{pxexpec}
% , \eqref{e77n}
and \eqref{25n}  that 
\begin{eqnarray}
\left\vert E(\hat{p}_{x})-p_{x}\right\vert 
% &= &
% \left\vert \int_{\R}e^{-isx}(\mathcal{F}F_Y)(s)hH(hs)ds-p_{x}\right\vert 
% \notag\\
&\leq &
\omega _{\mathcal{F}H}(h^{\varepsilon_{2}-\varepsilon _{1}})\left[ p_{x}+\delta _{F_Y}(x,h^{1-\varepsilon _{1}})\right]  +\left( 1+\left\Vert \mathcal{F}H\right\Vert _{C_b}\right) \delta _{F_Y}(x,h^{1-\varepsilon_{1}}) 
% \notag\\
% &+&
+
\phi _{\mathcal{F}H}(h^{-\varepsilon _{1}}).  \label{25nn}
\end{eqnarray}
\hfill$\square$
%%%%%%%%%%%%%%%%%%%%%%%%%%%%%%%%%%%%%%%%%%%

%%%%%%%%%%%%%%%%%%%%%%%%%%%%%%%%%
% Proof Theorem thm12
%%%%%%%%%%%%%%%%%%%%%%%%%%%%%%%%
\noindent \bf Theorem \ref{thm12}: \it Proof. \rm We have
$
V\left( \hat{p}_{x}\right) \leq \frac{1}{n}E\left(K_{h}^{2}\left( X-x\right) \right)
$.  Then,
\begin{equation*}
E\left(K_{h}^{2}\left( X-x\right) \right)=\int_{\R}\int_{\R}e^{-ix(s+t)}\gamma _{h}\left(s\right) \gamma _{h}\left( t\right) \mathcal{F} F_{Y}\left( s+t\right) \mathcal{F} F_Z\left( s+t\right) ds dt.
\end{equation*}
This is bounded as
$
E\left(K^{2}\left( X-x\right) \right) \leq \int_{\R}\int_{\R}\left\vert \gamma _{h}\left(
s\right) \gamma _{h}\left( t\right) \right\vert dsdt=\left(
\int_{\R}\left\vert \gamma _{h}\left( s\right) \right\vert ds\right) ^{2}.
$  It follows that $V\left( \widehat{p}_{x}\right) \leq \frac{1}{n}
u_{h }^{2}\rightarrow 0$.  This convergence together with Theorem \ref{thm11} completes the proof. \hfill$\square$.
%%%%%%%%%%%%%%%%%%%%%%%%%%%%%%%%%%%%%%%%%%

%%%%%%%%%%%%%%%%%%%%%%%%%%%%%%%%%
% Proof Theorem thm10j
%%%%%%%%%%%%%%%%%%%%%%%%%%%%%%%%
\noindent \bf Theorem \ref{thm10j}: \it Proof. \rm 
% \medskip
% \noindent
\textbf{(a)}
% a) 
Given that $|\Phi_Z(t)| \neq 0$, the assumptions on $H$ and continuity of $\Phi_Z$ at $t=0$, we have that for every $\epsilon \in (0,1)$ there exists $\beta>0$ such that  $\frac{1}{|\Phi_Z(t)|}\le \frac{1}{1-\epsilon}$ whenever $|t|\le \beta$.  Hence, for every $h>0$
\begin{align}\label{eq10j}
u_h &\le C_H h\int_{|t| \le \gamma/h}\frac{1}{|\Phi_Z(t)|}dt \le C_Hh\left( \frac{2\beta}{1-\epsilon} +2 \int_{\beta}^{ \gamma/h}\frac{1}{|\Phi_Z(t)|}dt\right)\\
& \le C_Hh\left( \frac{2\beta}{1-\epsilon} +2 C\int_{\beta}^{ \gamma/h}\exp(\tau t^\rho)dt\right) \mbox{ since $|\Phi_Z(t)| \asymp \exp (-\tau |t|^{\rho})$ as $|t| \to \infty$.}\notag
\end{align}
Since
$
 \int_{\beta}^{\gamma/h} e^{\tau t^{\rho}}\,dt
 \;\asymp\;\frac{e^{\tau(\gamma/h)^{\rho}}}{\tau\rho\,(\gamma/h)^{\rho-1}} .
$
we have
\[
 \frac{u_h}{\sqrt n}
 \le \frac{2\beta C_H}{1-\epsilon}\,\frac{h}{\sqrt n}
   +\frac{2C_H C}{\tau\rho}\,
    \frac{h}{(\gamma/h)^{\rho-1}}\,\frac{e^{\tau(\gamma/h)^{\rho}}}{\sqrt n}.
\]
The first term tends to $0$. For the second, the conditions on $h$ give
$\gamma/h\le(\log n/2\tau)^{1/\rho}$ for all large $n$, so
$\tau(\gamma/h)^{\rho}\le\tfrac12\log n$ and $e^{\tau(\gamma/h)^{\rho}}\le\sqrt n$.  Therefore
\[
 \frac{2C_H C}{\tau\rho}\,\frac{h}{(\gamma/h)^{\rho-1}}\,
     \frac{e^{\tau(\gamma/h)^{\rho}}}{\sqrt n}
 \;\le\;\frac{2C_H C}{\tau\rho\,\gamma^{\rho-1}}\,h^{\rho}\;\longrightarrow\;0
\]
since $h\to0$ and $\rho>0$. Thus $u_h/\sqrt n=o(1)$, including the boundary
case $\gamma/h=(\log n/2\tau)^{1/\rho}$.

\medskip
\noindent
\textbf{(b)}
% \noindent b) 
Since $|\Phi_Z(t)| \asymp |t|^{-\rho}$ as $|t| \to \infty$, we have from the arguments in part (a) that
\begin{align*}
\frac{u_h}{\sqrt{n}} & \le C_H\frac{h}{\sqrt{n}} \left( \frac{2 \beta}{1-\epsilon}+ 2  \int_{\beta}^{ \gamma/h}t^\rho dt \right)=2C_H\frac{h}{\sqrt{n}} \left(\frac{\beta}{1-\epsilon}-\frac{ \beta^{\rho+1}}{\rho+1}\right)+ 2C_H \frac{1}{\sqrt{n}h^\rho}\frac{\gamma^{\rho+1}}{\rho+1}.
\end{align*}
Hence, $\frac{u_h}{\sqrt{n}} =o(1)$ if $nh^{2\rho}\to \infty$.
\hfill$\square$

\noindent \bf Corollary \ref{coro2}: \it Proof. \rm
By the fact that 
$\phi_{\mathcal F H}(N)\le C\,N^{-k}$,
 $\omega_{\mathcal F H}(\varepsilon)\le C\,\varepsilon^{m}$,
 $\delta_{F_Y}(\eta)\le \psi(\eta)$ and $p_x=O(1)$, the three bias terms are
$\omega_{\mathcal F H}(h^{\varepsilon_2-\varepsilon_1})(p_x+\delta)=O(h^{m(\varepsilon_2-\varepsilon_1)})$,
$(1+\|\mathcal F H\|_{C_b})\,\delta_{F_Y}(h^{1-\varepsilon_1})=O(\psi(h^{1-\varepsilon_1}))$
and $\phi_{\mathcal F H}(h^{-\varepsilon_1})=O(h^{k\varepsilon_1})$, so the bias is
$B(h)=O\big(h^{c}+\psi(h^{1-\varepsilon_1})\big)$ with $c=\min\{m(\varepsilon_2-\varepsilon_1),k\varepsilon_1\}$.
For the variance, $u_h\le C_H h\int_{\beta<t\le\gamma/h}|\Phi_Z(t)|^{-1}dt$ gives,
by Lemma \ref{lem7a}, $\tfrac1n u_h^2\asymp n^{-1}h^{-2\rho}$ in the ordinary-smooth case
and $\tfrac1n u_h^2\asymp n^{-1}h^{2\rho}e^{2\tau(\gamma/h)^{\rho}}$ in the super-smooth case.

\medskip
\noindent
\textbf{(a)}
With $h_n=A(\log n)^{-1/\rho}$ and $A^{\rho}>4\tau\gamma^{\rho}$ one has
$(\gamma/h_n)^{\rho}=\gamma^{\rho}(\log n)/A^{\rho}$, so
$e^{2\tau(\gamma/h_n)^{\rho}}=n^{2\tau\gamma^{\rho}/A^{\rho}}$ with
$2\tau\gamma^{\rho}/A^{\rho}<\tfrac12$; hence the variance is $O(n^{-1/2})$ and
is dominated by the bias. Since $h_n^{c}\asymp(\log n)^{-c/\rho}$ and
$h_n^{1-\varepsilon_1}\asymp(\log n)^{-(1-\varepsilon_1)/\rho}$, the stated bound
follows, and the H\"older specialization gives $c=r$ and $(\log n)^{-2r/\rho}$.

\medskip
\noindent
\textbf{(b)}
With $\psi(\eta)=C\eta^{s}$ and the stated $\varepsilon_1,\varepsilon_2$,
$B(h)=O(h^{r})$; balancing $n^{-1}h^{-2\rho}=h^{2r}$ gives
$h_n\asymp n^{-1/\{2(r+\rho)\}}$ and $\mathrm{MSE}=O(h_n^{2r})=O(n^{-r/(r+\rho)})$.
\hfill $\square$

%%%%%%%%%%%%%%%%%%%%%%%%%%%%%%%%%%%%%%%%%
% Appendix
%%%%%%%%%%%%%%%%%%%%%%%%%%%%%%%%
% \section*{Appendix}
\section{Extensions to non-compact regularization kernels}\label{app:noncompH}

%%%%%%%%%%%%%%%%%%%%%%%%%%%%%
%% lemma
%%%%%%%%%%%%%%%%%%%%%%%%%%%%%
\begin{applemma}\label{lem6a_NoncompH} 
Suppose $H$ is not compactly supported and $C_H :=\underset{t \in \R}{\sup}|H(t)|<\infty$.
 % Assume that  $C_H :=\underset{t \in \R}{\sup}|H(t)|$ and $H$ does not have compact support.  
 In addition, assume that  for each $h>0$ and sufficiently small, there exists $t(h)>0$ such that $t(h) \to \infty $ as $h \to 0$ with $\underset{t> t(h)}{\int} \frac{|H(ht)|}{t|\Phi_Z(t)|}dt=o\left(\underset{\beta < t \le t(h)}{\int} \frac{|H(ht)|}{t|\Phi_Z(t)|}dt\right)$.  Then, for $\lambda \in (0,1)$ and $h \in (0,1]$ there exists $\beta>0$ such that 
$$
 v_{h,\lambda} =O\left( \log \lambda^{-1}+  \underset{ \beta < t\le t(h)}{\int} \frac{|H(ht)|}{t|\Phi_Z(t)|}dt\right).
 $$ 
\end{applemma}
% \noindent \bf Lemma \ref{lem6a_NoncompH}: 
\noindent
\it Proof. \rm  
For each $h>0$ and sufficiently small, there exists $t(h)>0$ such that $t(h) \to \infty$ as $h \to 0$ and given $C_H := \underset{\R}{\sup} |H(t)|$, using the second bound in equation \eqref{eq75} we have
 \begin{align*}
v_{h,\lambda} &\le \frac{2 \pi C_H}{1-\epsilon}+2\underset{ \pi \lambda \le |t|\le t(h)}{\int} \frac{|H(ht)|}{|t\Phi_Z(t)|}dt +2\underset{ |t|> t(h)}{\int} \frac{|H(ht)|}{|t\Phi_Z(t)|}dt.
\end{align*} 
For $\beta<t(h)$, and given that $H$ is even
\begin{align*}
\underset{ \pi \lambda \le |t|\le t(h)}{\int} \frac{|H(ht)|}{|t\Phi_Z(t)|}dt = \underset{ \pi \lambda \le |t|\le \beta}{\int} \frac{|H(ht)|}{|t\Phi_Z(t)|}dt+\underset{ \beta < |t|\le t(h)}{\int} \frac{|H(ht)|}{|t\Phi_Z(t)|}dt & \le \frac{2 C_H}{1- \epsilon}(\log \beta - \log \pi + \log \lambda^{-1} )\\
&+2\underset{ \beta < t\le t(h)}{\int} \frac{|H(ht)|}{|t\Phi_Z(t)|}dt.
\end{align*}
 Hence,
 \begin{align*}
v_{h,\lambda} &\le \frac{2 \pi C_H}{1-\epsilon}+\frac{4 C_H}{1-\epsilon}\left(\log \beta - \log \pi + \log \lambda^{-1} \right)+4\underset{ \beta < t\le t(h)}{\int} \frac{|H(ht)|}{t|\Phi_Z(t)|}dt+4\underset{ t> t(h)}{\int} \frac{|H(ht)|}{t|\Phi_Z(t)|}dt.
\end{align*} 
Then, given that  $\underset{ t> t(h)}{\int} \frac{|H(ht)|}{t|\Phi_Z(t)|}dt=o\left(\underset{ \beta < t\le t(h)}{\int} \frac{|H(ht)|}{t|\Phi_Z(t)|}dt\right)$, we have 
$
v_{h,\lambda} =O\left( \log \lambda^{-1}+  \underset{ \beta < t\le t(h)}{\int} \frac{|H(ht)|}{t|\Phi_Z(t)|}dt\right)
$.\hfill$\square$
 %%%%%%%%%%%%%%%%%%%%%%%%%%%%%%%%%%%%%%%%%%%

%%%%%%%%%%%%%%%%%%%%%%%%%%%%%
%% lemma
%%%%%%%%%%%%%%%%%%%%%%%%%%%%%
 \begin{applemma}\label{lem7a_NoncompH}
Let $0<\beta<\xi_n$ for $n \in \mathds{N}$ with $\xi_n \to \infty$ as $n \to \infty$ and $I(\xi_n):=\int_{\beta <t \le {\xi_n}}\frac{1}{t|\Phi_Z(t)|}dt$.  
If there exist $\tau,\rho>0$ such that
$|t|\,|\Phi_Z(t)|\asymp\exp(-\tau|t|^{\rho})$
 as $|t|\to\infty$, then
\[
I(\xi_n)\;\asymp\;\frac{e^{\tau\xi_n^{\rho}}}{\tau\rho\,\xi_n^{\rho-1}}
\;=\;O\!\left(\xi_n\,e^{\tau\xi_n^{\rho}}\right),
\]
and
 if $\xi_n^{\rho}=o(\log n)$,
then
 $I(\xi_n)=o\big(n^{1/2}\big)$.
\end{applemma}
\noindent
\it Proof. \rm 
Since $\dfrac{1}{t\,|\Phi_Z(t)|}\asymp e^{\tau t^{\rho}}$ for $t>\beta$, we have 
$I(\xi_n)\asymp\int_{\beta}^{\xi_n}e^{\tau
 t^{\rho}}\,dt$.
Writing
 $e^{\tau t^{\rho}}=\dfrac{1}{\tau\rho\,t^{\rho-1}}\dfrac{d}{dt}e^{\tau t^{\rho}}$
and
 integrating by parts,
\[
\int_{\beta}^{\xi_n}e^{\tau
 t^{\rho}}\,dt
=\frac{\xi_n^{\,1-\rho}}{\tau\rho}\,e^{\tau\xi_n^{\rho}}
-\frac{\beta^{\,1-\rho}}{\tau\rho}\,e^{\tau\beta^{\rho}}
-\frac{1-\rho}{\tau\rho}\int_{\beta}^{\xi_n}t^{-\rho}e^{\tau
 t^{\rho}}\,dt ,
\]
where
 the constant term is $O(1)$ and the last integral is
$o\big(\xi_n^{\,1-\rho}e^{\tau\xi_n^{\rho}}\big)$.
 Hence
$\int_{\beta}^{\xi_n}e^{\tau
 t^{\rho}}\,dt\sim
\dfrac{\xi_n^{\,1-\rho}}{\tau\rho}\,e^{\tau\xi_n^{\rho}}
=\dfrac{e^{\tau\xi_n^{\rho}}}{\tau\rho\,\xi_n^{\rho-1}}$,
which
 is the stated order. Equivalently, apply L'H\^opital to the ratio
$\int_{\beta}^{\xi_n}e^{\tau
 t^{\rho}}dt\big/\big(\tfrac{\xi_n^{1-\rho}}{\tau\rho}e^{\tau\xi_n^{\rho}}\big)$,
whose
 derivative quotient is $\big(1+\tfrac{1-\rho}{\tau\rho}\xi_n^{-\rho}\big)^{-1}\to1$.
Finally,
 if $\xi_n^{\rho}=o(\log n)$ then $e^{\tau\xi_n^{\rho}}=e^{o(\log n)}=n^{o(1)}$,
while
 $\xi_n^{\,1-\rho}$ is at most $\big((\log n)^{1/\rho}\big)^{|1-\rho|}$, i.e.
$\mathrm{polylog}(n)=n^{o(1)}$;
 therefore $I(\xi_n)=n^{o(1)}$, and
$n^{o(1)}/n^{1/2}=e^{o(\log
 n)-\frac12\log n}\to0$ gives $I(\xi_n)=o(n^{1/2})$.
\hfill$\square$
%%%%%%%%%%%%%%%%%%%%%%%%%%%%%%%%%%%%%%%%%%%%

%%%%%%%%%%%%%%%%%%%%%%%%%%%%%%%%%%
%% Theorem 11
%%%%%%%%%%%%%%%%%%%%%%%%%%%%%%%%%
\begin{apptheorem}\label{thm11a}
Suppose $H$ is not compactly supported and
$C_H :=\underset{t \in \R}{\sup}|H(t)|<\infty$.
% Let $C_H :=\underset{t \in \R}{\sup}|H(t)|$ and assume $H$ is not compactly supported.  
Suppose there exist positive constants $\tau, \rho, \theta, \epsilon$ such that 
\begin{equation}\label{eq77a}
|t|| \Phi_Z(t)| \asymp \exp \left(-\tau |t|^{\rho} \right) \mbox{ and } H(t)=O\left( \exp (-\theta |t|^{\rho+\epsilon})\right).
\end{equation}
If $\lambda=\exp(-o(\sqrt{n}))$ and $ h^{\rho(\frac{\rho}{\epsilon} +1)} \log n \to \infty$, then $\frac{1}{\sqrt{n}}v_{h,\lambda}=o(1)$.
% \iffalse
% \noindent b) Suppose there exist positive constants $\rho, \beta,  t_0$ and $\delta>1$ such that 
% \begin{equation}\label{eq77b}
% | \Phi_Z(t)| \asymp |t|^{-(\rho+1)},\, H(t)=O\left( |t|^{ \frac{\delta}{\delta-1} (\rho+1)}\right) \mbox{ and } \int_{\beta<t<t_0}\frac{1}{t|\Phi_Z(t)|}dt<\infty.
% \end{equation}
% Then, if $\lambda=\exp(-o(\sqrt{n}))$ and $  \to \infty$, $\frac{1}{\sqrt{n}}v_{h,\lambda}=o(1)$.
% \fi
\end{apptheorem}
%%%%%%%%%%%%%%%%%%%%%%%%%%%%%%%%%%%%%%
%%%%%%%%%%%%%%%%%%%%%%%%%%%%%%%%%
%% Proof Theorem thm11a
%%%%%%%%%%%%%%%%%%%%%%%%%%%%%%%%%%
% \noindent \bf Theorem \ref{thm11a}: 
\noindent
\it Proof. \rm Let $t_0$ be such that $\beta<t_0<t(h) := \left(\frac{2\tau}{\theta h^{\rho+\epsilon}} \right)^{1/\epsilon}$ and note that $t(h) \to \infty$ as $h \to 0$.  Now,
\begin{equation}
\int_{\beta<t \le t(h)}\frac{|H(ht)|}{t|\Phi_Z(t)|}dt=\int_{\beta<t \le t_0}\frac{|H(ht)|}{t|\Phi_Z(t)|}dt+\int_{t_0<t \le t(h)}\frac{|H(ht)|}{t|\Phi_Z(t)|}dt,
\end{equation}
where the first integral on the right-hand side is finite by the assumption that $|t|| \Phi_Z(t)| \asymp \exp \left(-\tau |t|^{\rho} \right)$ and the uniform bound on $H$.  For the second integral, we note that given equation \eqref{eq77a}, there exists $C>0$ such that 
\begin{equation*}
\frac{|H(ht)|}{t|\Phi_Z(t)|} \le C \exp(\psi_h(t)) \mbox{ where $\psi_h(t)=\tau t^\rho -\theta (th)^{\rho+\epsilon}$,}
\end{equation*}
and consequently $\int_{t_0<t \le t(h)}\frac{|H(ht)|}{t|\Phi_Z(t)|}dt \le C \int_{t_0<t \le t(h)}\exp(\psi_h(t)) dt$.  If $t \le \frac{t(h)}{2^{2/\epsilon}}$ then $\psi_h(t)\ge \frac{\tau t^\rho}{2}$ and
\begin{equation}\label{eq80a}
\int_{t_0 < t \le t(h) }\exp(\psi_h(t)) dt \ge \int_{t_0 < t \le  \frac{t(h)}{2^{2/\epsilon}} }\exp(\psi_h(t)) dt \ge \int_{  t_0 < t \le \frac{t(h)}{2^{2/\epsilon}} }\exp\left(\frac{\tau t^\rho}{2}\right) dt \to \infty \mbox{ as $h \to 0$.}
\end{equation}
Now, note that $t>t(h)$ implies that $\psi_h(t) <-\tau t^\rho$, hence 
\begin{equation}\label{eq81a}
\int_{t > t(h)}\frac{|H(ht)|}{t|\Phi_Z(t)|}dt \le C \int_{t > t(h)}\exp(\psi_h(t)) dt\le C \int_{t > t(h)}\exp(-\tau t^\rho) dt \le C \int_{t > t_0}\exp(-\tau t^\rho) dt<\infty.
\end{equation}
Equations \eqref{eq80a} and \eqref{eq81a} imply that $\underset{t> t(h)}{\int} \frac{|H(ht)|}{t|\Phi_Z(t)|}dt=o\left(\underset{\beta < t \le t(h)}{\int} \frac{|H(ht)|}{t|\Phi_Z(t)|}dt\right)$.  Hence, 
% by part (b) of Lemma \ref{lem6a}
by Lemma \ref{lem6a_NoncompH}
we have that for some $C>0$
\begin{equation}\label{eq74n}
\frac{1}{\sqrt{n}}v_{h,\lambda} \le C\left( \frac{1}{\sqrt{n}} \log \lambda^{-1} + \frac{C_H}{\sqrt{n}} I(t(h))\right),
\end{equation}
where $I(t(h)):=\int_\beta^{t(h)}\frac{1}{t|\Phi(t)|}dt$.  

Under
 \eqref{eq77a} we have $|t|\,|\Phi_Z(t)|\asymp\exp(-\tau|t|^{\rho})$, hence
$\dfrac{1}{t\,|\Phi_Z(t)|}\asymp
 e^{\tau t^{\rho}}$, so by Lemma~\ref{lem7a_NoncompH}
the
 term $I(t(h))=\int_{\beta}^{t(h)}\frac{1}{t|\Phi_Z(t)|}\,dt$ appearing in
\eqref{eq74n}
 satisfies
\[
I\big(t(h)\big)\;\asymp\;\frac{e^{\tau
 t(h)^{\rho}}}{\tau\rho\,t(h)^{\rho-1}} .
\]
Since
 $t(h)=\big(2\tau/(\theta h^{\rho+\epsilon})\big)^{1/\epsilon}$, we have
$t(h)^{\rho}\asymp
 h^{-\rho(\rho/\epsilon+1)}$, and the assumption
$h^{\rho(\rho/\epsilon+1)}\log
 n\to\infty$ is precisely
$h^{-\rho(\rho/\epsilon+1)}=o(\log
 n)$, hence $t(h)^{\rho}=o(\log n)$. Therefore
$e^{\tau
 t(h)^{\rho}}=n^{o(1)}$, and as $t(h)^{\,1-\rho}$ is polylogarithmic in $n$,
$I(t(h))=n^{o(1)}=o(\sqrt
 n)$. Consequently the second term on the right-hand side
of
 \eqref{eq74n} obeys $\frac{C_H}{\sqrt n}I(t(h))\to0$.  Together with
$\lambda=\exp(-o(\sqrt
 n))$, the entire right-hand side of
\eqref{eq74n}
 tends to zero, so $\dfrac{1}{\sqrt n}\,v_{h,\lambda}\to0$.
\hfill$\square$
%%%%%%%%%%%%%%%%%%%%%%%%%%%%%%%%%%%%%%%%%%%%

%%%%%%%%%%%%%%%%%%%%%%%%%
% Theorem 5
%%%%%%%%%%%%%%%%%%%%%
\begin{apptheorem}\label{thm:FYx_MSEbnd_NoncompH}
Suppose the conditions stated in Theorem \ref{thm7} hold and $H$ does not have compact support.
Let $C_H:=\underset{t \in \R}{\sup}|H(t)|$ and assume that for each sufficiently small $h>0$, there exists $t(h)>0$ such that $t(h) \to \infty $ as $h \to 0$ with $\underset{t> t(h)}{\int} \frac{|H(ht)|}{t|\Phi_Z(t)|}dt=o\left(\underset{\beta < t \le t(h)}{\int} \frac{|H(ht)|}{t|\Phi_Z(t)|}dt\right)$.  Then, for $\lambda \in (0,1)$ and $h \in (0,1]$ there exist $C, \,\beta>0$ such that 
\begin{align*}
MSE(\FYhat(x)) &\le \frac{1}{n} \frac{C}{ 4\pi^{2}}  \left( \log \lambda^{-1}+ \underset{ \beta < t\le t(h)}{\int} \frac{|H(ht)|}{t|\Phi_Z(t)|}dt\right)^2+ \left[\phi _{G}(h^{\delta -1})+(1+\left\Vert
G\right\Vert _{C_b})\omega_{F_Y} (x,h^{\delta }) \right.\\
&\left.+(2+\left\Vert G\right\Vert _{\infty})F_{Y}(x+1-1/\lambda ) \right]^2.
\end{align*}
\end{apptheorem}
\noindent
\it Proof. \rm 
Follows immediately from Theorems \ref{thm7} and \ref{thm8} with the bound on $v_{h,\lambda}$ obtained in Lemma \ref{lem6a_NoncompH}.
\hfill$\square$
%%%%%%%%%%%%%%%%%%%%%%%%%%%%%%%%%%%%%%%%%%%%

%%%%%%%%%%%%%%%%%%%%%%%%%%%%%
%% lemma
%%%%%%%%%%%%%%%%%%%%%%%%%%%%%
\begin{applemma}\label{lem6ab_NoncompH} 
% Suppose $\supp H \subset [-\gamma,\gamma]$ for some $\gamma >0$ and let $C_H:=\underset{t \in [-\gamma,\gamma]}{\sup}|H(t)|$.  Then, there exists $\beta>0$ such that for all $h\in(0,\gamma/\beta)$, 
% $$
%  v_{h}(x,y) =O\left( \underset{ \beta < t\le\gamma/h}{\int} \frac{1}{t|\Phi_Z(t)|}dt\right)
%  $$ 
%  \noindent b) 
Suppose $H$ does not have compact support and $C_H :=\underset{t \in \R}{\sup}|H(t)|<\infty$.  Assume that for each $h>0$ and sufficiently small, there exists $t(h)>0$ such that $t(h) \to \infty $ as $h \to 0$ with $\underset{t> t(h)}{\int} \frac{|H(ht)|}{t|\Phi_Z(t)|}dt=o\left(\underset{\beta < t \le t(h)}{\int} \frac{|H(ht)|}{t|\Phi_Z(t)|}dt\right)$.  Then, for $h \in (0,1]$ there exists $\beta>0$ such that 
$$
 v_{h}(x,y) =O\left(  \underset{ \beta < t\le t(h)}{\int} \frac{|H(ht)|}{t|\Phi_Z(t)|}dt\right).
 $$ 
\end{applemma}
\noindent \it Proof. \rm 
For each $h>0$ and sufficiently small, there exists $t(h)>\beta>0$ such that $t(h) \to \infty$ as $h \to 0$ and given $C_H := \underset{\R}{\sup} |H(t)|$, we have
 \begin{align*}
v_{h}(x,y) &\le \frac{2 C_H(y-x) \beta}{1-\epsilon}+4  \underset{ \beta < t\le t(h)}{\int} \frac{|H(ht)|}{t|\Phi_Z(t)|}dt+4\underset{ t> t(h)}{\int} \frac{|H(ht)|}{t|\Phi_Z(t)|}dt.
\end{align*} 
Then, given that  $\underset{ t> t(h)}{\int} \frac{|H(ht)|}{t|\Phi_Z(t)|}dt=o\left(\underset{ \beta < t\le t(h)}{\int} \frac{|H(ht)|}{t|\Phi_Z(t)|}dt\right)$, we have 
$
v_{h}(x,y) =O\left(  \underset{ \beta < t\le t(h)}{\int} \frac{|H(ht)|}{t|\Phi_Z(t)|}dt\right)
$. \hfill$\square$
%%%%%%%%%%%%%%%%%%%%%%%%%%%%%%%%%%%%%%%%%

%%%%%%%%%%%%%%%%%%%%%%%%%%%%%%%%%%%%%%%%%%%%

%%%%%%%%%%%%%%%%%%%%%%%%%%%%%%%%%%
%% Theorem 11ab
%%%%%%%%%%%%%%%%%%%%%%%%%%%%%%%%%
\begin{apptheorem}\label{thm11ab}
Suppose $H$ is not compactly supported and $C_H :=\underset{t \in \R}{\sup}|H(t)|<\infty$.
% Let $C_H :=\underset{t \in \R}{\sup}|H(t)|$ and assume $H$ is not compactly supported.  
Suppose there exist positive constants $\tau, \rho, \theta, \epsilon$ such that equation \eqref{eq77a} holds.  Then, if $ h^{\rho(\frac{\rho}{\epsilon} +1)} \log n \to \infty$, $\frac{1}{\sqrt{n}}v_{h}(x,y)=o(1)$.
\end{apptheorem}
%%%%%%%%%%%%%%%%%%%%%%%%%%%%%%%%%
% Proof Theorem thm11ab
%%%%%%%%%%%%%%%%%%%%%%%%%%%%%%%%
\noindent \it Proof. \rm The proof is exactly as that of Theorem \ref{thm11a}, where we can conclude here that
\begin{equation}\label{eq74nab}
\frac{1}{\sqrt{n}}v_{h}(x,y) \le \frac{C}{\sqrt{n}} I(t(h)),
\end{equation}
with $I(t(h)):=\int_\beta^{t(h)}\frac{1}{t|\Phi(t)|}dt$.  By
Lemma~\ref{lem7a_NoncompH},
 the term $I(t(h))$ in \eqref{eq74nab} is $n^{o(1)}=o(\sqrt n)$
under
 $t(h)^{\rho}=o(\log n)$, so $\dfrac{1}{\sqrt n}\,v_{h,\lambda}(x,y)\to0$.
\hfill$\square$  
%%%%%%%%%%%%%%%%%%%%%%%%%%%%%%%%%%

%%%%%%%%%%%%%%%%%%%%%%%%%%%%%%%%%%
%% Theorem 
%%%%%%%%%%%%%%%%%%%%%%%%%%%%%%%%%
\begin{apptheorem}\label{thm:FYxy_MSEbnd_NoncompH}
Suppose the conditions stated in Theorem \ref{thm13n} hold. 
% Under the conditions stated in Theorem \ref{thm13n}, we have:\\
If $C_H:=\underset{t \in \R}{\sup}|H(t)|$ and $H$ does not have compact support, assume that for each $h>0$ and sufficiently small, there exists $t(h)>0$ such that $t(h) \to \infty $ as $h \to 0$ with $\underset{t> t(h)}{\int} \frac{|H(ht)|}{t|\Phi_Z(t)|}dt=o\left(\underset{\beta < t \le t(h)}{\int} \frac{|H(ht)|}{t|\Phi_Z(t)|}dt\right)$.  Then, for $h \in (0,1]$ there exist $C, \,\beta>0$ such that 
\begin{align*}
MSE(\FYhat(x,y)) &\le \frac{1}{n} \frac{C}{ 4\pi^{2}} \left( \underset{ \beta < t\le t(h)}{\int} \frac{|H(ht)|}{t|\Phi_Z(t)|}dt\right)^2+ \left[ \phi _{G}(h^{\delta
-1})+(1+\left\Vert G\right\Vert _{\infty})(\omega_{F_Y}(x,h^{\delta })+\omega_{F_Y}
(y,h^{\delta })) \right]^2.
\end{align*}
\end{apptheorem}
\noindent \it Proof. \rm
Follows immediately from Theorems \ref{thm13n} and \ref{thm8a} with the bound on $v_{h,\lambda}$ obtained in Lemma \ref{lem6ab_NoncompH}.
\hfill$\square$  
%%%%%%%%%%%%%%%%%%%%%%%%%%%%%%%%%%

% \clearpage
% \newpage
%%%%%%%%%%%%%%%%%%%%%%%%%%%%%%%%%%%%%%%%%
% Appendix
%%%%%%%%%%%%%%%%%%%%%%%%%%%%%%%%
\section{Simulation study details and additional results}\label{app:sims}

This appendix contains 
detailed descriptions of the kink distributions and estimators, and additional simulation results.
% full simulation results and detailed descriptions of the estimators and kink distributions.

\subsection{Kink Distribution Definitions}\label{app:FYkink_def}

The three non-smooth distributions are denoted by $K_1$, $K_2$, and $K_3$. 
The first two are constructed by piecewise reweighting the standard normal
density, while the third is an asymmetric piecewise-uniform distribution.

Let $\Phi$ denote the standard normal distribution function. For a partition
$
-\infty=a_0<a_1<\cdots<a_J=\infty
$
and nonnegative weights $c_1,\ldots,c_J$, with at least one strictly positive
weight, define
\[
F_Y^{\mathrm{(K)}}\left(
    x;[a_j]_{j=0}^{J},[c_j]_{j=1}^{J}
\right)
\coloneqq
\frac{1}{C}
\sum_{j=1}^{J} c_j
\left[
    \Phi\left(
        \min\{\max\{x,a_{j-1}\},a_j\}
    \right)
    -\Phi(a_{j-1})
\right],
\]
where
$
C\coloneqq
\sum_{j=1}^{J}c_j
\left\{
    \Phi(a_j)-\Phi(a_{j-1})
\right\}.
$
Thus, on the interval $(a_{j-1},a_j)$, the density is proportional to
$c_j\phi(x)$, where $\phi$ denotes the standard normal density.

The one-kink distribution $Y\sim K_1$ is defined by
$
F_Y(x)
=
F_Y^{\mathrm{(K)}}\left(
    x;[a_j]_{j=0}^{2},[c_j]_{j=1}^{2}
\right),
$
with
$
(a_0,a_1,a_2)=(-\infty,0,\infty),
$ and $
(c_1,c_2)=(3,1).
$
Its normalizing constant is $C=2$, and its density has a discontinuity at
$x=0$.

The gap-kink distribution $Y\sim K_2$ is defined by
$
F_Y(x)
=
F_Y^{\mathrm{(K)}}\left(
    x;[a_j]_{j=0}^{3},[c_j]_{j=1}^{3}
\right),
$
with
$
(a_0,a_1,a_2,a_3)=(-\infty,-1,1,\infty),
$ and $
(c_1,c_2,c_3)=(1,0,1).
$
Its normalizing constant is $C=2\Phi(-1)$. Here the distribution function is constant and equal to $1/2$ on $[-1,1]$, with kink points at $-1$ and $1$.

To define the third distribution, let
$
-\infty<b_0<b_1<\cdots<b_L<\infty
$
be finite interval endpoints and let $p_1,\ldots,p_L$ be nonnegative
probability masses satisfying $\sum_{\ell=1}^{L}p_\ell=1$. Define the
piecewise-uniform distribution function
\[
F_Y^{\mathrm{(U)}}\left(
    x;[b_\ell]_{\ell=0}^{L},[p_\ell]_{\ell=1}^{L}
\right)
\coloneqq
\sum_{\ell=1}^{L}
p_\ell
\frac{
    \min\{\max\{x,b_{\ell-1}\},b_\ell\}-b_{\ell-1}
}{
    b_\ell-b_{\ell-1}
}.
\]
Then, the asymmetric kink distribution $Y\sim K_3$ is defined by
$
F_Y(x)
=
F_Y^{\mathrm{(U)}}\left(
    x;[b_\ell]_{\ell=0}^{3},[p_\ell]_{\ell=1}^{3}
\right),
$
with
\[
(b_0,b_1,b_2,b_3)
=
\left(-2,-\frac{1}{2},\frac{1}{4},2\right),
\qquad
(p_1,p_2,p_3)
=
\left(\frac{1}{5},\frac{3}{5},\frac{1}{5}\right).
\]
The corresponding density levels are
$
\left(
    \frac{2}{15},
    \frac{4}{5},
    \frac{4}{35}
\right),
$
and the interior kink points are $-1/2$ and $1/4$.

%%%%%%%%%%%%%%%%%%%%%%%%%%%%%%%%%%%%%%%%%%%%%%%%
\subsection{Estimator implementation details}\label{app:SimsEstImplDet}
%%%%%%%%%%%%%%%%%%%%%%%%%%%%%%%%%%%%%%%%%%%%%%%%

%%%%%%%%%%%%%%%%%%%%%%%%
\subsubsection{$\FDaRe$ description}\label{app:FYx_FDaRe}
%%%%%%%%%%%%%%%%%%%%%%%%
We implement the adaptive estimator of
\citet[Section~2.2]{Dattner2013}.
For $\lambda>0$ define the unprojected candidate estimator
\[
    \widetilde F_{\lambda}(x)
    =
    \frac{1}{2}
    -
    \frac{1}{n}\sum_{j=1}^{n}I_{\lambda}(X_j,x)
,
\quad \text{where} \quad 
    I_{\lambda}(X_j,x)
    =
    \frac{1}{\pi}
    \int_{0}^{\lambda}
    \frac{1}{t}
    \operatorname{Im}
    \left\{
        \frac{e^{it(X_j-x)}}{\Phi_Z(t)}
    \right\}dt.
\]
Let $\Lambda$ be a grid of candidate $\lambda$ values.
For each $\lambda\in \Lambda$, the estimator $\hat{F}_{\lambda}$ is associated with the interval
        $$
            Q_\lambda 
        \coloneqq
            \left[
                \hat{F}_{\lambda}(x)
                -
                K_Z \hat{\sigma}_\lambda \sqrt{\frac{\log(n)}{n}}
                ,
                \;
                \hat{F}_{\lambda}(x)
                +
                K_Z \hat{\sigma}_\lambda \sqrt{\frac{\log(n)}{n}}
            \right]
        $$
    where 
    $
    K_Z 
    = 
    0.0275 
    +
    0.3074
    \sigma_Z
    $, 
    and 
        $
        \textstyle
         \hat{\sigma}_\lambda
         \coloneqq
         \sqrt{
         \sampavg
         I_\lambda(X_j,x)^2
         }
         .
        $
Then $\lambda$ is selected separately at each evaluation
point \(x\) by the interval-intersection rule
\[
    \widehat\lambda
    =
    \min\left\{
        \lambda\in\Lambda:
        \bigcap_{\substack{\mu\in\Lambda\\ \mu\geq\lambda}}
        Q_{\mu}\neq\varnothing
    \right\}.
\]
This is used to obtain the final estimator, which is projected onto the unit interval
\[
    \widehat F_{DR}(x)
    =
    \min\left\{1,\max\right\{0,\widetilde F_{\widehat\lambda}(x)\left\}\right\}.
\]

We use the candidate grid
    $
    \Lambda
    = 
    \{0.01,0.06,\ldots,4.91,4.96\}
    .$
Dattner and Reiser use a grid extending to \(9.96\). 
We found that imposing the smaller
upper limit of $4.96$ improved performance and allowed us to closely replicate Table~3 from \citet{Dattner2013}.
Extending the grid to
\(9.96\) produced extreme variance under normal measurement errors and caused the estimator to perform poorly.
In our simulations no selected $\lambda$ values
occurred at the upper endpoint of the restricted grid.

%%%%%%%%%%%%%%%%%%%%%%%%
\subsubsection{$\FHaLa$ description}\label{app:FYx_FHaLa}
%%%%%%%%%%%%%%%%%%%%%%%%
$\FHaLa$ is the normal-reference implementation of the
distribution estimator in \citet{Hall2008}. We make several slight modifications to accommodate asymmetric error distributions and errors with nonzero means. These reduce exactly to the representations in \citet{Hall2008} under symmetric zero-mean errors.
% denoted by
% $\FHaLa$. 

To accommodate errors with nonzero means, let
$\mu_Z=\mathbb{E}[Z]$ and define
\[
    X_j^\circ=X_j-\mu_Z,
    \qquad
    \Phi_{Z^\circ}(t)=e^{-it\mu_Z}\Phi_Z(t).
\]
Then \(X_j^\circ=Y_j+Z_j^\circ\), where \(Z_j^\circ=Z_j-\mu_Z\), so the
target remains \(F_Y\).

For a bandwidth \(h>0\), define the estimator
\[
\widetilde F_{\mathrm{HL}}(x;h)
=
\frac{1}{2}
-
\frac{1}{\pi n}
\sum_{j=1}^{n}
\int_{0}^{1/h}
\operatorname{Im}
\left\{
e^{it(X_j^\circ-x)}
\frac{K^{\mathrm{Ft}}(ht)}
     {\Phi_{Z^\circ}(t)}
\right\}
\frac{1}{t}dt,
\]
where
$
    K^{\mathrm{Ft}}(t)
    =(1-t^2)^2\mathbf{1}\{|t|\leq1\}.
$
The complex-valued form above permits
implementation with asymmetric errors and reduces to the sine representation in
\citet[Appendix~A.1]{Hall2008} when $F_Z$ is symmetric. 
As with the other distribution estimators 
% in the simulations,
we project the
final estimate onto the unit interval:
\[
    \FHaLa(x;h)
    =
    \min\left\{1,\max\left\{0,
        \widetilde F_{\mathrm{HL}}(x;h)
    \right\}\right\}.
\]

The bandwidth is selected separately for each simulated sample using the
normal-reference MISE criterion from \citet[Section~4.2]{Hall2008}. 
Let \(\mathcal H_{\mathrm{HL}}\) be the grid of candidate $h$ values, which contains 161 logarithmically spaced values between \(0.01\) and \(2\).
The selected bandwidth is
\[
    \widehat h_{\mathrm{HL}}
=
    \arg\min_{h\in\mathcal H_{\mathrm{HL}}}
    \left\{
        \frac{I(h)}{n}
        +
         \frac{h^4}
         {\sqrt{\pi}\,\widehat\sigma_Y^3}
    \right\},
\quad \text{where} \quad
        2\pi I(h)
=
    \int_{\mathbb R}
    \frac{1}{t^2}
    \left|
        1-
        \frac{K^{\mathrm{Ft}}(ht)}
             {\Phi_{Z^\circ}(t)}
    \right|^2dt
,
\]
and
$\widehat\sigma_Y^2 = \widehat \sigma_X^2 - \sigma_Z^2$, 
where $\widehat \sigma_X^2$ is the sample variance and $\sigma_Z^2$ is the known error variance. 

We evaluate the modulus-squared expression for $I(h)$ for every error
distribution. For symmetric errors, $\Phi_{Z^\circ}(t)$ is real, so the
modulus squared reduces to the ordinary square used by
\citet{Hall2008}. For the asymmetric Gamma errors,
$\Phi_{Z^\circ}(t)$ is complex, and we use this squared complex modulus form to
obtain a real, nonnegative bandwidth criterion.

%%%%%%%%%%%%%%%%%%%%%%%%
\subsubsection{$\FYhatCV(x)$ description}%
% \subsubsection{Cross-Validation Procedure}
\label{app:FYx_CV}
%%%%%%%%%%%%%%%%%%%%%%%%
We select $(h,\lambda)$ using localized, variance-penalized reconvolution
cross-validation.
Note that \eqref{decon} implies
    \begin{equation*}
    \begin{aligned}
        \widehat{F}_{Y}(x)
&=
    \frac{1}{2\pi }\int_{\mathds{R}}
    e^{-ixs}
    \left(
        \frac{1}{n}\sum_{j=1}^{n}
        e^{iX_{j}s}
    \right)
    % \alpha _{h,\lambda}\left( s\right) 
    \left(
        \frac{e^{is/\lambda }-1}{is}
    \right)
    \frac{H(hs)}{\Phi _{Z}(s)}
    ds
    .
    \end{aligned}
    \end{equation*}
Thus, 
    $
    (\mathcal{F}\FYhat)(s)
    \cdot
    \Phi_{Z}(s)
    =
    \left(
        \frac{1}{n}\sum_{j=1}^{n}
        e^{iX_{j}s}
    \right)
    % \alpha _{h,\lambda}\left( s\right) 
    \left(
        \frac{e^{is/\lambda }-1}{is}
    \right)
    % \frac{H(hs)}{\Phi _{Z}(s)}
    H(hs)
    $.
This 
and
$\Phi_{X}=\Phi _{Y}\Phi_{Z}$, 
motivates the following reconvolution approximation of $F_X(t)$
    $$
    \widehat F_{X,h,\lambda}(t)
    =
        \frac{1}{n}\sum_{j=1}^{n}
        \frac{1}{2\pi }\int_{\mathds{R}}e^{i\left(
        X_{j}-t\right)s}
        % q_{\lambda}(s)
        \left(
            \frac{e^{is/\lambda }-1}{is}
        \right)
        H(hs)
        % \alpha _{h,\lambda}\left( s\right)
        ds
    .
    $$

Using this, we now construct the $K$-fold  cross-validation criterion.
Let $\{I_k\}_{k=1}^K$ be index sets which form a partition of $\{1,\ldots,n\}$, and let $I_{-k}=\{1,\ldots,n\}\setminus I_k$.
For each candidate pair $(h,\lambda)$, and each fold $k\in\{1,\ldots,K\}$, 
define
the training-sample reconvolution estimator
    $$
        \widehat F_{X,h,\lambda}^{(-k)}(t)
    =
        \frac{1}{|I_{-k}|}\sum_{j\in I_{-k}}
        % \frac{1}{n}\sum_{j=1}^{n}
        \frac{1}{2\pi }\int_{\mathds{R}}e^{i\left(
        X_{j}-t\right)s}
        % \alpha _{h,\lambda}\left( s\right)
        % q_{\lambda}(s)
        \left(
            \frac{e^{is/\lambda }-1}{is}
        \right)
        H(hs)
        ds
    .
    $$
The validation target is the empirical distribution function of the validation subsample
    $$
    \widetilde F_X^{(k)}(t)=\frac{1}{|I_k|}\sum_{j\in I_k} \chi_{\{X_j\leq t\}}.
    $$

$\FYhatCV(x)$ estimates $F_Y(x)=P(Y\leq x)$, but validation is performed on the observed $X$-sample. Therefore, instead of comparing 
$\widetilde F_X^{(k)}$
and
$ \widehat F_{X,h,\lambda}^{(-k)}$ 
around $x$,
the loss is localized
around the corresponding location in the observed $X$ space
$$
c_x=x+E[Z].
$$
Then we construct a finite grid
$\{u_{\ell k}\}_{\ell=1}^{m_k}$ of observed-space validation points around $c_x$.
In the simulations, $\{u_{\ell k}\}_{\ell=1}^{m_k}$ contains the target observed-space point $c_x$, and empirical quantiles of the validation-fold
$X$-sample at up to 51 equally spaced probability levels between $0.05$
and $0.95$.
The loss is localized around $c_x$ using weights
    $$
    w_x(u)=\exp\left\{-\frac{1}{2}\left(\frac{u-c_x}{b_X}\right)^2\right\},
    $$ 
where in the simulations, 
$
    b_X=
    \max\left\{
0.15\widehat\sigma_X,\,
\min\left[
0.50\widehat\sigma_X,\,
0.30\widehat\sigma_X(n/500)^{-1/5}
\right]
\right\},
    $
and 
    $\hat{\sigma}_X$ denotes the sample standard deviation of $X$.

The variance penalized reconvolution CV criterion is defined as
    $$
\begin{aligned}
CV(h,\lambda;x)
=
\frac1K\sum_{k=1}^K
\frac{
% \sum_{u\in\mathcal U_k} 
% \sum_{j\in I_k}
\sum_{\ell=1}^{m_k}
w_x(u_{\ell k})
\left[
\widehat F_{X,h,\lambda}^{(-k)}(u_{\ell k})-\widetilde F_X^{(k)}(u_{\ell k})
\right]^2
}{
% \sum_{j\in I_k} 
\sum_{\ell=1}^{m_k}
w_x(u_{\ell k})
}
+
\cvvarpen \frac{1}{n}\frac{v_{h,\lambda}^2}{4 \pi^2}
,
\end{aligned}
$$
where  $v_{h,\lambda}$ is defined as in Theorem \ref{thm8} and $\cvvarpen\geq 0$ is the variance penalty scaling parameter.
In the simulations we use $\cvvarpen=0.0025$.
The cross-validation tuning parameters are then
$$
(h_{\mathrm{cv}}(x),\lambda_{\mathrm{cv}}(x))
=
\arg\min_{(h,\lambda)\in\mathcal H_{\mathrm{cv}}\times\Lambda_{\mathrm{cv}}}
CV(h,\lambda;x),
$$
where $\mathcal H_{\mathrm{cv}}$ and $\Lambda_{\mathrm{cv}}$ are finite candidate grids. The corresponding cross-validation estimator projected onto $[0,1]$ is
$$
\FYhatCV(x)
% \widehat F_{Y,\mathrm{cv}}(x)
=
% \widehat F_{Y,h_{\mathrm{cv}}(x),\lambda_{\mathrm{cv}}(x)}(x).
\min\left\{1,\max\left\{0,
        \widehat F_{Y,h_{\mathrm{cv}}(x),\lambda_{\mathrm{cv}}(x)}(x)
    \right\}\right\}.
$$
In the simulations, $K=3$, and the candidate grids $\mathcal H_{cv}$ and $\Lambda_{cv}$ 
% are centered around the asymptotic-rate parameters $(h_{AR},\lambda_{AR})$.
% The candidate grids 
are constructed as fixed multiples of the asymptotic-rate pair $(h_{\mathrm{AR}},\lambda_{\mathrm{AR}})$ used for $\FYhatAR$.

%%%%%%%%%%%%%%%%%%%%%%%%
\subsubsection{$\widehat F_Y^{CV}(x,y)$ description}
\label{app:FYxy_CV}
%%%%%%%%%%%%%%%%%%%%%%%%

To choose $h$ for the direct interval-probability estimator, we use a
$K$-fold reconvolution cross-validation procedure analogous to the
procedure used for $\widehat F_Y^{CV}(x)$.
Let $\{I_k\}_{k=1}^K$ be
index sets which form a partition of $\{1,\ldots,n\}$, and let
$I_{-k}=\{1,\ldots,n\}\setminus I_k$.
Using similar reasoning as Appendix \ref{app:FYx_CV}, for a candidate bandwidth $h$ and $b>a$
reconvolving the direct deconvolution estimator with
the known distribution of $Z$ gives the training-fold estimator
\[
    \widehat F_{X,h}^{(-k)}(a,b)
    =
    \frac{1}{|I_{-k}|}
    \sum_{j\in I_{-k}}
    \frac{1}{2\pi}
    \int_{\mathbb R}
        e^{i(X_j-a)s}
        \left(
            \frac{1-e^{-i(b-a)s}}{is}
        \right)
        H(hs)\,ds .
\]
The validation target is the empirical observed interval probability
\[
    \widetilde F_X^{(k)}(a,b)
    =
    \frac{1}{|I_k|}
    \sum_{j\in I_k}
    \chi_{\{a<X_j\leq b\}}.
\]

For each validation fold $k$, we construct a finite grid 
$\{c_{\ell k}\}_{\ell=1}^{m_k}$ 
of observed-space interval centers. 
Specifically, $\{c_{\ell k}\}_{\ell=1}^{m_k}$  contains
empirical quantiles of the validation-fold $X$-sample at up to 21 equally
spaced probability levels between $0.05$ and $0.95$, together with the
target observed-space center
\[
    c_{xy} = \frac{x+y}{2}+E[Z].
\]
For each $c_{\ell k}$ we form the observation space interval
\[
    a_{\ell k}=c_{\ell k}-\frac{y-x}{2},
    \qquad
    b_{\ell k}=c_{\ell k}+\frac{y-x}{2}.
\]
The loss is localized around the target observed-space center $c_{xy}$
using weights
\[
    w_{xy}(c)
    =
    \exp\left\{
        -\frac{1}{2}
        \left(\frac{c-c_{xy}}{b_X}\right)^2
    \right\},
\]
where, in the simulations,
\[
    b_X
    =
    \max\left\{
        0.25\widehat\sigma_X,\,
        \min\left(
            0.80\widehat\sigma_X,\,
            0.50\widehat\sigma_X(n/500)^{-1/5}
        \right)
    \right\},
\]
and $\widehat\sigma_X$ denotes the sample standard deviation of the
observed $X$-sample.
Specifically, the reconvolution CV criterion is 
\[
    CV_{xy}(h)
    =
    \frac{1}{K}
    \sum_{k=1}^K
    \frac{
        \sum_{\ell=1}^{m_k}
        w_{xy}(c_{\ell k})
        \left[
            \widehat F_{X,h}^{(-k)}(a_{\ell k},b_{\ell k})
            -
            \widetilde F_X^{(k)}(a_{\ell k},b_{\ell k})
        \right]^2
    }{
        \sum_{\ell=1}^{m_k} w_{xy}(c_{\ell k})
    } 
    +
    \cvvarpen
        \frac{v_h(x,y)^2}{4\pi^2 n}
    ,
\]
where $v_h(x,y)^2/(4\pi^2 n)$ is the variance bound from Theorem \ref{thm8a},
% where $v_h(x,y)$ is the variance bound from Theorem \ref{thm8a}, 
and in the simulations we set $\cvvarpen=0.01$.
By comparing 
$\widehat F_{X,h}^{(-k)}$ and
$ \widetilde F_X^{(k)}$
at the points $(a_{\ell k},b_{\ell k})$,
the CV criterion evaluates the reconvolved estimator on
intervals in the observed $X$ space with the same width as the target interval in the latent $Y$ space, but centered
at a collection of locations in the validation sample, with more weight given to intervals centered near $c_{xy}$.
% included by construction.

Then for a candidate grid $\mathcal H_{cv}$, the cross-validation estimator is
\[
    \widehat F_Y^{CV}(x,y)
    =
    % \widehat F_{Y,h_{cv}(x,y)}(x,y)
    \min\left\{1,\max\left\{0,
        \widehat F_{Y,h_{cv}(x,y)}(x,y)
    \right\}\right\}
    ,
    \quad
    \text{where}
    \quad
    h_{cv}(x,y)
    =
    \arg\min_{h\in\mathcal H_{cv}}
    CV_{xy}(h)
    .
\]
% $$
% \FYhatCV(x)
% % \widehat F_{Y,\mathrm{cv}}(x)
% =
% % \widehat F_{Y,h_{\mathrm{cv}}(x),\lambda_{\mathrm{cv}}(x)}(x).
% \min\left\{1,\max\left\{0,
%         \widehat F_{Y,h_{\mathrm{cv}}(x),\lambda_{\mathrm{cv}}(x)}(x)
%     \right\}\right\}.
% $$
In the simulations, $K=3$. The candidate grid $\mathcal H_{cv}$ contains values 
around the asymptotic-rate bandwidth $h_{AR}$ evenly spaced on a logarithmic scale.

% \newpage
%%%%%%%%%%%%%%%%%%%%%%%%%%%%%%%%%%%%%%%%%%%%%%%%
\subsection{Additional Simulation Results}\label{app:Full_sim_results}
%%%%%%%%%%%%%%%%%%%%%%%%%%%%%%%%%%%%%%%%%%%%%%%%

%%%%%%%%%%%%%%%%%%%%%%%%
\subsubsection{$F_Y(x)$ simulations}
% \label{app:FYxSims}
%%%%%%%%%%%%%%%%%%%%%%%%
{
\begin{table}[H]
    \caption{
    Sensitivity to the CV variance-penalty coefficient $\cvvarpen$ for estimating $F_Y(x)$.
        % Monte Carlo MSE$\times 100$ and variance$\times 100$ (in parentheses) for estimators of $F_Y(x,y)$.  
        }\label{tab:FYx_CV_sensitivity}
    \adjustbox{max width=\textwidth, center = \textwidth}{ 
    \footnotesize
    % Requires \usepackage{booktabs}
% Displayed MSE x 100; displayed variance x 100.
% Superscript * marks the CV specification used in the main tables.
% Boldface marks the smallest unrounded MSE in each row.
\begin{tabular}{@{}lcccccc@{}}
\toprule
& \multicolumn{3}{c}{$\FYhatCV(x)$} & $\FYhatAR(x)$ & $\FDaRe(x)$ & $\FHaLa(x)$ \\
\cmidrule(lr){2-4}
& $\cvvarpen=0.005$ & $\cvvarpen=0.0025^{*}$ & $\cvvarpen=0.001$ & & & \\
\midrule
\multicolumn{1}{l}{\rule[-.10ex]{0pt}{3.25ex}\!Smooth $F_Y$} &  &  &  &  &  &  \\
\quad $n=200$ & 0.090 (0.077) & 0.091 (0.079) & 0.093 (0.081) & 0.076 (0.069) & \textbf{0.075 (0.070)} & 0.087 (0.087) \\
\quad $n=1000$ & 0.018 (0.017) & 0.018 (0.017) & 0.019 (0.018) & 0.017 (0.015) & \textbf{0.015 (0.014)} & 0.018 (0.018) \\
\quad $n=5000$ & 0.004 (0.004) & 0.004 (0.004) & 0.004 (0.004) & 0.004 (0.003) & \textbf{0.003 (0.003)} & 0.004 (0.004) \\
\midrule
\multicolumn{1}{l}{\rule[-.10ex]{0pt}{3.25ex}\!Non-smooth $F_Y$} &  &  &  &  &  &  \\
\quad $n=200$ & 0.352 (0.145) & 0.343 (0.152) & 0.337 (0.161) & 0.500 (0.090) & 0.383 (0.105) & \textbf{0.247 (0.111)} \\
\quad $n=1000$ & 0.119 (0.028) & 0.113 (0.032) & \textbf{0.111 (0.041)} & 0.280 (0.018) & 0.231 (0.020) & 0.141 (0.022) \\
\quad $n=5000$ & 0.064 (0.007) & \textbf{0.061 (0.007)} & 0.069 (0.025) & 0.178 (0.004) & 0.184 (0.004) & 0.101 (0.005) \\
\bottomrule
\end{tabular}%

    }

    \tabnote{Entries report Monte Carlo MSE$\times100$, with variance$\times100$ in
parentheses. The superscript $*$ identifies the cross-validation
specification reported in Tables \ref{tab:FYx_main}, \ref{tab:FYx_app_smth}, and \ref{tab:FYx_app_nonsmth}. Boldface identifies the
estimator with the smallest unrounded MSE in each row.}
\end{table}
}

{
\setlength{\tabcolsep}{9pt}
\begin{table}[H]
    \caption{
        Monte Carlo MSE$\times 100$ and variance$\times 100$ (in parentheses) for estimators of $F_Y(x)$ at continuously differentiable evaluation points.
        % Empirical root MSE and empirical variance$\times 10$ (in parenthesis) for estimating $F_Y(x)$ with smooth distributions. 
        }\label{tab:FYx_app_smth}
    \adjustbox{max width=\textwidth, center = \textwidth}{ 
    \footnotesize
    % Requires \usepackage{booktabs}
% Displayed MSE x 100; displayed variance x 100.
% The difference between MSE and variance is squared bias.
% Boldface marks the smallest unrounded MSE in each row.
\begin{tabular}{lcccc}
\toprule
&
$\FYhatCV(x)$ &
$\FYhatAR(x)$ &
$\FDaRe(x)$ &
$\FHaLa(x)$ \\
\midrule
\multicolumn{1}{l}{\rule[-.10ex]{0pt}{3.25ex}\!${Y\sim N(0,1)}$, $Z\sim N\!\left(0,\frac{1}{5}\right)\!$} &  &  &  &  \\
\quad $n=200$ & 0.094 (0.082) & 0.079 (0.076) & \textbf{0.075 (0.072)} & 0.091 (0.091) \\
\quad $n=1000$ & 0.018 (0.017) & 0.017 (0.015) & \textbf{0.014 (0.013)} & 0.018 (0.018) \\
\quad $n=5000$ & 0.004 (0.004) & 0.004 (0.003) & \textbf{0.003 (0.003)} & 0.004 (0.004) \\
\multicolumn{1}{l}{\rule[-.10ex]{0pt}{3.25ex}\!${Y\sim N(0,1)}$, $Z\sim \Gamma\!\left(2,\frac{1}{5\sqrt{2}}\right)\!$} &  &  &  &  \\
\quad $n=200$ & 0.086 (0.074) & 0.070 (0.065) & \textbf{0.067 (0.063)} & 0.079 (0.079) \\
\quad $n=1000$ & 0.017 (0.016) & 0.016 (0.015) & \textbf{0.014 (0.013)} & 0.017 (0.016) \\
\quad $n=5000$ & 0.004 (0.004) & 0.004 (0.003) & \textbf{0.003 (0.003)} & 0.004 (0.004) \\
\multicolumn{1}{l}{\rule[-.10ex]{0pt}{3.25ex}\!${Y\sim N(0,1)}$, $Z\sim \mathrm{S}\Gamma\!\left(2,\frac{1}{5\sqrt{2}}\right)\!$} &  &  &  &  \\
\quad $n=200$ & 0.096 (0.083) & 0.079 (0.075) & \textbf{0.076 (0.072)} & 0.096 (0.096) \\
\quad $n=1000$ & 0.018 (0.016) & 0.016 (0.015) & \textbf{0.014 (0.013)} & 0.018 (0.018) \\
\quad $n=5000$ & 0.004 (0.003) & 0.003 (0.003) & \textbf{0.003 (0.002)} & 0.004 (0.004) \\
\midrule
\multicolumn{1}{l}{\rule[-.10ex]{0pt}{3.25ex}\!${Y\sim \Gamma\big(3,1/\sqrt{3}\big)}$, $Z\sim N\!\left(0,\frac{1}{5}\right)\!$} &  &  &  &  \\
\quad $n=200$ & 0.096 (0.083) & \textbf{0.081 (0.073)} & 0.082 (0.076) & 0.091 (0.090) \\
\quad $n=1000$ & 0.019 (0.017) & 0.018 (0.014) & \textbf{0.016 (0.014)} & 0.018 (0.017) \\
\quad $n=5000$ & 0.005 (0.005) & 0.006 (0.003) & \textbf{0.004 (0.003)} & 0.005 (0.004) \\
\multicolumn{1}{l}{\rule[-.10ex]{0pt}{3.25ex}\!${Y\sim \Gamma\big(3,1/\sqrt{3}\big)}$, $Z\sim \Gamma\!\left(2,\frac{1}{5\sqrt{2}}\right)\!$} &  &  &  &  \\
\quad $n=200$ & 0.087 (0.074) & 0.071 (0.061) & \textbf{0.071 (0.065)} & 0.078 (0.077) \\
\quad $n=1000$ & 0.020 (0.018) & 0.018 (0.015) & \textbf{0.017 (0.015)} & 0.018 (0.018) \\
\quad $n=5000$ & 0.004 (0.004) & 0.004 (0.003) & \textbf{0.004 (0.003)} & 0.004 (0.004) \\
\multicolumn{1}{l}{\rule[-.10ex]{0pt}{3.25ex}\!${Y\sim \Gamma\big(3,1/\sqrt{3}\big)}$, $Z\sim \mathrm{S}\Gamma\!\left(2,\frac{1}{5\sqrt{2}}\right)\!$} &  &  &  &  \\
\quad $n=200$ & 0.090 (0.079) & \textbf{0.075 (0.066)} & 0.076 (0.070) & 0.088 (0.088) \\
\quad $n=1000$ & 0.019 (0.018) & 0.018 (0.015) & \textbf{0.016 (0.015)} & 0.019 (0.019) \\
\quad $n=5000$ & 0.004 (0.004) & 0.004 (0.003) & \textbf{0.004 (0.003)} & 0.004 (0.004) \\
\bottomrule
\end{tabular}%

    }
    
    \tabnote{Boldface identifies the estimator with the smallest empirical MSE in each row.
    Results grouped by $Y$ and $Z$ distribution. For each $Y$ design, empirical MSE and variance are averaged over the evaluation points then multiplied by 100.
    % and the reported RMSE is the square root of the resulting average MSE.
    }
\end{table}
}

{
\setlength{\tabcolsep}{12pt}
\begin{table}[H]
    \caption{
        Monte Carlo MSE$\times 100$ and variance$\times 100$ (in parentheses) for estimators of $F_Y(x)$ at non-differentiable evaluation points.
        }\label{tab:FYx_app_nonsmth}
    \adjustbox{max width=\textwidth, center = \textwidth}{ 
    \footnotesize
    % Requires \usepackage{booktabs}
% Displayed MSE x 100; displayed variance x 100.
% The difference between MSE and variance is squared bias.
% Boldface marks the smallest unrounded MSE in each row.
\begin{tabular}{lcccc}
\toprule
&
$\FYhatCV(x)$ &
$\FYhatAR(x)$ &
$\FDaRe(x)$ &
$\FHaLa(x)$ \\
\midrule
\multicolumn{1}{l}{\rule[-.10ex]{0pt}{3.25ex}\!${Y\sim K_1}$, $Z\sim N\!\left(0,\frac{1}{5}\right)\!$} &  &  &  &  \\
\quad $n=200$ & 0.220 (0.127) & 0.230 (0.084) & 0.233 (0.096) & \textbf{0.163 (0.103)} \\
\quad $n=1000$ & 0.083 (0.024) & 0.141 (0.017) & 0.127 (0.018) & \textbf{0.080 (0.020)} \\
\quad $n=5000$ & \textbf{0.049 (0.008)} & 0.100 (0.004) & 0.086 (0.004) & 0.054 (0.005) \\
\multicolumn{1}{l}{\rule[-.10ex]{0pt}{3.25ex}\!${Y\sim K_1}$, $Z\sim \Gamma\!\left(2,\frac{1}{5\sqrt{2}}\right)\!$} &  &  &  &  \\
\quad $n=200$ & 0.229 (0.149) & 0.276 (0.087) & 0.257 (0.106) & \textbf{0.166 (0.109)} \\
\quad $n=1000$ & \textbf{0.054 (0.026)} & 0.126 (0.016) & 0.123 (0.017) & 0.067 (0.018) \\
\quad $n=5000$ & \textbf{0.019 (0.006)} & 0.066 (0.004) & 0.090 (0.004) & 0.039 (0.004) \\
\multicolumn{1}{l}{\rule[-.10ex]{0pt}{3.25ex}\!${Y\sim K_1}$, $Z\sim \mathrm{S}\Gamma\!\left(2,\frac{1}{5\sqrt{2}}\right)\!$} &  &  &  &  \\
\quad $n=200$ & 0.253 (0.156) & 0.311 (0.093) & 0.288 (0.110) & \textbf{0.173 (0.123)} \\
\quad $n=1000$ & 0.054 (0.027) & 0.125 (0.017) & 0.124 (0.018) & \textbf{0.053 (0.022)} \\
\quad $n=5000$ & \textbf{0.019 (0.006)} & 0.067 (0.003) & 0.090 (0.003) & 0.030 (0.004) \\
\midrule
\multicolumn{1}{l}{\rule[-.10ex]{0pt}{3.25ex}\!${Y\sim K_2}$, $Z\sim N\!\left(0,\frac{1}{5}\right)\!$} &  &  &  &  \\
\quad $n=200$ & 0.490 (0.163) & 0.545 (0.109) & 0.501 (0.120) & \textbf{0.342 (0.123)} \\
\quad $n=1000$ & \textbf{0.216 (0.032)} & 0.381 (0.024) & 0.334 (0.025) & 0.240 (0.027) \\
\quad $n=5000$ & \textbf{0.150 (0.008)} & 0.298 (0.005) & 0.269 (0.005) & 0.191 (0.005) \\
\multicolumn{1}{l}{\rule[-.10ex]{0pt}{3.25ex}\!${Y\sim K_2}$, $Z\sim \Gamma\!\left(2,\frac{1}{5\sqrt{2}}\right)\!$} &  &  &  &  \\
\quad $n=200$ & 0.427 (0.162) & 0.611 (0.091) & 0.487 (0.104) & \textbf{0.312 (0.104)} \\
\quad $n=1000$ & \textbf{0.128 (0.035)} & 0.344 (0.022) & 0.326 (0.023) & 0.211 (0.024) \\
\quad $n=5000$ & \textbf{0.051 (0.008)} & 0.199 (0.005) & 0.270 (0.005) & 0.161 (0.005) \\
\multicolumn{1}{l}{\rule[-.10ex]{0pt}{3.25ex}\!${Y\sim K_2}$, $Z\sim \mathrm{S}\Gamma\!\left(2,\frac{1}{5\sqrt{2}}\right)\!$} &  &  &  &  \\
\quad $n=200$ & 0.448 (0.189) & 0.634 (0.114) & 0.513 (0.129) & \textbf{0.276 (0.140)} \\
\quad $n=1000$ & \textbf{0.119 (0.034)} & 0.343 (0.022) & 0.327 (0.023) & 0.155 (0.026) \\
\quad $n=5000$ & \textbf{0.049 (0.007)} & 0.198 (0.004) & 0.270 (0.004) & 0.117 (0.005) \\
\midrule
\multicolumn{1}{l}{\rule[-.10ex]{0pt}{3.25ex}\!${Y\sim K_3}$, $Z\sim N\!\left(0,\frac{1}{5}\right)\!$} &  &  &  &  \\
\quad $n=200$ & 0.373 (0.142) & 0.554 (0.082) & 0.385 (0.097) & \textbf{0.294 (0.103)} \\
\quad $n=1000$ & \textbf{0.187 (0.055)} & 0.378 (0.016) & 0.240 (0.018) & 0.193 (0.019) \\
\quad $n=5000$ & \textbf{0.139 (0.008)} & 0.298 (0.004) & 0.194 (0.004) & 0.147 (0.005) \\
\multicolumn{1}{l}{\rule[-.10ex]{0pt}{3.25ex}\!${Y\sim K_3}$, $Z\sim \Gamma\!\left(2,\frac{1}{5\sqrt{2}}\right)\!$} &  &  &  &  \\
\quad $n=200$ & 0.318 (0.133) & 0.660 (0.069) & 0.383 (0.085) & \textbf{0.263 (0.087)} \\
\quad $n=1000$ & \textbf{0.091 (0.029)} & 0.343 (0.015) & 0.240 (0.017) & 0.153 (0.018) \\
\quad $n=5000$ & \textbf{0.038 (0.008)} & 0.188 (0.003) & 0.195 (0.004) & 0.099 (0.004) \\
\multicolumn{1}{l}{\rule[-.10ex]{0pt}{3.25ex}\!${Y\sim K_3}$, $Z\sim \mathrm{S}\Gamma\!\left(2,\frac{1}{5\sqrt{2}}\right)\!$} &  &  &  &  \\
\quad $n=200$ & 0.327 (0.149) & 0.675 (0.079) & 0.397 (0.096) & \textbf{0.232 (0.108)} \\
\quad $n=1000$ & \textbf{0.086 (0.029)} & 0.341 (0.015) & 0.239 (0.017) & 0.114 (0.020) \\
\quad $n=5000$ & \textbf{0.033 (0.008)} & 0.186 (0.003) & 0.194 (0.003) & 0.072 (0.004) \\
\bottomrule
\end{tabular}%

    }
    
    \tabnote{Boldface identifies the estimator with the smallest empirical MSE in each row.
    Results grouped by $Y$ and $Z$ distribution. For each $Y$ design, empirical MSE and variance are averaged over the evaluation points then multiplied by 100.
    % and the reported RMSE is the square root of the resulting average MSE.
    }
\end{table}
}

\newpage
%%%%%%%%%%%%%%%%%%%%%%%%
\subsubsection{$F_Y(x,y)$ simulations}
% \label{app:FYxSims}
%%%%%%%%%%%%%%%%%%%%%%%%

{
\begin{table}[H]
    \caption{
    Sensitivity to the CV variance-penalty coefficient $\cvvarpen$ for estimating $F_Y(x,y)$.
        }\label{tab:FYxy_CV_sensitivity}
    \adjustbox{max width=\textwidth, center = \textwidth}{ 
    \footnotesize
    % Requires \usepackage{booktabs}
% Displayed MSE x 100; displayed variance x 100.
% Superscript * marks the CV specification used in the main tables.
% Boldface marks the smallest unrounded MSE in each row.
\begin{tabular}{@{}lcccccc@{}}
\toprule
& \multicolumn{3}{c}{$\FYhatCV(x,y)$} & $\FYhatAR(x,y)$ & $\FDaRe(x,y)$ & $\FHaLa(x,y)$ \\
\cmidrule(lr){2-4}
& $\cvvarpen=0.0125$ & $\cvvarpen=0.010^{*}$ & $\cvvarpen=0.005$ & & & \\
\midrule
\multicolumn{1}{l}{\rule[-.10ex]{0pt}{3.25ex}\!Smooth $F_Y$} &  &  &  &  &  &  \\
\quad $n=200$ & 0.054 (0.047) & 0.054 (0.048) & 0.056 (0.051) & 0.047 (0.041) & \textbf{0.042 (0.035)} & 0.062 (0.061) \\
\quad $n=1000$ & 0.013 (0.011) & 0.013 (0.011) & 0.013 (0.012) & 0.012 (0.010) & \textbf{0.010 (0.007)} & 0.014 (0.013) \\
\quad $n=5000$ & 0.004 (0.003) & 0.004 (0.003) & 0.004 (0.003) & 0.004 (0.003) & \textbf{0.003 (0.002)} & 0.004 (0.003) \\
\midrule
\multicolumn{1}{l}{\rule[-.10ex]{0pt}{3.25ex}\!Non-smooth $F_Y$} &  &  &  &  &  &  \\
\quad $n=200$ & 0.303 (0.117) & 0.297 (0.120) & 0.286 (0.129) & 0.411 (0.074) & 0.402 (0.086) & \textbf{0.255 (0.099)} \\
\quad $n=1000$ & 0.132 (0.029) & 0.129 (0.030) & \textbf{0.121 (0.035)} & 0.230 (0.017) & 0.233 (0.018) & 0.152 (0.021) \\
\quad $n=5000$ & 0.080 (0.007) & 0.078 (0.008) & \textbf{0.076 (0.019)} & 0.146 (0.004) & 0.176 (0.004) & 0.108 (0.005) \\
\bottomrule
\end{tabular}%

    }

    \tabnote{Entries report Monte Carlo MSE$\times100$, with variance$\times100$ in
parentheses. The superscript $*$ identifies the cross-validation
specification reported in Tables \ref{tab:FYxy_main}, \ref{tab:FYxy_app_smth}, and \ref{tab:FYxy_app_nonsmth}. Boldface identifies the
estimator with the smallest unrounded MSE in each row.}
\end{table}
}

{
\setlength{\tabcolsep}{9pt}
\begin{table}[H]
    \caption{Monte Carlo MSE$\times 100$ and variance$\times 100$ (in parentheses) for estimators of $F_Y(x,y)$ for the smooth $F_Y$ designs.}\label{tab:FYxy_app_smth}
    \adjustbox{max width=\textwidth, center = \textwidth}{ 
    \footnotesize
    % Requires \usepackage{booktabs}
% Displayed MSE x 100; displayed variance x 100.
% The difference between MSE and variance is squared bias.
% Boldface marks the smallest unrounded MSE in each row.
\begin{tabular}{lcccc}
\toprule
&
$\FYhatCV(x,y)$ &
$\FYhatAR(x,y)$ &
$\FDaRe(x,y)$ &
$\FHaLa(x,y)$ \\
\midrule
\multicolumn{1}{l}{\rule[-.10ex]{0pt}{3.25ex}\!${Y\sim N(0,1)}$, $Z\sim N\!\left(0,\frac{1}{5}\right)\!$} &  &  &  &  \\
\quad $n=200$ & 0.056 (0.049) & 0.048 (0.044) & \textbf{0.047 (0.037)} & 0.064 (0.063) \\
\quad $n=1000$ & 0.013 (0.011) & \textbf{0.012 (0.010)} & 0.012 (0.007) & 0.014 (0.013) \\
\quad $n=5000$ & 0.004 (0.003) & 0.004 (0.002) & \textbf{0.003 (0.002)} & 0.004 (0.003) \\
\multicolumn{1}{l}{\rule[-.10ex]{0pt}{3.25ex}\!${Y\sim N(0,1)}$, $Z\sim \Gamma\!\left(2,\frac{1}{5\sqrt{2}}\right)\!$} &  &  &  &  \\
\quad $n=200$ & 0.055 (0.048) & \textbf{0.045 (0.041)} & 0.049 (0.036) & 0.055 (0.055) \\
\quad $n=1000$ & 0.012 (0.010) & \textbf{0.011 (0.010)} & 0.012 (0.007) & 0.012 (0.011) \\
\quad $n=5000$ & 0.003 (0.003) & 0.003 (0.003) & 0.003 (0.002) & \textbf{0.003 (0.003)} \\
\multicolumn{1}{l}{\rule[-.10ex]{0pt}{3.25ex}\!${Y\sim N(0,1)}$, $Z\sim \mathrm{S}\Gamma\!\left(2,\frac{1}{5\sqrt{2}}\right)\!$} &  &  &  &  \\
\quad $n=200$ & 0.056 (0.048) & \textbf{0.047 (0.042)} & 0.047 (0.034) & 0.071 (0.070) \\
\quad $n=1000$ & 0.013 (0.011) & \textbf{0.012 (0.011)} & 0.012 (0.007) & 0.015 (0.014) \\
\quad $n=5000$ & 0.003 (0.003) & 0.003 (0.003) & \textbf{0.003 (0.002)} & 0.003 (0.003) \\
\midrule
\multicolumn{1}{l}{\rule[-.10ex]{0pt}{3.25ex}\!${Y\sim \Gamma\big(3,1/\sqrt{3}\big)}$, $Z\sim N\!\left(0,\frac{1}{5}\right)\!$} &  &  &  &  \\
\quad $n=200$ & 0.054 (0.047) & 0.049 (0.042) & \textbf{0.038 (0.037)} & 0.060 (0.058) \\
\quad $n=1000$ & 0.014 (0.012) & 0.013 (0.010) & \textbf{0.008 (0.007)} & 0.014 (0.013) \\
\quad $n=5000$ & 0.005 (0.004) & 0.005 (0.003) & \textbf{0.003 (0.002)} & 0.005 (0.004) \\
\multicolumn{1}{l}{\rule[-.10ex]{0pt}{3.25ex}\!${Y\sim \Gamma\big(3,1/\sqrt{3}\big)}$, $Z\sim \Gamma\!\left(2,\frac{1}{5\sqrt{2}}\right)\!$} &  &  &  &  \\
\quad $n=200$ & 0.054 (0.047) & 0.047 (0.040) & \textbf{0.035 (0.034)} & 0.058 (0.057) \\
\quad $n=1000$ & 0.013 (0.012) & 0.013 (0.011) & \textbf{0.009 (0.008)} & 0.014 (0.013) \\
\quad $n=5000$ & 0.003 (0.003) & 0.003 (0.003) & \textbf{0.003 (0.002)} & 0.003 (0.003) \\
\multicolumn{1}{l}{\rule[-.10ex]{0pt}{3.25ex}\!${Y\sim \Gamma\big(3,1/\sqrt{3}\big)}$, $Z\sim \mathrm{S}\Gamma\!\left(2,\frac{1}{5\sqrt{2}}\right)\!$} &  &  &  &  \\
\quad $n=200$ & 0.052 (0.047) & 0.044 (0.038) & \textbf{0.034 (0.033)} & 0.067 (0.066) \\
\quad $n=1000$ & 0.013 (0.011) & 0.013 (0.010) & \textbf{0.009 (0.007)} & 0.015 (0.014) \\
\quad $n=5000$ & 0.004 (0.003) & 0.004 (0.003) & \textbf{0.003 (0.002)} & 0.004 (0.004) \\
\bottomrule
\end{tabular}%

    }
    
    \tabnote{Boldface identifies the estimator with the smallest empirical MSE in each row.
    Results grouped by $Y$ and $Z$ distribution. For each $Y$ design, empirical MSE and variance are averaged over the evaluation points then multiplied by 100.
    }
\end{table}
}

{
\setlength{\tabcolsep}{12pt}
\begin{table}[H]
    \caption{Monte Carlo MSE$\times 100$ and variance$\times 100$ (in parentheses) for estimators of $F_Y(x,y)$ for the non-smooth $F_Y$ designs.}\label{tab:FYxy_app_nonsmth}
    \adjustbox{max width=\textwidth, center = \textwidth}{ 
    \footnotesize
    % Requires \usepackage{booktabs}
% Displayed MSE x 100; displayed variance x 100.
% The difference between MSE and variance is squared bias.
% Boldface marks the smallest unrounded MSE in each row.
\begin{tabular}{lcccc}
\toprule
&
$\FYhatCV(x,y)$ &
$\FYhatAR(x,y)$ &
$\FDaRe(x,y)$ &
$\FHaLa(x,y)$ \\
\midrule
\multicolumn{1}{l}{\rule[-.10ex]{0pt}{3.25ex}\!${Y\sim K_1}$, $Z\sim N\!\left(0,\frac{1}{5}\right)\!$} &  &  &  &  \\
\quad $n=200$ & 0.172 (0.093) & 0.159 (0.070) & 0.207 (0.082) & \textbf{0.132 (0.088)} \\
\quad $n=1000$ & 0.071 (0.021) & 0.091 (0.016) & 0.098 (0.018) & \textbf{0.065 (0.019)} \\
\quad $n=5000$ & \textbf{0.041 (0.007)} & 0.062 (0.004) & 0.053 (0.004) & 0.042 (0.004) \\
\multicolumn{1}{l}{\rule[-.10ex]{0pt}{3.25ex}\!${Y\sim K_1}$, $Z\sim \Gamma\!\left(2,\frac{1}{5\sqrt{2}}\right)\!$} &  &  &  &  \\
\quad $n=200$ & 0.183 (0.100) & 0.179 (0.071) & 0.232 (0.084) & \textbf{0.138 (0.090)} \\
\quad $n=1000$ & \textbf{0.052 (0.021)} & 0.069 (0.015) & 0.094 (0.017) & 0.052 (0.017) \\
\quad $n=5000$ & \textbf{0.020 (0.005)} & 0.034 (0.004) & 0.055 (0.004) & 0.030 (0.004) \\
\multicolumn{1}{l}{\rule[-.10ex]{0pt}{3.25ex}\!${Y\sim K_1}$, $Z\sim \mathrm{S}\Gamma\!\left(2,\frac{1}{5\sqrt{2}}\right)\!$} &  &  &  &  \\
\quad $n=200$ & 0.190 (0.109) & 0.186 (0.074) & 0.241 (0.087) & \textbf{0.135 (0.101)} \\
\quad $n=1000$ & 0.054 (0.025) & 0.073 (0.017) & 0.098 (0.018) & \textbf{0.046 (0.021)} \\
\quad $n=5000$ & \textbf{0.020 (0.005)} & 0.034 (0.004) & 0.056 (0.003) & 0.023 (0.004) \\
\midrule
\multicolumn{1}{l}{\rule[-.10ex]{0pt}{3.25ex}\!${Y\sim K_2}$, $Z\sim N\!\left(0,\frac{1}{5}\right)\!$} &  &  &  &  \\
\quad $n=200$ & 0.312 (0.109) & 0.241 (0.084) & 0.266 (0.085) & \textbf{0.212 (0.106)} \\
\quad $n=1000$ & 0.151 (0.031) & 0.156 (0.022) & 0.160 (0.021) & \textbf{0.134 (0.025)} \\
\quad $n=5000$ & \textbf{0.092 (0.010)} & 0.117 (0.005) & 0.122 (0.004) & 0.100 (0.005) \\
\multicolumn{1}{l}{\rule[-.10ex]{0pt}{3.25ex}\!${Y\sim K_2}$, $Z\sim \Gamma\!\left(2,\frac{1}{5\sqrt{2}}\right)\!$} &  &  &  &  \\
\quad $n=200$ & 0.310 (0.125) & 0.248 (0.083) & 0.265 (0.088) & \textbf{0.209 (0.103)} \\
\quad $n=1000$ & 0.122 (0.030) & 0.128 (0.020) & 0.155 (0.018) & \textbf{0.117 (0.022)} \\
\quad $n=5000$ & \textbf{0.060 (0.008)} & 0.072 (0.005) & 0.122 (0.004) & 0.086 (0.005) \\
\multicolumn{1}{l}{\rule[-.10ex]{0pt}{3.25ex}\!${Y\sim K_2}$, $Z\sim \mathrm{S}\Gamma\!\left(2,\frac{1}{5\sqrt{2}}\right)\!$} &  &  &  &  \\
\quad $n=200$ & 0.315 (0.141) & 0.259 (0.095) & 0.278 (0.099) & \textbf{0.206 (0.134)} \\
\quad $n=1000$ & 0.112 (0.035) & 0.126 (0.021) & 0.158 (0.019) & \textbf{0.093 (0.026)} \\
\quad $n=5000$ & \textbf{0.045 (0.009)} & 0.073 (0.005) & 0.122 (0.004) & 0.066 (0.005) \\
\midrule
\multicolumn{1}{l}{\rule[-.10ex]{0pt}{3.25ex}\!${Y\sim K_3}$, $Z\sim N\!\left(0,\frac{1}{5}\right)\!$} &  &  &  &  \\
\quad $n=200$ & \textbf{0.474 (0.147)} & 0.780 (0.067) & 0.696 (0.088) & 0.477 (0.092) \\
\quad $n=1000$ & \textbf{0.317 (0.047)} & 0.563 (0.015) & 0.445 (0.017) & 0.365 (0.019) \\
\quad $n=5000$ & \textbf{0.281 (0.007)} & 0.451 (0.003) & 0.351 (0.003) & 0.290 (0.005) \\
\multicolumn{1}{l}{\rule[-.10ex]{0pt}{3.25ex}\!${Y\sim K_3}$, $Z\sim \Gamma\!\left(2,\frac{1}{5\sqrt{2}}\right)\!$} &  &  &  &  \\
\quad $n=200$ & \textbf{0.358 (0.124)} & 0.815 (0.059) & 0.710 (0.080) & 0.434 (0.082) \\
\quad $n=1000$ & \textbf{0.144 (0.029)} & 0.434 (0.014) & 0.447 (0.016) & 0.285 (0.017) \\
\quad $n=5000$ & \textbf{0.077 (0.007)} & 0.239 (0.004) & 0.353 (0.003) & 0.194 (0.004) \\
\multicolumn{1}{l}{\rule[-.10ex]{0pt}{3.25ex}\!${Y\sim K_3}$, $Z\sim \mathrm{S}\Gamma\!\left(2,\frac{1}{5\sqrt{2}}\right)\!$} &  &  &  &  \\
\quad $n=200$ & 0.363 (0.134) & 0.828 (0.063) & 0.722 (0.085) & \textbf{0.349 (0.100)} \\
\quad $n=1000$ & \textbf{0.138 (0.033)} & 0.430 (0.015) & 0.443 (0.017) & 0.208 (0.020) \\
\quad $n=5000$ & \textbf{0.070 (0.009)} & 0.238 (0.004) & 0.351 (0.003) & 0.141 (0.004) \\
\bottomrule
\end{tabular}%

    }
    
    \tabnote{Boldface identifies the estimator with the smallest empirical MSE in each row.
    Results grouped by $Y$ and $Z$ distribution. For each $Y$ design, empirical MSE and variance are averaged over the evaluation points then multiplied by 100.
    }
\end{table}
}

\newpage
%%%%%%%%%%%%%%%%%%%%%%%
\subsubsection{$p_x$ simulations}
% \label{app:FYxSims}
%%%%%%%%%%%%%%%%%%%%%%%

{
\setlength{\tabcolsep}{16pt}
\begin{table}[H]
    \caption{
        Sensitivity to the GL tuning coefficients for estimating $p_x$.
        % Monte Carlo MSE$\times 100$ and variance$\times 100$ (in parentheses) for estimators of jumps $p_x$. 
        }\label{tab:jump_GL_sensitivity}
    \adjustbox{max width=\textwidth, center = \textwidth}{ 
    \footnotesize
    % Requires \usepackage{booktabs}
% Displayed MSE x 100; displayed variance x 100.
\begin{tabular}{llccc}
\toprule
& $(\kappa_A,\kappa_V)$ & $p_x=0.01$ & $p_x=0.10$ & $p_x=0.20$ \\
\midrule
\multicolumn{1}{l}{\rule[-2.2ex]{0pt}{4.5ex}\!$Z\!\sim\! N\!\left(0,\frac{1}{5}\right)\!$} &  &  &  &  \\
\quad $n=200$ & $(.05,.25)^{*}$ & 2.626 (0.127) & 2.358 (0.155) & 1.944 (0.189) \\
 & $(.05,.15)$ & 2.577 (0.131) & 2.312 (0.161) & 1.909 (0.196) \\
 & $(.025,.25)$ & \textbf{2.432 (0.128)} & \textbf{2.165 (0.158)} & \textbf{1.790 (0.192)} \\
\addlinespace[1.0ex]
\quad $n=1000$ & $(.05,.25)^{*}$ & 1.737 (0.038) & 1.525 (0.051) & 1.250 (0.059) \\
 & $(.05,.15)$ & 1.691 (0.041) & 1.490 (0.054) & 1.220 (0.063) \\
 & $(.025,.25)$ & \textbf{1.659 (0.040)} & \textbf{1.450 (0.052)} & \textbf{1.188 (0.061)} \\
\addlinespace[1.0ex]
\quad $n=5000$ & $(.05,.25)^{*}$ & 1.235 (0.019) & 1.089 (0.022) & 0.896 (0.023) \\
 & $(.05,.15)$ & \textbf{1.201 (0.020)} & 1.055 (0.023) & 0.869 (0.025) \\
 & $(.025,.25)$ & 1.202 (0.020) & \textbf{1.054 (0.022)} & \textbf{0.866 (0.024)} \\
\midrule
\multicolumn{1}{l}{\rule[-2.2ex]{0pt}{4.5ex}\!$Z\!\sim\! \Gamma\!\left(2,\frac{1}{5\sqrt{2}}\right)\!$} &  &  &  &  \\
\quad $n=200$ & $(.05,.25)^{*}$ & 0.922 (0.199) & 0.871 (0.300) & 0.852 (0.440) \\
 & $(.05,.15)$ & 0.866 (0.211) & 0.835 (0.334) & \textbf{0.848 (0.492)} \\
 & $(.025,.25)$ & \textbf{0.809 (0.223)} & \textbf{0.811 (0.363)} & 0.858 (0.520) \\
\addlinespace[1.0ex]
\quad $n=1000$ & $(.05,.25)^{*}$ & 0.479 (0.068) & 0.425 (0.095) & 0.371 (0.131) \\
 & $(.05,.15)$ & 0.445 (0.074) & 0.397 (0.105) & \textbf{0.353 (0.149)} \\
 & $(.025,.25)$ & \textbf{0.409 (0.079)} & \textbf{0.375 (0.114)} & 0.354 (0.164) \\
\addlinespace[1.0ex]
\quad $n=5000$ & $(.05,.25)^{*}$ & 0.274 (0.032) & 0.239 (0.041) & 0.202 (0.047) \\
 & $(.05,.15)$ & 0.256 (0.033) & 0.226 (0.048) & 0.195 (0.056) \\
 & $(.025,.25)$ & \textbf{0.233 (0.037)} & \textbf{0.210 (0.051)} & \textbf{0.185 (0.058)} \\
\midrule
\multicolumn{1}{l}{\rule[-2.2ex]{0pt}{4.5ex}\!$Z\!\sim\! \mathrm{S}\Gamma\!\left(2,\frac{1}{5\sqrt{2}}\right)\!$} &  &  &  &  \\
\quad $n=200$ & $(.05,.25)^{*}$ & 0.830 (0.225) & 0.967 (0.484) & \textbf{1.132 (0.793)} \\
 & $(.05,.15)$ & 0.778 (0.235) & \textbf{0.961 (0.544)} & 1.177 (0.895) \\
 & $(.025,.25)$ & \textbf{0.727 (0.250)} & 0.972 (0.580) & 1.256 (0.967) \\
\addlinespace[1.0ex]
\quad $n=1000$ & $(.05,.25)^{*}$ & 0.482 (0.101) & 0.472 (0.190) & \textbf{0.544 (0.353)} \\
 & $(.05,.15)$ & 0.453 (0.106) & \textbf{0.455 (0.207)} & 0.566 (0.405) \\
 & $(.025,.25)$ & \textbf{0.433 (0.116)} & 0.468 (0.231) & 0.609 (0.442) \\
\addlinespace[1.0ex]
\quad $n=5000$ & $(.05,.25)^{*}$ & 0.264 (0.037) & 0.261 (0.089) & \textbf{0.279 (0.174)} \\
 & $(.05,.15)$ & 0.247 (0.040) & \textbf{0.250 (0.098)} & 0.286 (0.200) \\
 & $(.025,.25)$ & \textbf{0.227 (0.047)} & 0.261 (0.117) & 0.308 (0.221) \\
\bottomrule
\end{tabular}%

    }
    
    \tabnote{Entries report Monte Carlo MSE$\times100$, and variance$\times100$ in parentheses. 
        Boldface identifies the GL specification with the smallest unrounded empirical MSE within each measurement-error distribution, sample size, and jump-size column.
        The superscript $*$ identifies the specification reported in Tables \ref{tab:jump_main}, and \ref{tab:jump_appendix}. 
        For each measurement-error distribution,
        jump size, and sample size, entries are averaged over the three jump
        designs $J_1,J_2,J_3$, with the $J_3$ entries first averaged over the jump
        locations $x=-2,0,2$. 
        % For $Y\sim J_1$ and $Y\sim J_2$, the jump is estimated at $x=0$. 
        % For $Y\sim J_3$, MSE and variance are each averaged over the three jump locations
        % $x=-2,0,2$. 
% Boldface identifies the estimator with the smallest empirical MSE in each row.
        % For
        % $Y\sim J_1$ and $Y\sim J_2$, the jump is estimated at $x=0$. For
        % $Y\sim J_3$, RMSE is computed by first averaging MSE over the three jump locations
        % $x=-2,0,2$ then taking the square root. Variance entries are averaged
        % over the same jump locations and multiplied by 10.
        }
\end{table}
}
 % \tabnote{Boldface identifies the estimator with
 %        the smallest unrounded empirical MSE in each row.
 %    For each measurement-error distribution,
 %    jump size, and sample size, entries are averaged over the three jump
 %    designs $J_1,J_2,J_3$, with the $J_3$ entries first averaged over the jump
 %    locations $x=-2,0,2$. 
 %    % and the square root is then taken. Variance entries are
 %    % averaged over the same designs and jump locations. 
 %    Full DGP-by-DGP results
 %    are reported in Table \ref{tab:jump_appendix} in Appendix~\ref{app:Full_sim_results}.
 %    }

{
\begin{table}[H]
    \caption{ 
        Monte Carlo MSE$\times 100$ and variance$\times 100$ (in parentheses) for estimators of jumps $p_x$. 
        }\label{tab:jump_appendix}
    \adjustbox{max width=\textwidth, center = \textwidth}{ 
    \footnotesize
    % Requires \usepackage{booktabs}
% Displayed MSE x 100; displayed variance x 100.
\begin{tabular}{lcccccc}
\toprule
& \multicolumn{2}{c}{$p_x = 0.01$} & \multicolumn{2}{c}{$p_x = 0.10$} & \multicolumn{2}{c}{$p_x = 0.20$} \\
\cmidrule(lr){2-3}
\cmidrule(lr){4-5}
\cmidrule(lr){6-7}
& $\pxhatGL$ & $\pxhatAR$ & $\pxhatGL$ & $\pxhatAR$ & $\pxhatGL$ & $\pxhatAR$ \\
\midrule
\multicolumn{1}{l}{\rule[-.10ex]{0pt}{3.25ex}\!$Y\sim J_1$, $Z\!\sim\! N\!\left(0,\frac{1}{5}\right)\!$} &  &  &  &  &  &  \\
\quad $n=200$ & \textbf{4.389 (0.191)} & 5.439 (0.122) & \textbf{4.272 (0.209)} & 5.123 (0.142) & \textbf{3.826 (0.253)} & 4.467 (0.157) \\
\quad $n=1000$ & \textbf{3.013 (0.058)} & 4.167 (0.026) & \textbf{2.887 (0.075)} & 3.920 (0.031) & \textbf{2.593 (0.088)} & 3.438 (0.034) \\
\quad $n=5000$ & \textbf{2.148 (0.034)} & 3.306 (0.006) & \textbf{2.086 (0.037)} & 3.139 (0.008) & \textbf{1.890 (0.039)} & 2.764 (0.008) \\
\multicolumn{1}{l}{\rule[-.10ex]{0pt}{3.25ex}\!$Y\sim J_1$, $Z\!\sim\! \Gamma\!\left(2,\frac{1}{5\sqrt{2}}\right)\!$} &  &  &  &  &  &  \\
\quad $n=200$ & \textbf{1.437 (0.374)} & 1.556 (1.001) & \textbf{1.463 (0.550)} & 2.239 (1.923) & \textbf{1.544 (0.843)} & 2.916 (2.715) \\
\quad $n=1000$ & \textbf{0.732 (0.125)} & 0.748 (0.488) & \textbf{0.700 (0.179)} & 1.019 (0.919) & \textbf{0.671 (0.253)} & 1.307 (1.229) \\
\quad $n=5000$ & 0.432 (0.066) & \textbf{0.397 (0.256)} & \textbf{0.412 (0.086)} & 0.574 (0.512) & \textbf{0.381 (0.095)} & 0.650 (0.612) \\
\multicolumn{1}{l}{\rule[-.10ex]{0pt}{3.25ex}\!$Y\sim J_1$, $Z\!\sim\! \mathrm{S}\Gamma\!\left(2,\frac{1}{5\sqrt{2}}\right)\!$} &  &  &  &  &  &  \\
\quad $n=200$ & \textbf{1.243 (0.356)} & 1.384 (0.905) & \textbf{1.480 (0.711)} & 2.418 (2.179) & \textbf{1.704 (1.077)} & 3.657 (3.424) \\
\quad $n=1000$ & \textbf{0.778 (0.179)} & 0.920 (0.582) & \textbf{0.775 (0.298)} & 1.260 (1.102) & \textbf{0.879 (0.517)} & 1.926 (1.820) \\
\quad $n=5000$ & \textbf{0.419 (0.063)} & 0.434 (0.284) & \textbf{0.444 (0.148)} & 0.817 (0.725) & \textbf{0.472 (0.264)} & 1.190 (1.153) \\
\midrule
\multicolumn{1}{l}{\rule[-.10ex]{0pt}{3.25ex}\!$Y\sim J_2$, $Z\!\sim\! N\!\left(0,\frac{1}{5}\right)\!$} &  &  &  &  &  &  \\
\quad $n=200$ & \textbf{1.856 (0.106)} & 1.859 (0.080) & 1.786 (0.146) & \textbf{1.696 (0.109)} & 1.518 (0.182) & \textbf{1.348 (0.129)} \\
\quad $n=1000$ & \textbf{1.182 (0.032)} & 1.399 (0.017) & \textbf{1.102 (0.047)} & 1.263 (0.025) & \textbf{0.917 (0.056)} & 1.001 (0.030) \\
\quad $n=5000$ & \textbf{0.809 (0.013)} & 1.095 (0.004) & \textbf{0.763 (0.017)} & 1.001 (0.006) & \textbf{0.638 (0.020)} & 0.796 (0.007) \\
\multicolumn{1}{l}{\rule[-.10ex]{0pt}{3.25ex}\!$Y\sim J_2$, $Z\!\sim\! \Gamma\!\left(2,\frac{1}{5\sqrt{2}}\right)\!$} &  &  &  &  &  &  \\
\quad $n=200$ & 0.780 (0.089) & \textbf{0.364 (0.249)} & \textbf{0.720 (0.174)} & 0.896 (0.833) & \textbf{0.691 (0.275)} & 1.354 (1.322) \\
\quad $n=1000$ & 0.408 (0.027) & \textbf{0.167 (0.115)} & 0.374 (0.048) & \textbf{0.357 (0.341)} & \textbf{0.314 (0.080)} & 0.603 (0.591) \\
\quad $n=5000$ & 0.231 (0.010) & \textbf{0.084 (0.060)} & 0.203 (0.018) & \textbf{0.169 (0.156)} & \textbf{0.172 (0.030)} & 0.234 (0.229) \\
\multicolumn{1}{l}{\rule[-.10ex]{0pt}{3.25ex}\!$Y\sim J_2$, $Z\!\sim\! \mathrm{S}\Gamma\!\left(2,\frac{1}{5\sqrt{2}}\right)\!$} &  &  &  &  &  &  \\
\quad $n=200$ & \textbf{0.669 (0.161)} & 0.922 (0.634) & \textbf{0.927 (0.472)} & 1.950 (1.796) & \textbf{1.077 (0.747)} & 3.258 (3.104) \\
\quad $n=1000$ & \textbf{0.377 (0.066)} & 0.520 (0.352) & \textbf{0.398 (0.154)} & 0.907 (0.851) & \textbf{0.470 (0.305)} & 1.478 (1.447) \\
\quad $n=5000$ & 0.216 (0.024) & \textbf{0.202 (0.147)} & \textbf{0.223 (0.078)} & 0.559 (0.531) & \textbf{0.255 (0.179)} & 0.987 (0.985) \\
\midrule
\multicolumn{1}{l}{\rule[-.10ex]{0pt}{3.25ex}\!$Y\sim J_3$, $Z\!\sim\! N\!\left(0,\frac{1}{5}\right)\!$} &  &  &  &  &  &  \\
\quad $n=200$ & \textbf{1.632 (0.086)} & 1.823 (0.063) & \textbf{1.016 (0.110)} & 1.019 (0.088) & 0.489 (0.133) & \textbf{0.444 (0.120)} \\
\quad $n=1000$ & \textbf{1.017 (0.025)} & 1.331 (0.012) & \textbf{0.586 (0.031)} & 0.685 (0.019) & 0.241 (0.034) & \textbf{0.230 (0.026)} \\
\quad $n=5000$ & \textbf{0.748 (0.011)} & 1.082 (0.003) & \textbf{0.418 (0.011)} & 0.572 (0.004) & \textbf{0.160 (0.011)} & 0.193 (0.006) \\
\multicolumn{1}{l}{\rule[-.10ex]{0pt}{3.25ex}\!$Y\sim J_3$, $Z\!\sim\! \Gamma\!\left(2,\frac{1}{5\sqrt{2}}\right)\!$} &  &  &  &  &  &  \\
\quad $n=200$ & 0.550 (0.134) & \textbf{0.549 (0.382)} & \textbf{0.430 (0.178)} & 0.913 (0.865) & \textbf{0.321 (0.201)} & 1.292 (1.289) \\
\quad $n=1000$ & \textbf{0.298 (0.053)} & 0.319 (0.207) & \textbf{0.201 (0.058)} & 0.466 (0.441) & \textbf{0.127 (0.061)} & 0.538 (0.534) \\
\quad $n=5000$ & 0.158 (0.021) & \textbf{0.152 (0.108)} & \textbf{0.103 (0.019)} & 0.205 (0.200) & \textbf{0.054 (0.014)} & 0.228 (0.225) \\
\multicolumn{1}{l}{\rule[-.10ex]{0pt}{3.25ex}\!$Y\sim J_3$, $Z\!\sim\! \mathrm{S}\Gamma\!\left(2,\frac{1}{5\sqrt{2}}\right)\!$} &  &  &  &  &  &  \\
\quad $n=200$ & \textbf{0.578 (0.157)} & 0.796 (0.552) & \textbf{0.495 (0.270)} & 1.481 (1.381) & \textbf{0.617 (0.554)} & 2.575 (2.559) \\
\quad $n=1000$ & \textbf{0.290 (0.059)} & 0.336 (0.236) & \textbf{0.241 (0.117)} & 0.842 (0.802) & \textbf{0.283 (0.237)} & 1.466 (1.457) \\
\quad $n=5000$ & \textbf{0.157 (0.022)} & 0.167 (0.119) & \textbf{0.116 (0.041)} & 0.455 (0.443) & \textbf{0.108 (0.081)} & 0.769 (0.765) \\
\bottomrule
\end{tabular}%

    }
    
    \tabnote{Boldface identifies the estimator with the smallest empirical MSE within each measurement-error distribution, sample size, and jump-size column.
        For
        $Y\sim J_1$ and $Y\sim J_2$, the jump is estimated at $x=0$. For
        $Y\sim J_3$, MSE and variance are each averaged over the three jump locations
        $x=-2,0,2$. 
        % then taking the square root. Variance entries are averaged
        % over the same jump locations and multiplied by 10.
        }
\end{table}
}

% %%%%%%%%%%%%%%%%%%%%%%%%
% \subsubsection{$F_Y(x,y)$ simulations}\label{app:FYxySims}
% %%%%%%%%%%%%%%%%%%%%%%%%

\end{appendices}
\clearpage

\setlength{\baselineskip}{12pt}
\bibliographystyle{elsart-harv}
\bibliography{library2}

\end{document}